\documentclass[12pt]{article}
\usepackage{graphicx}
\usepackage{amssymb}
\usepackage{latexsym}
\usepackage[mathscr]{euscript}

\usepackage[all]{xypic}

\newtheorem{prop}{Proposition}[section]
\newtheorem{rema}{Remark}[section]
\newtheorem{defi}{Definition}[section]

\newtheorem{theo}{Theorem}[section]

\newcommand{\iN}{\hbox{ {\leaders\hrule\hskip.2cm}{\vrule height .22cm} }}

\newcommand{\I}{\mathbb{I}}
\newcommand{\A}{\mathbb{A}}
\newcommand{\N}{\mathbb{N}}
\newcommand{\AI}{\Bbb{A}\!\Bbb{I}}

\begin{document}
\title{An introduction to
supermanifolds and supersymmetry}
\author{Fr{\'e}d{\'e}ric H{\'e}lein\\
Institut de Math{\'e}matiques de Jussieu, UMR 7586\\
Universit{\'e} de Paris,\\
B{\^a}timent Sophie Germain, 8 place Aur{\'e}lie Nemours\\
75013 Paris, France}

\maketitle
\textbf{Abstract}
\emph{The aim of these notes (which were partially covered in lectures given at the Peyresq Summer School on 17--22 June, 2002) is to give an introduction to some mathematical aspects of supersymmetry. Some (hopefully) original point of view are added, by using and developing some results in https://arxiv.org/abs/math-ph/0603045, where maps of supermanifolds are analyzed in details and, also, in the presentation of the super Minkowski spaces in dimensions 3, 4, 6 and 10.}\\

\noindent
The aim of these notes (which were partially presented in a talk given at the Peyresq Summer School on 17--22 June, 2002) is to give an introduction to some mathematical aspects of supersymmetry. Although supersymmetry has a greater and greater impact on mathematics today, this subject remains not very well-known among mathematicians, despite the fact that several mathematical theories has been built for three decades (and many good texts are available now). I believe that, beside the technical complexity of the subject, this is due to the psychological difficulty to feel confortable with a manifold with `odd', i.e. anticommutating coordinates. Hence one of my motivations in writing this text was precisely to try to remove some of these psychological obstacles and I hope to help the interested reader in this direction. Most notions expounded here are standard and can be found in many other references (see for instance \cite{DEF,DeligneMorgan,Freed,Rogers07}). However I also added some (hopefully) original point of view, by using and developping some results in \cite{helein}, where maps of supermanifolds are analyzed in details and also in the presentation of the super Minkowski spaces in dimensions 3, 4, 6 and 10 in Section \ref{superpoincare}. Two different theories exists (see Section \ref{firstdraft}) and I choosed to follow the one inspired by algebraic geometry (which goes back to F. Berezin). I introduce some terminology, speaking of `skeletal' supermanifolds and of supermanifolds `with flesh' (D. Freed writes in \cite{Freed} `with fuzz') for pedagogical reasons, since we find both kinds of animals in the litterature under the same name. During the maturation and the preparation of theses notes, I have been very much inspired by \cite{DeligneMorgan,Freed} and I beneficiated from several discusssions with D. Bennequin.

\noindent

\section{Introduction}
\subsection{Motivations}
Supergeometry was invented by physicists and mathematicians at the beginning of the seventies as a mathematical framework for the notion of \emph{supersymmetry}. This notion merged out from two questions:
\begin{enumerate}
\item Does there exists a symmetry which exchanges Bosons and Fermions ?
\item Are they non trivial extensions of the Poincar{\'e} group, the symmetry group of Special Relativity ?
\end{enumerate}
It could be tempting a priori to answer `no' to both questions. Indeed first it is recognized for a long time that, in the theory of quantum fields, Bosons are created and annihilated by commutative operators, whereas to Fermions correspond anticommuting operators, so that it is not straightforward to figure out how these two families of operators could be exchanged. Second a `no go' theorem of S. Coleman and J. Mandula \cite{coleman} answers negatively to the second question. Note also that a connection between the two questions can be expected, since Bosons and Fermions can also be distinguished by the fact that their fields are transformed in two different ways by the action of the Lorentz group.

However it turns out that, in some enlarged conceptual framework, the answers to both questions are positive and rely on the same ideas. The key was the introduction by A. Salam and J. Strathdee of \emph{anticommuting} coordinate variables (also called \emph{Grassmann} variables), which one could heuristically think as classical limits of Fermionic operators. This leads to a theory of \emph{superspaces} or \emph{supermanifolds}, proposed by J. Wess and B. Zumino and developped by F. Berezin, B. DeWitt, A. Rogers in which the `classical limit' of the symmetry between Bosons and Fermions admits a geometric formulation. In this framework one can also extend the Lie algebra of the Poincar{\'e} group into a super Lie algebra, as done by Y.A. Gol'fand and E.P. Likhtman \cite{golfand} and D. Volkov and V. Akulov \cite{volkov}. The mathematical theory of Lie super algebra and Lie super groups was then constructed by V. Kac \cite{kac}. 

\subsection{Supermanifolds}\label{defiprem}
\subsubsection{A first draft}\label{firstdraft}
We first need a mathematical definition of supermanifolds. Actually several theories are possible, depending on the point of view we start with and on the applications we have in mind. We hence have different, sometimes non equivalent, notions of supermanifolds which however all share the same underlying idea. The  two main approaches are:
\begin{enumerate}
\item either we define supermanifolds indirectly by defining their (super)algebras of functions. This point of view, inspired by algebraic geometry, was developped by F. Berezin, D. Le{\u\i}tes, Y. Manin.
\item or we define supermanifolds as sets with a suitable topology. This point of view, inspired by the differential geometry, was introduced by B. DeWitt and A. Rogers (see section \ref{dewittrogers}).
\end{enumerate}
In both cases the definition of a supermanifold can be decomposed in two steps: first one defines the analogues $\mathbb{R}^{m|k}$ (for $m,k\in \mathbb{N}$) of $\mathbb{R}^m$ and their open subsets and, second, we fix the rules for gluing together open subsets of $\mathbb{R}^{m|k}$ to get a supermanifold of dimension $m|k$. However variants exist in both approaches. Here we will follow mainly the first point of view. We start by defining a basic version of it, that we propose to call `\emph{skeletal supermanifolds}'.

\begin{defi}[Skeletal $\mathbb{R}^{m|k}$]\label{skeletal} The skeletal super space $\mathbb{R}^{m|k}$ is a `geometric object' indirectly defined through its ring $(\mathcal{C}^\infty(\mathbb{R}^{m|k}),+,\cdot)$ of smooth functions. This ring is spanned algebraically over functions in\footnote{This can be rephrased by saying that $\mathcal{C}^\infty(\mathbb{R}^{m|k})$ is a \emph{module} over $\mathcal{C}^\infty(\mathbb{R}^m)$.} $\mathcal{C}^\infty(\mathbb{R}^m)$ by $k$ generators $\theta^1,\cdots ,\theta^k \in \mathcal{C}^\infty(\mathbb{R}^{m|k})$ which satisfy
\begin{equation}\label{supercommute}
\forall i,j,\quad \theta^i\theta^j + \theta^j\theta^i = 0
\end{equation}
and such that 
\begin{equation}\label{freealgebra}
\theta^1\cdots \theta^k\neq 0.
\end{equation}
In other words, $\mathcal{C}^\infty(\mathbb{R}^{m|k}):= \mathcal{C}^\infty(\mathbb{R}^m)[\theta^1,\cdots ,\theta^k]$, i.e. any smooth function $F\in \mathcal{C}^\infty(W)$ can be written in an unique way as the polynomial
\[
F = \sum_{j=0}^k\sum_{1\leq i_1<\cdots <i_j\leq k}f_{i_1\cdots i_j}\theta^{i_1} \cdots \theta^{i_j},
\]
where each coefficient $f_{i_1\cdots i_j}$ belongs to $\mathcal{C}^\infty(\mathbb{R}^m)$.

Similarly an \textbf{open subset} $\Omega$ of $\mathbb{R}^{m|k}$ is defined through its ring of functions $\mathcal{C}^\infty(\Omega)$ which is defined to be $\mathcal{C}^\infty(|\Omega|)[\theta^1,\cdots
,\theta^k]$, where $|\Omega|$ is an open subset of $\mathbb{R}^m$.
\end{defi}
Functions $\theta^i$ are the \emph{odd coordinate functions} on $\mathbb{R}^{m|k}$. Note that (\ref{supercommute}) implies in particular that $(\theta^i)^2 = 0$, $\forall i$, i.e. each $\theta^i$ is nilpotent. We say that the even dimension of $\mathbb{R}^{m|k}$ is $m$ and its odd dimension is $k$ or, equivalentely, that the dimension of $\mathbb{R}^{m|k}$ is $m|k$. One can summarize the properties of $\mathcal{C}^\infty(\mathbb{R}^{m|k})$ by introducing the following algebraic notions.

\begin{defi}[super vector space] A real \emph{super vector space} $V$ is a real $\mathbb{Z}_2$-graded vector space, i.e. a vector space over $\mathbb{R}$ which admits a decomposition $V = V^0\oplus V^1$. Vectors in $V^0$ are called \emph{even}, vectors in $V^1$ are called \emph{odd}.

If $x\in V^0\cup V^1$, its \emph{parity} or its \emph{grading} $\hbox{gr}(x)\in \mathbb{Z}_2$ is $0$ if $x\in V^0$ and $1$ if $x\in V^1$.
\end{defi}

\begin{defi}[Super algebra] A real \emph{super algebra} $A$ is a real super vector space $A= A^0\oplus A^1$ endowed with an associative bilinear multiplication\footnote{One can alternatively view the multiplication law as a linear map from $A\otimes A$ to $A$.} law $A\times A\longrightarrow A$ which maps $A^a\times A^b$ to $A^c$, for all $a,b,c \in \{0,1\}$ such that $a+b = c$ mod 2.

A super algebra is \textbf{super commutative} if $\forall a,b \in\{0,1\}$, $\forall f\in A^a$, $\forall g\in A^b$, $fg = (-1)^{ab}gf$.
\end{defi}
We see immediately that $\mathcal{C}^\infty(\mathbb{R}^{m|k})$ is a super algebra, and that $\mathcal{C}^\infty(\mathbb{R}^{m|k})^0$ (respectively $\mathcal{C}^\infty(\mathbb{R}^{m|k})^1$) is the subspace of even (respectively odd) polynomials in the $\theta^i$'s. Relation (\ref{supercommute}) means that this super algebra is super commutative\footnote{Moreover (\ref{freealgebra}) garantees that it is the \emph{free} super commutative super algebra spanned by $\theta^1,\cdots, \theta^k$.}. Examples of super algebras are the exterior algebra (or the Grassmann algebra) $\Lambda^*V$ of a vector space $V$ or the set of sections of the exterior bundle $\Lambda^*E$ of a vector bundle over a manifold $\mathcal{M}$ (actually the latter example is very close to the notion of a supermanifold, see below).

With the definition of open subsets of $\mathbb{R}^{m|k}$ at hand, one can define a \emph{skeletal supermanifold} $\mathcal{M}$ of dimension $m|k$ as follows: it is an ordinary manifold $|\mathcal{M}|$ of dimension $m$ endowed with a sheaf of super algebras $\mathcal{O}_\mathcal{M}$ (the sheaf of functions) over $|\mathcal{M}|$ such that, for any open subset $|U|$ of $|\mathcal{M}|$ with a chart $|U|\longrightarrow |\Omega|\subset \mathbb{R}^m$, the superalgebra $\Gamma(|U|,\mathcal{O}_{|U|})$ of sections of $\mathcal{O}_\mathcal{M}$ over $|U|$ is isomorphic to some $\mathcal{C}^\infty(\Omega) = \mathcal{C}^\infty(|\Omega|)[\theta^1,\cdots
,\theta^k]$, where $\Omega$ is an open subset of $\mathbb{R}^{m|k}$. Note that the algebra $\mathcal{C}^\infty(|U|)\simeq \mathcal{C}^\infty(|\Omega|)$ of smooth functions on the ordinary open subset $|U|$ coincides with $\mathcal{C}^\infty(\Omega)/ \mathcal{J}_\Omega$, where $\mathcal{J}_\Omega$ is the ideal of nilpotent elements of $\mathcal{C}^\infty(\Omega)$. This allows to recover the sheaf of smooth functions $\mathcal{O}_{|\mathcal{M}|}$ from $\mathcal{O}_{\mathcal{M}}$ and hence the manifold $|\mathcal{M}|$. We shall not developp this description further, since most of the examples studied in this text requires only to understand $\mathbb{R}^{m|k}$ (see \cite{Manin} for details).

Skeletal supermanifolds $\mathcal{M}$ of dimension $m|k$ can be constructed starting from a rank $k$ real vector bundle $E$ over an $m$-dimensional ordinary manifold $|\mathcal{M}|$. We just set that the sheaf of functions on $\mathcal{M}$ is isomorphic to the set $\Gamma(|\mathcal{M}|,\Lambda^*E)$ of smooth sections of the exterior algebras bundle $\Lambda^*E$ over $|\mathcal{M}|$. We denote by $S(|\mathcal{M}|,E) = \mathcal{M}$ this supermanifold and call it a \emph{split supermanifold} (see e.g. \cite{Rogers07}). A theorem by M. Batchelor \cite{batchelor79} asserts that any supermanifold has a split structure.

We chosed the terminology `skeletal' because, as we shall see in Section \ref{toy}, if we use naively Definition \ref{skeletal} for studying physical models we run relatively quickly into inconsistencies. Hence we will need to refine this notion and to put `flesh' on the skeletal super manifolds. However (in the algebro-geometric theory) a super manifold with flesh can be completely recovered by knowing its skeletal part.

\subsubsection{Related algebraic notions}\label{otheralgebras}

\noindent \textbf{Clifford algebras} ---
Let $V$ be a real vector space endowed with a symmetric bilinear form $B(\cdot,\cdot)$ and let us define the \emph{Clifford algebra} $C(V)$ associated to $(V,B(\cdot,\cdot))$. We consider the free tensor algebra $\bigotimes V:= \bigoplus_{j=0}^\infty V^{\otimes j}$ (where $V^{\otimes 0}:= \mathbb{R}$) and the ideal $\mathcal{I}$ in $\bigotimes V$ spanned by $\{x\otimes y + y\otimes x + 2B(x,y)|\,x,y\in V\}$. We let $C(V)$ to be the quotient of $\bigotimes V$ by $\mathcal{I}$. If $(e_1,\cdots ,e_k)$ is a basis of $V$, then as a vector space, $C(V)$ is spanned over $\mathbb{R}$ by all products $e_{i_1}\cdots e_{i_j}$, where $0\leq j\leq k$ and $1\leq i_1<\cdots <i_j\leq k$, where we write $xy = x\otimes y$ for shortness (with the convention that, for $j=0$, $e_{i_1}\cdots e_{i_j} = 1$). As an algebra, $C(V)$ is spanned by $1, e_1,\cdots ,e_k$ assuming the conditions $e_ie_j + e_je_i = -2B(e_i,e_j)$. If for instance $B=0$, $C(V)$ is then nothing but the Grassmann algebra $\mathbb{R}[e_1,\cdots ,e_k]$. A Clifford algebra is naturally a super algebra with $C(V) = C(V)^0 \oplus C(V)^1$, where $C(V)^0$ (respectively $C(V)^1$) is spanned by products $e_{i_1}\cdots e_{i_j}$ of an even (respectively odd) number of factors. It is however not super commutative, unless $B=0$.

A simple example is the following: let $\mathbb{L}$ be a real Euclidean vector space of dimension 1 (i.e. isomorphic to... $\mathbb{R}$) and note $\epsilon$ a normed basis of $\mathbb{L}$ (i.e. $\langle \epsilon,\epsilon\rangle = 1$). As a vector space, the Clifford algebra $C(\mathbb{L})$ is isomorphic to $\{ a+b\epsilon|\, a,b\in \mathbb{R}\}$ and, because of the relation $\epsilon^2 =-1$, we have the algebra isomorphism $C(\mathbb{L}) \simeq \mathbb{C}$. It looks like a complicated way to define $\mathbb{C}$, however $C(\mathbb{L})$ comes with a natural grading $C(\mathbb{L}) = C(\mathbb{L})^0 \oplus C(\mathbb{L})^1$, where $C(\mathbb{L})^0 = \mathbb{R}$ and $C(\mathbb{L})^1 = \mathbb{R}\epsilon$ (i.e. $\epsilon$ is odd). We shall use this construction in Section \ref{superpoincare}.

Clifford algebras play an important role in physical applications since, roughly speaking, the quantization of (observable) functions on an odd supermanifold are elements of a Clifford algebra (the Hilbert space of quantum states being a spinorial representation of the Clifford representation). A Clifford algebra can so be seen as a deformation quantization of a super commutative super algebra described to first order by the bilinear form $B$ (which can be interpreted as a Poisson bracket, see \cite{KostantSternberg}).\\

\noindent \textbf{Lie super algebras} ---
A super Lie algebra $\mathfrak{g}$ is a super vector space $\mathfrak{g} = \mathfrak{g}^0 \oplus \mathfrak{g}^1$ endowed with a bilinear product law $[\cdot,\cdot]: \mathfrak{g}\times \mathfrak{g}\longrightarrow \mathfrak{g}$ called \emph{super Lie bracket} which satisfies the following properties. First it is \emph{super skewsymmetric}: $\forall x,y\in \mathfrak{g}^0 \cup \mathfrak{g}^1$,
\begin{equation}\label{Lieskew}
[x,y] + (-1)^{\hbox{gr}(x)\hbox{gr}(y)}[y,x] = 0,
\end{equation}
and second it satisfies a \emph{super Jacobi identity}: $\forall x,y,z\in \mathfrak{g}^0 \cup \mathfrak{g}^1$
\begin{equation}\label{Liejacobi}
(-1)^{\hbox{gr}(x)\hbox{gr}(z)}[x,[y,z]] +
(-1)^{\hbox{gr}(y)\hbox{gr}(x)}[y,[z,x]] +
(-1)^{\hbox{gr}(z)\hbox{gr}(y)}[z,[x,y]]  = 0.
\end{equation}
In particular a super Lie algebra is not a super algebra since it is not associative (the latter property is replaced by the super Jacobi identity). A convenient way to take into account the sign rules in these identities consists in tensoring $\mathfrak{g}$ with a Grassmann algebra $\Lambda_L = \mathbb{R}[\eta^1,\cdots ,\eta^L]$ (and assuming that the odd variables $\eta^i$ supercommute with elements in $\mathfrak{g}$): for any $x,y,z\in \mathfrak{g}^0 \cup \mathfrak{g}^1$, we choose `coefficients' $\alpha,\beta, \gamma\in \Lambda_L$ such that $\alpha x$, $\beta y$ and $\gamma z$ are all even (equivalentely $\hbox{gr}(\alpha) = \hbox{gr}(x)$, etc.). Then (\ref{Lieskew}) and (\ref{Liejacobi}) are equivalent to the usual relations $[\alpha x,\beta y] + [\beta y,\alpha x] = 0$ and $[\alpha x,[\beta y,\gamma z]] + [\beta y ,[ \gamma z,\alpha x]] + [ \gamma z,[\alpha x ,\beta y]] = 0$ of a Lie algebra (see \S \ref{supertime} for an example). In other words $\left(\Lambda_L\otimes \mathfrak{g}\right)^0 = \left(\Lambda_L^0\otimes \mathfrak{g}^0\right) \oplus \left(\Lambda_L^1\otimes \mathfrak{g}^1\right)$ is a Lie algebra.

Alternatively the super Jacobi identity (\ref{Liejacobi}) can be interpreted as follows:
\begin{enumerate}
\item relation (\ref{Lieskew}) for $x,y\in \mathfrak{g}^0$ and (\ref{Liejacobi}) for $x,y,z\in \mathfrak{g}^0$ means that $(\mathfrak{g}^0,[\cdot,\cdot])$ is a Lie algebra;
\item relation (\ref{Liejacobi}) for $x,y\in \mathfrak{g}^0$ and $z\in \mathfrak{g}^1$ means that, $\forall x\in \mathfrak{g}^0$, the linear map
\[
\begin{array}{cccc}
\hbox{ad}_x: & \mathfrak{g}^1 & \longrightarrow & \mathfrak{g}^1\\
& z & \longmapsto & [x,z]
\end{array}
\]
satisfies $\hbox{ad}_{[x,y]}(z) = \hbox{ad}_x\left(\hbox{ad}_y(z)\right) - \hbox{ad}_y\left(\hbox{ad}_x(z)\right)$, i.e. $\mathfrak{g}^0\longrightarrow \hbox{End}(\mathfrak{g}^1)$, $x\longmapsto \hbox{ad}_x$ is a representation of $\mathfrak{g}^0$;
\item relation (\ref{Liejacobi}) for $x\in \mathfrak{g}^0$ and $y,z\in \mathfrak{g}^1$ means that the (symmetric) bilinear map
 \[
\begin{array}{ccc}
\mathfrak{g}^1\times \mathfrak{g}^1  & \longrightarrow & \mathfrak{g}^0\\
(y,z) & \longmapsto & [y,z]
\end{array}
\]
is $\mathfrak{g}^0$-equivariant (hence the adjoint representation of $\mathfrak{g}^0$ on $\mathfrak{g}^1$ is somehow a `square root' of the adjoint representation of $\mathfrak{g}^0$ on $\mathfrak{g}^0$);
\item relation (\ref{Liejacobi}) for $x,y,z\in \mathfrak{g}^1$ is equivalent to $[x,[x,x]] = 0$, $\forall x\in \mathfrak{g}^1$.
\end{enumerate}
The classical Lie algebras are of course examples of super Lie algebra (with $\mathfrak{g}^1 = \{0\}$). Some non trivial Lie algebras come from differential Geometry. For instance consider an usual manifold $\mathcal{M}$ and let $\Omega^*(\mathcal{M})$ be the space of smooth sections of the exterior bundle $\Lambda^*T^*\mathcal{M}$ of the cotangent bundle $T^*\mathcal{M}$ (i.e. smooth exterior differential forms on $\mathcal{M}$). Let $\xi$ be a smooth vector field on $\mathcal{M}$. Then there exists three natural operators acting on $\Omega^*(\mathcal{M})$: the exterior differential $d$, the interior product by $\xi$, $\iota_\xi$, and the Lie derivative $\hbox{Lie}_\xi$. The space $\mathfrak{g}$ spanned by these three operators is a Lie super algebra, where $\mathfrak{g}^0$ is spanned by $\hbox{Lie}_\xi$ and $\mathfrak{g}^1$ is spanned by $d$ and  $\iota_\xi$. This is a way to summarize the commutation relations $[\hbox{Lie}_\xi,d] = [\hbox{Lie}_\xi,\iota_\xi] = 0$, $[d,d] = 2d d = 0$, $[\iota_\xi,\iota_\xi] = 2\iota_\xi\iota_\xi = 0$ and the well-known Cartan identity $[d, \iota_\xi] = d\iota_\xi + \iota_\xi d = \hbox{Lie}_\xi$, an anti-commutator. Another example is the space of multivectors (i.e. sections of the exterior product of the tangent bundle $T\mathcal{M}$) equipped with the Schouten bracket (a generalization of the Lie bracket of vector fields), see e.g. \cite{Tulczyjew}. Then the parity of a multivector field is $1$ plus its degree mod 2.

Super Lie algebra play also an important role in supersymmetric theories since they precisely encode the supersymmetry of a given physical model. Among the most interesting superalgebras are the extensions of the (trivial) Lie algebra of translations of the Minkowski space-time (this is then the super Lie algebra of super translations of space-time) or of the Poincar{\'e} Lie algebra (called the super Poincar{\'e} Lie algebra). We present examples of such Lie super algebras in Section \ref{superpoincare}. Lie super algebras are naturally associated with Lie super groups, which are particular examples of supermanifolds. However a Lie super algebra as defined here is a skeletal object, which needs some extra `flesh' for the construction of the corresponding Lie group (what we would call a Lie algebra with flesh corresponds actually to a \emph{super Lie module}, see \cite{Rogers07}). 

We refer to \cite{Freed,Rogers07} for quickly accessible expositions of the theory of super Lie algebras and to \cite{DeligneMorgan,Deligne,vara,kac} for more details. 

\section{The superparticle: a first overview}\label{toy}
Recall that supersymmetry was introduced as a hypothetical 
symmetry of a quantum field theory which would exchange Bosons with
Fermions. However most quantum field 
descriptions of a particle are connected with a classical field
description, i.e.\,a variational problem. So we can expect that the supersymmetry can be read at this classical level as a symmetry group which acts on solutions of the Euler--Lagrange equations of some variational problem: we shall see that this is indeed true.

\subsection{The model}

A simple example is a superparticle evolving in a Riemannian manifold
$\mathcal{N}$. A first (and relatively rough) description of this system
is through a pair $(x,\psi)$, where 
\begin{itemize}
\item $x$ is a smooth map from $\mathbb{R}$ to $\mathcal{N}$
\item $\psi$ is a section of the pull-back bundle $x^*\left(\Pi T\mathcal{M}\right)$, i.e.\,for all 
$t\in \mathbb{R}$, $\psi(t)$ belongs to $\Pi T_{x(t)}\mathcal{N}$. Here $\Pi
      T_{x(t)}\mathcal{N}$ is the space modelled on 
$T_{x(t)}\mathcal{N}$ with opposite parity, i.e.\,odd. This means that
      the set of smooth functions on 
$\Pi T_{x(t)}\mathcal{N}$ is $\mathbb{R}[\alpha^1,\cdots ,\alpha^n]\simeq
      \Lambda^*T_{x(t)}^*\mathcal{N}$, 
where $(\alpha^1,\cdots ,\alpha^n)$ is a basis of $T_{x(t)}^*\mathcal{N}$.
\end{itemize}
In other words $(x,\psi)$ can be understood as a map from $\mathbb{R}$ to the split manifold $S(\mathcal{N}, T^*\mathcal{N})$ (up to some difficulties discussed at the end of \S \ref{example1}).

Here we have used the notation $\Pi T_{x(t)}\mathcal{N}$, where $\Pi$ is
a vector space functor which reverses the parity (odd versus
even) of a given $\mathbb{Z}_2$ graded vector space. This notation is useful because of its concision but one should keep in mind that $\Pi$ is never a
super algebra morphism. 

The Lagrangian for the superparticle is
\[
\mathcal{L}[x,\psi]:= \int_{\mathbb{R}}\left( {1\over 2}|\dot{x}|^2 +
{1\over 2}\langle \psi,\nabla_{\dot{x}}\psi\rangle \right) dt,
\]
where, if $\psi(t) = \psi^i(t){\partial \over \partial x^i}$,
$\nabla_{\dot{x}}\psi(t) = \left( {d\psi^i\over dt}(t) +
  \Gamma^i_{jk}(x(t))\dot{x}^j(t)\psi^k(t)\right) 
{\partial \over \partial x^i}$ (the $\Gamma^i_{jk}$'s being
Christoffel's symbols). We already see here 
that the $\psi^i$'s should be anticommuting quantities: if not then we
would have 
${1\over 2}\langle \psi,\nabla_{\dot{x}}\psi\rangle = {1\over
  4}{d\over dt}\langle \psi,\psi\rangle$,  
so that the second term in the Lagrangian would be unuseful.

The Euler--Lagrange system of equations is
\begin{equation}\label{EL}
\left\{ \begin{array}{ccl}
\nabla_{\dot{x}}\dot{x} & = & {1\over 2}R(\psi,\psi)\dot{x}\\
\nabla_{\dot{x}}\psi & = & 0.
\end{array}\right.
\end{equation}
Again $R(\psi,\psi)\dot{x} =
{R(x)_{ijk}}^l\psi^i\psi^j\dot{x}^k{\partial \over \partial x^l}$ 
does not necessarily vanish precisely because $\psi^i$ and $\psi^j$
anticommute. When contemplating this Euler--Lagrange system we see a
difficulty due to our too naive approach. Indeed in the
first Euler--Lagrange equation 
$\ddot{x}^l + \Gamma^l_{jk}(x)\dot{x}^j\dot{x}^k = {1\over
2}{R(x)_{ijk}}^l\psi^i\psi^j\dot{x}^k$, the left hand side is real valued whereas the right hand side contains the bilinear quantity $\psi^i\psi^j$. However $\psi^i\psi^j$ cannot be real (although it is even) because it is nilpotent (it squares to zero). Hence
\begin{rema}\label{firstinconsistency}
The coordinates $x^i$ of the map $x$ cannot be real valued because of (\ref{EL}).
\end{rema}

The action $\mathcal{L}$ enjoys a particular symmetry (which corresponds actually to a \emph{supersymmetry}) as follows: we consider the transformation $1 - \eta\tau_Q$ acting on the field $(x,\psi)$ defined by:
\begin{equation}\label{1.transfo}
\left\{ \begin{array}{ccl}
x & \longmapsto & x - \eta \psi\\
\psi & \longmapsto & \psi + \eta \dot{x}.
\end{array}\right.
\end{equation}
Here an important point is that $\eta$ which plays the role of an infinitesimal parameter is odd. Otherwise this transformation would
exchange odd fields with even ones. Note however that $x$ cannot not be mapped into an ordinary map by (\ref{1.transfo}) because the components of $\eta\psi$ cannot be real; hence this confirms the conclusion claimed in Remark \ref{firstinconsistency}. The `vector' $(- \eta\psi(t),\eta \dot{x}(t))$ can be interpreted as a vector tangent to $\Pi T\mathcal{N}$ to $(x(t),\psi(t))$. Let us compute the effect of this infinitesimal transformation on the action $\mathcal{L}$: we need to compute the term $\eta\tau_Q\iN \delta\mathcal{L}(x,\psi)$ in the development $\mathcal{L}(x - \eta \psi, \psi + \eta \dot{x}) = \mathcal{L}(x,\psi)
- \eta\tau_Q\iN \delta\mathcal{L}(x,\psi)$. Observing that (\ref{1.transfo})
implies that 
\[ 
\left\{ \begin{array}{ccl}
\dot{x} & \longmapsto & \dot{x} - \eta \nabla_{\dot{x}}\psi\\
\nabla_{\dot{x}}\psi & \longmapsto & \nabla_{\dot{x}}\psi + \eta
\nabla_{\dot{x}}\dot{x}, 
\end{array}\right.
\]
we obtain that
\[
\begin{array}{ccl}
\eta\tau_Q\iN \delta\mathcal{L}(x,\psi) & = & \displaystyle 
\int_{\mathbb{R}}\left( \left\langle \dot{x},\eta
    \nabla_{\dot{x}}\psi\right\rangle 
+ {1\over 2} \left\langle - \eta\dot{x},\nabla_{\dot{x}}\psi\right\rangle
+ {1\over 2} \left\langle
  \psi,-\eta\nabla_{\dot{x}}\dot{x}\right\rangle\right) dt \\
 & = & \displaystyle {1\over 2} \eta \int_{\mathbb{R}}{d\over dt}
\left\langle \psi,\dot{x}\right\rangle dt \\
 & = & \displaystyle {1\over 2} \eta\left[ \left\langle
\psi(+\infty),\dot{x}(+\infty)\right\rangle
- \left\langle \psi(-\infty),\dot{x}(-\infty)\right\rangle \right].
\end{array}
\]
This quantity does not vanish in general, but is exact. So the transformation (\ref{1.transfo}) should formally change infinitesimally a solution of the Euler--Lagrange equations into another one\footnote{The question whether this infinitesimal transformation can be exponentiated is however more delicate and should be done in terms of a map defined on $\mathbb{R}^{1|1}$, see \cite{Rogers07}}. 

This symmetry gives rise to a conserved quantity (a version of Noether theorem): let us introduce a smooth compactly supported function $\chi\in \mathcal{C}^\infty_0(\mathbb{R})$ and, instead of (\ref{1.transfo}), apply to $(x,\psi)$ the infinitesimal transformation
\begin{equation}\label{1.transfobis}
\left\{ \begin{array}{ccl}
x & \longmapsto & x - \eta \chi\psi\\
\psi & \longmapsto & \psi + \eta \chi\dot{x},
\end{array}\right.
\end{equation}
with
\[
\left\{ \begin{array}{ccl}
\dot{x} & \longmapsto & \dot{x} - \eta
\left(\chi\nabla_{\dot{x}}\psi+\dot{\chi}\psi\right) \\
\nabla_{\dot{x}}\psi & \longmapsto & \nabla_{\dot{x}}\psi + \eta
\left(\chi\nabla_{\dot{x}}\dot{x} 
+ \dot{\chi}\dot{x}\right).
\end{array}\right.
\]
Then we have
\[
\begin{array}{ccl}
(\chi\eta\tau_Q)\iN \delta\mathcal{L}(x,\psi) & = & \displaystyle 
\int_{\mathbb{R}}\chi\left( \left\langle \dot{x},\eta
    \nabla_{\dot{x}}\psi\right\rangle 
+ {1\over 2} \left\langle -\eta\dot{x},\nabla_{\dot{x}}\psi\right\rangle
+ {1\over 2} \left\langle
  \psi,-\eta\nabla_{\dot{x}}\dot{x}\right\rangle\right) dt \\
 & &\displaystyle \ + \int_{\mathbb{R}}\left(\left\langle \dot{x},\eta
     \dot{\chi}\psi\right\rangle 
+ {1\over 2}\left\langle \psi, -\eta \dot{\chi}\dot{x}\right\rangle\right) dt\\
 & = & \displaystyle {1\over 2} \eta \int_{\mathbb{R}}\chi
{d\over dt}\left\langle \psi,\dot{x}\right\rangle dt 
- {3\over 2}\eta\int_{\mathbb{R}}\chi {d\over dt}\left\langle
  \psi,\dot{x}\right\rangle dt \\
 & = & \displaystyle - \eta\int_{\mathbb{R}}\chi {d\over dt}\left\langle
   \psi,\dot{x}\right\rangle dt  . 
\end{array}
\]
In all these computations we have used that $\eta$, $\psi^i$ and
$\dot{\psi}^i$ anticommute. 
We hence conclude that the quantity $\left\langle
\psi,\dot{x}\right\rangle$ is conserved, i.e. Noether's theorem holds here\footnote{in a way similar to the fact that the time translation symmetry
$t\longmapsto t+\varepsilon$ induces the fields transformation $(x,\psi)\longmapsto (x - \varepsilon \dot{x},\psi - \varepsilon
\nabla_{\dot{x}}\psi)$ which is related, through Noether's theorem, to the conservation of the energy ${1\over 2}|\dot{x}|^2 + {1\over 2} \left\langle \psi, \nabla_{\dot{x}}\psi\right\rangle$.}, hence the symmetry which generates our fields transformation should have a geometrical interpretation. Eventually this `Noether's charge' has an interesting physical meaning, since the corresponding quantum operator is nothing but the Dirac operator on $\mathcal{N}$ (see \cite{Alvarez,Freed,Rogers07,Witten}).

Also we can interpret the transformation acting on the
set of fields 
\[
Q:= -\eta\tau_Q: (x,\psi) \longmapsto (-\eta\psi, \eta\dot{x}) 
\]
as a tangent vector field to the infinite dimensional set of
fields $(x,\psi)$: 
\[
Q(x,\psi) = \int_{\mathbb{R}}\eta \left( -\psi(t){\partial \over \partial
    x(t)} + \dot{x}(t){\partial \over \partial \psi(t)}\right) dt 
\]
It is interesting to choose two different anticommuting variables
$\eta_1$ and $\eta_2$ and to compute the commutator of the `vector fields'
$Q_1:= -\eta_1\tau_Q$ and $Q_2:= -\eta_2\tau_Q$. This can be done by computing
the action of $Q_1\circ Q_2$:
\[
\begin{array}{ccccc}
& Q_2 & & Q_1 & \\
\left( \begin{array}{c}x\\ \psi\end{array}\right)
& \longmapsto & \left( \begin{array}{c}-\eta_2\psi\\
    \eta_2\dot{x}\end{array}\right) 
& \longmapsto & \left( \begin{array}{c}-\eta_1(\eta_2\dot{x})\\
\eta_1(-\eta_2\nabla_{\dot{x}}\psi) \end{array}\right)
= -\eta_1\eta_2\left( \begin{array}{c}\dot{x}\\
    \nabla_{\dot{x}}\psi\end{array}\right), 
\end{array}
\]
and the action of $Q_2\circ Q_1$ (vice-versa). Then by using the fact that the Levi--Civita connection $\nabla$ is torsion-free, we can write that
$[Q_1,Q_2] (x,\psi) = Q_1\left( Q_2(x,\psi)\right) - Q_2\left(
  Q_1(x,\psi)\right)$ and deduce that
\[
[\eta_1\tau_Q,\eta_2\tau_Q](x,\psi) = [Q_1,Q_2](x,\psi) =
- 2\eta_1\eta_2\nabla_{\dot{x}}(x,\psi). 
\]
Factoring out by $\eta_1$ and $\eta_2$ (and taking into account the fact that the odd numbers $\eta_i$ anticommute also with the odd operator $\tau_Q$) we deduce that 
\[
[\tau_Q,\tau_Q](x,\psi) = 2\nabla_{\dot{x}}(x,\psi).
\]
Here the notation $[\tau_Q,\tau_Q]$ does not recover a commutator, but
an anticommutator ($[\tau_Q,\tau_Q] = 2\tau_Q\tau_Q$). We conclude that the action of $\tau_Q$ on fields should be related somehow to
an infinitesimal geometric symmetry which behaves like the square root of the time translation generator $2{d\over dt}$.\\

\subsection{The superfield formulation}
As we shall see in more details in Section \ref{particlebis} there is an alternative, more elegant way to picture geometrically this symmetry by viewing the multiplet $(x,\psi)$ as the components of a single \emph{superfield} $\Phi = x + \theta\psi$, from the supertime $\mathbb{R}^{1|1}$ on which the ring of functions is $\mathcal{C}^\infty(\mathbb{R})[\theta]$ (i.e.\,$\theta$ is here an odd coordinate) to $\mathcal{N}$. Then the Lagrangian action can be written as 
$\mathcal{L}[\Phi]:=
-{1\over 2} \int \int_{\mathbb{R}^{1|1}}dt\,d\theta \left\langle D\Phi,{\partial \Phi\over \partial t}\right\rangle$, 
where $D:= {\partial \over\partial \theta} - \theta{\partial \over
  \partial t}$. 
And the transformation (\ref{1.transfo}) results from an infinitesimal
translation in $\mathbb{R}^{1|1}$ by the vector $\eta\tau_Q$, where $\tau_Q:= {\partial
\over\partial \theta} +\theta{\partial \over \partial t}$. However another limitation of our temporary definition of a supermanifold is that:
\begin{rema}\label{secondinconsistency}
The superfield $\Phi$ does not belong to $\mathcal{C}^\infty(\mathbb{R}^{1|1},\mathcal{N})$ according to the definition of maps in $\mathcal{C}^\infty(\mathbb{R}^{1|1})$, since $\psi$ is odd.
\end{rema}

\subsection{Conclusion}
We have seen a first example of a supersymmetric model, an opportunity to test the concept of a skeletal supermanifold, and some points must be clarified:
\begin{enumerate}
\item to give a sense to (\ref{EL}) and (\ref{1.transfobis}), which requires in our example to change the mathematical definition of $x$ and also symmetrically of $\psi$ (Remark \ref{firstinconsistency});
\item to extend the definition of a map on a supermanifold in such a way that the superfield $\Phi = x + \theta \psi$ makes sense (Remark \ref{secondinconsistency});
\item a last question, not yet discussed: can we figure out what is a point in a supermanifold ?
\end{enumerate}
Let us first consider the second point: one of the more natural ideas which comes in mind to answer it would be to assume that the components $\psi^i$ of $\psi$ are not $\mathbb{R}$-valued but takes values in, say $\mathbb{R}[\eta_1]$, where $\eta_1$ is an odd parameter which is different from $\theta$ (to ensure that $\theta\eta_1 \neq 0$). However this would imply that the right hand side of the first equation of (\ref{EL}) is ${1\over 2}R(\psi,\psi)\dot{x} =0$, because $\eta_1\eta_1 = 0$. This would solve the first point, but in a stupid way and the superparticle equation would have little interest since the equation on $x$ would not be coupled to $\psi$. Hence a refined solution is to assume that the components of $\psi$ take values in, say $\mathbb{R}[\eta_1,\eta_2]$, where $\theta, \eta_1,\eta_2$ are three different odd parameters. Then the right hand side of (\ref{EL}) would be proportional to $\eta_1\eta_2$. This means that the components of $x$ should not be real but with values in $\mathbb{R}[\eta_1\eta_2]$.

Now we observe that there are no a priori reason for using only two odd parameters $\eta_1,\eta_2$. Hence we can answer the two first questions by assuming that all the fields $x$ and $\psi$ are $\mathbb{R}[\eta_1,\cdots ,\eta_L]$-valued, where $L$ is arbitrary, possibly infinite and that $x$ is an even polynomial on $(\eta_1,\cdots ,\eta_L)$, whereas $\psi$ is an odd polynomial. There are different ways to implement this idea. We here privilege the point of view inspired by algebraic geometry, i.e. to view any odd parameter as a coordinate function on some `space'. Hence the ring $\mathbb{R}[\eta_1,\cdots ,\eta_L]$ has to be thought as the ring of functions on some space $\mathbb{R}^{0|L}$ and the field $\Phi = x + \theta\psi$ as a map on $\mathbb{R}^{1|1}\times \mathbb{R}^{0|L}\simeq \mathbb{R}^{1|1+L}$.  Then the conditions that $x$ (respectively $\psi$) is an even (respectively odd) function of the $\eta^i$'s can be encapsulated by requiring that $\Phi$ is an \emph{even} function of $(\theta,\eta^1,\cdots ,\eta^L)$.

All that show that we need a definition of the set $\mathcal{C}^\infty(\mathbb{R}^{1|1}\times \mathbb{R}^{0|L},\mathcal{N})^0$ of \emph{even} maps from $\mathbb{R}^{1|1}\times \mathbb{R}^{0|L}$ to $\mathcal{N}$. It should be the set of $\Phi$ such that, for any local coordinate $y:\mathcal{N}\supset U\longrightarrow \mathbb{R}$, one can make sense of $y\circ \Phi$ and such that $y\circ \Phi\in \mathcal{C}^\infty(\mathbb{R}^{1|1}\times \mathbb{R}^{0|L},\mathbb{R})^0$. A precise statement of this property will be given in \S \ref{morphismproperty}.

More generally the mathematical description of a $\sigma$-model\footnote{This includes the important case of superstrings, where $\mathcal{M}$ is a supermanifold of even dimension 2.} on a supermanifold $\mathcal{M}$, i.e. a \emph{field} defined on $\mathcal{M}$ and with values on a (possibly super) manifold $\mathcal{N}$ will be obtained through an even smooth map from $\mathcal{M}\times \mathbb{R}^{0|L}$ to $\mathcal{N}$, where the choice of the value of $L$ has to be precised and its meaning can be interpreted in different ways.

Hence we are led to another definition of a supermanifold, that we propose to call `supermanifolds with flesh', see \S \ref{newdef}.

This change of paradigm changes also the content of the last question, which relies on our intuitive picture of a supermanifold. We invite the Reader to build his own vision, after reading the next sections (see also \cite{Freed,helein}).

\subsection{A new definition}\label{newdef}
\begin{defi}\label{fleshy}
A \emph{supermanifold with flesh} is a product $\mathcal{M}^{m|k}\times \mathbb{R}^{0|L}$, where $\mathcal{M}^{m|k}$ is a skeletal supermanifold and $L\in \mathbb{N}$ is `sufficiently large'.

A \emph{map with flesh} on the supermanifold $\mathcal{M}^{m|k}$ and with values in an ordinary manifold $\mathcal{N}^{n|\ell}$ is an \emph{even} smooth map from $\mathcal{M}^{m|k}\times \mathbb{R}^{0|L}$ to $\mathcal{N}^{n|\ell}$.
\end{defi}
We postpone the precise definition of what is an \emph{even} smooth map from $\mathcal{M}^{m|k}\times \mathbb{R}^{0|L}$ to $\mathcal{N}^{n|\ell}$ to the next paragraph (\ref{morphismproperty}). Moreover we stayed vague about the the precise value of $L$ and there are several ways to implement its role.
\begin{itemize}
\item Either one let $L$ to be infinite and we precise some topology: this point of view is connected to the theory of $G^\infty$-functions and $G^\infty$-manifolds developped by B. DeWitt and A. Rogers and which will be briefly expounded in Section \ref{dewittrogers}. 
\item Or one looks for the minimal value of $L$ which is needed in order to make the computations consistent (we need at least $L>k$): this is connected to the theory of $GH^\infty$-functions (an intermediate between the $G^\infty$ and $H^\infty$ theories, see Section \ref{dewittrogers}), see \cite{rogers86}. 
\item Or we let the value of $L$ to be free and arbitrary: it means that a map with flesh $\Phi: \mathcal{M}^{m|k}\longrightarrow
\mathcal{N}^{n|\ell}$ represents the collection of all even maps $\Phi:\mathcal{M}^{m|k}\times \mathbb{R}^{0|L}\longrightarrow \mathcal{N}^{n|\ell}$, for $L\in \mathbb{N}$. Equivalentely, as explained in the solution of the Fall Problem 2 in \cite{FP2}, a `map' $\Phi:\mathcal{M}^{m|k}\longrightarrow \mathcal{N}^{n|\ell}$ should be considered as a functor from the category of odd vector spaces $\mathbb{R}^{0|L}$ (or more generally superspaces) to even skeletal maps 
$\Phi:\mathcal{M}^{m|k}\times \mathbb{R}^{0|L}\longrightarrow \mathcal{N}^{n|\ell}$.
\end{itemize}
The strategy to understand maps with flesh between supermanifolds is then based on two steps:
\begin{enumerate}
\item first define smooth maps (and in particular even ones) between skeletal supermanifolds. This is the subject of the next Section;
\item then specialize the preceding analysis to maps on the skeletal super manifold $\mathcal{M}^{m|k}\times \mathbb{R}^{0|L}$ and deduce a description of maps with flesh on $\mathcal{M}^{m|k}$.
\end{enumerate}
Note that we may adapt Definition \ref{fleshy} of a map between supermanifolds to sections of fiber bundles or connections.

\section{Maps between skeletal supermanifolds}
In this section we define maps between skeletal supermanifolds $\mathcal{M}^{m|q}$ and $\mathcal{N}^{n|\ell}$ (of respective dimensions $m|q$ and $n|\ell$ and where, in the following applications, we may have $q=k+L$ and $\mathcal{M}^{m|q} = \widetilde{\mathcal{M}}^{m|k}\times \mathbb{R}^{0|L}$). The strategy (as in \cite{Freed,DeligneMorgan}) is the following: rather than defining a map $\Phi: \mathcal{M}^{m|q}\longrightarrow \mathcal{N}^{n|\ell}$ directly we set how it acts on functions by the pull-back (or composition) operation $f\longmapsto \Phi^*f = f\circ \Phi$. We will essentially use the fact that $\Phi^*$ should be a morphism between the two superalgebras $\mathcal{C}^\infty(\mathcal{M}^{m|q},\mathbb{R})$ and $\mathcal{C}^\infty(\mathcal{N}^{n|\ell},\mathbb{R})$.

\subsection{The morphism property}\label{morphismproperty}
Given two skeletal supermanifolds $\mathcal{M}^{m|q}$ and $\mathcal{N}^{n|\ell}$, a smooth map $\Phi: \mathcal{M}^{m|q}\longrightarrow \mathcal{N}^{n|\ell}$, is defined by duality through the induced morphism of algebras 
\[
\begin{array}{cccc}
\Phi^*: & \left(\mathcal{C}^\infty (\mathcal{N}^{n|\ell},\mathbb{R}),+,.\right) &
\longrightarrow & \left(\mathcal{C}^\infty (\mathcal{M}^{m|q},\mathbb{R}),+,.\right)\\
& A & \longmapsto & \Phi^*A,\end{array}
\]
where we can think secretly that $\Phi^*A = A\circ \Phi$. Recall also that we denote by $\mathcal{C}^\infty (\mathcal{M}^{m|q},\mathbb{R})^0$ and $\mathcal{C}^\infty (\mathcal{N}^{n|\ell},\mathbb{R})^0$ the subalgebras of even elements of respectively $\mathcal{C}^\infty (\mathcal{M}^{m|q},\mathbb{R})$ and $\mathcal{C}^\infty (\mathcal{N}^{n|\ell},\mathbb{R})$. Then we will say that $\Phi^*$ is an \emph{even morphism} (or equivalentely that $\Phi:\mathcal{M}^{m|q}\longrightarrow \mathcal{N}^{n|\ell}$ is an even map) if $\Phi^*$ maps $\mathcal{C}^\infty (\mathcal{N}^{n|\ell},\mathbb{R})^0$ to $\mathcal{C}^\infty (\mathcal{M}^{m|q},\mathbb{R})^0$ and maps $\mathcal{C}^\infty (\mathcal{N}^{n|\ell},\mathbb{R})^1$ to $\mathcal{C}^\infty (\mathcal{M}^{m|q},\mathbb{R})^1$.

The important thing is to check the morphism property, i.e.\,that
\begin{equation}\label{1.trivial}
\Phi^*(1) = 1,\quad\hbox{where 1 is the constant equal to one (the unit
  in algebraic terms)} 
\end{equation}
\begin{equation}\label{1.somme}
\forall \lambda,\mu\in \mathbb{R},\forall A, B\in \mathcal{C}^\infty
(\mathcal{N}^{n|\ell},\mathbb{R}),\quad 
\Phi^*(\lambda A + \mu B) = \lambda \Phi^*A + \mu \Phi^*B
\end{equation}
and
\begin{equation}\label{1.produit}
\forall A, B\in \mathcal{C}^\infty (\mathcal{N}^{n|\ell},\mathbb{R}),\quad
\Phi^*(AB) = (\Phi^*A)(\Phi^*B).
\end{equation}
These conditions impose severe constraints \cite{Freed}. In the
following we analyze their consequences through some examples.

\subsection{Some examples}
The goal of this paragraph is to experiment maps from a skeletal supermanifold according to the previous definition.

\subsubsection{Maps $\Phi:\mathbb{R}^{m|0}\longrightarrow \mathbb{R}^{0|\ell}$}\label{example1}
(From an \textbf{even} vector space to an \textbf{odd} vector space.)
This amounts to look at morphisms $\Phi^*$ from $\mathbb{R}[\theta^1,\cdots
,\theta^\ell]$ to $\mathcal{C}^\infty (\mathbb{R}^m)$. Any function $F\in
\mathbb{R}[\theta^1,\cdots ,\theta^\ell]$ writes
\[
F = F_\emptyset + \sum_{i=1}^\ell F_i\theta^i + {1\over 2} \sum_{i_1,i_2=1}^\ell F_{i_1i_2}\theta^{i_1}\theta^{i_2} + \cdots = \sum_{j=0}^\ell\sum_{1\leq i_1,\cdots ,i_j\leq \ell}F_{i_1\cdots i_j}\theta^{i_1} \cdots \theta^{i_j},
\] 
where the $F_{i_1\cdots i_j}$'s are real constants. Thus, because of
(\ref{1.somme}), it suffices to characterize all pull-back images $\Phi^*(\theta^{i_1} \cdots \theta^{i_j})$. For $j=0$, (\ref{1.trivial}) implies that $\Phi^*F_\emptyset= F_\emptyset$. For $j\geq 1$ we remark that $(\theta^{i_1} \cdots \theta^{i_j})^2 = 0$. We deduce by (\ref{1.produit}) that $\left(\Phi^*(\theta^{i_1} \cdots \theta^{i_j})\right)^2 = \Phi^*\left( (\theta^{i_1} \cdots \theta^{i_j})^2\right) = 0$ and hence, since $\Phi^*(\theta^{i_1} \cdots \theta^{i_j}) \in \mathcal{C}^\infty (\mathbb{R}^m)$, this forces $\Phi^*(\theta^{i_1} \cdots \theta^{i_j}) =0$.
\begin{figure}[h]\label{evenodd}
\begin{center}
\includegraphics[scale=0.5]{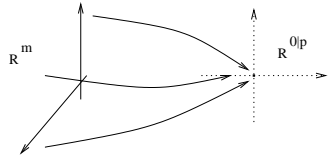}
\caption{A map from $\mathbb{R}^m$ to $\mathbb{R}^{0|p}$}
\end{center}
\end{figure}
Hence $\Phi^*F = F_\emptyset = F(0)$, i.e.\,the pull-back image of $F$ by
$\Phi$ is a constant function, whose value is equal to $F_\emptyset$, which can be interpreted as the value of F `at $0\in \mathbb{R}^{0|\ell}$'. In other words $\Phi$ looks like a constant function and its constant `value', if it would make sense, is $0$; we can hence conclude that $\Phi\equiv 0$. This reflects the fact that $0$ is the only `classical' point in $\mathbb{R}^{0|\ell}$.

The study of morphisms of the type $\Phi:\mathbb{R}^{m|0}\longrightarrow \mathbb{R}^{n|\ell}$ is completely similar: one finds that such morphisms are of the form $\Phi^*F = F\circ \varphi$, where $\varphi:\mathbb{R}^m\longrightarrow \mathbb{R}^n$ is an ordinary smooth map. In particular if $\Phi = x + \theta\psi$ maps $\mathbb{R}$ to $\mathbb{R}^{1|1}$ or, more generally, to $\Pi T\mathcal{M}$, the morphism property forces $\psi = 0$.

\subsubsection{Maps $\Phi:\mathbb{R}^{0|1}\longrightarrow \mathcal{N}^{n|0}\simeq \mathcal{N}$}\label{example2}
(From an `\textbf{odd} line' to an (\textbf{even}) standard manifold.)
This the opposite situation. We look for morphisms $\Phi^*:\mathcal{C}^\infty (\mathcal{N})\longrightarrow \mathbb{R}[\theta]$.
Any such map is characterized by two functionals $a,b: \mathcal{C}^\infty
(\mathcal{N})\longrightarrow \mathbb{R}$ such that, $\forall f\in \mathcal{C}^\infty
(\mathcal{N})$, $\Phi^*f = a(f) + b(f)\theta$. Condition
(\ref{1.somme}) implies that $a$ and $b$ are linear functionals. Then Condition (\ref{1.produit}) amounts to
\begin{equation}\label{1.point}
a(fg) = a(f) a(g),
\end{equation}
\begin{equation}\label{1.vecteur}
b(fg) = a(f) b(g) + b(f)a(g).
\end{equation}
The first relation (\ref{1.point}) implies that there exists some point
$m\in \mathcal{N}$ such that $a(f) = f(m)$, $\forall f$. Then the second
one (\ref{1.vecteur}) 
reads $b(fg) = f(m)b(g) + b(f)g(m)$, which 
implies\footnote{assuming that $\Phi^*$ is continuous with respect to the
  $\mathcal{C}^1$ topology} that 
$b$ is a derivation. So that there exists $\xi\in T_m\mathcal{N}$ such that
$b(f) = df_m(\xi)$, $\forall f$. Thus
\begin{figure}[h]\label{oddeven}
\begin{center}
\includegraphics[scale=0.5]{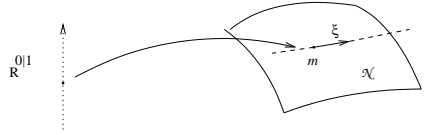}
\caption{A map from $\mathbb{R}^{0|1}$ to $\mathcal{N}$}
\end{center}
\end{figure}
\[
\Phi^*f = f(m) + df_m(\xi)\theta,\quad
\forall f\in \mathcal{C}^\infty (\mathcal{N}).
\]
Hence $\Phi$ is characterized by a point in the tangent bundle $T\mathcal{N}$.

\subsubsection{Maps $\Phi:\mathbb{R}^{0|2}\longrightarrow \mathcal{N}^{n|0}\simeq \mathcal{N}$}\label{example3}
(From an `\textbf{odd} plane' to an (\textbf{even}) standard smooth manifold.)
The analysis is similar. We analyze morphisms $\Phi^*:\mathcal{C}^\infty (\mathcal{N})\longrightarrow \mathbb{R}[\theta^1,\theta^2]$: they are characterized by four linear functionals $a,b_1,b_2,c:\mathcal{C}^\infty (\mathcal{N})\longrightarrow \mathbb{R}$ such that $\forall f \in \mathcal{C}^\infty (\mathcal{N})$, $\Phi^*f = a(f) + b_1(f)\theta^1 +  b_2(f)\theta^2 + c(f)\theta^1\theta^2$. Relation (\ref{1.produit}) for two functions $f,g\in \mathcal{C}^\infty(\mathcal{N})$ gives that
\[
\Phi^*(fg) = a(fg) + b_1(fg)\theta^1 +  b_2(fg)\theta^2 +
c(fg)\theta^1\theta^2
\]
should be equal to
\[
\begin{array}{rcl}
\Phi^*f\Phi^*g & = & \left( a(f) + b_1(f)\theta^1 +  b_2(f)\theta^2 +
  c(f)\theta^1\theta^2\right) \\
& & \times \left( a(g) + b_1(g)\theta^1 +  b_2(g)\theta^2 + c(g)\theta^1\theta^2\right) \\
 & = & a(f)a(g) + \left( a(f)b_1(g) + b_1(f)a(g)\right)\theta^1 +
 \left( a(f)b_2(g) + b_2(f)a(g)\right)\theta^2 \\
 &  & \ + \left( a(f)c(g) + c(f)a(g) + b_1(f)b_2(g) -
   b_2(f)b_1(g)\right)\theta^1\theta^2 . 
\end{array}
\]
Again, by identifying coefficients of $1$, $\theta^1$ and $\theta^2$, we deduce that there exists some point $m\in \mathcal{N}$ such that
$a(f) = f(m)$ and 
there exist two vectors $\xi_1,\xi_2\in T_m\mathcal{N}$ such that $b_1(f)
= df_m(\xi_1)$ and  
$b_2(f) = df_m(\xi_2)$. However by identifying the
$\theta^1\theta^2$ coefficients we obtain: 
\[
c(fg) = f(m)c(g) + c(f)g(m) + df_m(\xi_1)dg_m(\xi_2) - df_m(\xi_2)dg_m(\xi_1).
\]
From $fg = gf$ we deduce that the left hand side should be symmetric
in $f$ and $g$. However the right hand side is symmetric in $f$ and $g$ only if
\[
(df\wedge dg)_m(\xi_1,\xi_2) = 0,\quad \forall f,g\in \mathcal{C}^\infty
(\mathcal{N}). 
\]
This is possible only if $\xi_1$ and $\xi_2$ are linearly dependant\footnote{Such a conclusion does not look very natural: it is related to the fact that we did not assume that $\Phi$ is even.},
i.e.\,$\exists \xi\in T_m\mathcal{N}$, $\exists \lambda_1,\lambda_2\in \mathbb{R}$, s.t.\,$\xi_1 = \lambda_1\xi$ and $\xi_2 = \lambda_2\xi$. If so we then conclude for $c$ that $\exists \zeta \in T_m\mathcal{N}$ such that $c(f) = df_m(\zeta)$. Hence
\[
\Phi^*f = f(m) +  (\lambda_1\theta^1 + \lambda_2\theta^2)df_m(\xi) +
\theta^1\theta^2df_m(\zeta).
\]

\subsection{The structure of a map of an open subset of $\mathbb{R}^{m|q}$}
We here generalize and precise the observations of the previous section. These results were obtained in \cite{helein}. We assume in the following that $|\Omega|$ is an open subset of $\mathbb{R}^m$ and $\mathcal{N}$ is an ordinary manifold of dimension $n$.

First we introduce the following notations: for any positive integer $j$ we let $\I^q(j):= \{(i_1,\cdots i_j)\in [\![1,q]\!]^j| i_1<\cdots <i_j\}$, we denote by $I=(i_1,\cdots i_{j})\in \I^q(j)$ a multi-index  and we write $\eta^I:= \eta^{i_1}\cdots \eta^{i_j}$. We use also the convention $\I^q(0) = \{\emptyset \}$. We let $\I^q:= \cup_{j=0}^q\I^q(j)$, $\I^q_0:= \cup_{j=0}^{[q/2]}\I^q(2j)$, $\I^q_1:= \cup_{j=0}^{[(q-1)/2]}\I^q(2j+1)$ and $\I^q_2:= \cup_{j=1}^{[q/2]}\I^q(2j)$.

We denote by $\pi: |\Omega|\times \mathcal{N}\longrightarrow \mathcal{N}$ the canonical projection map and consider the vector bundle $\pi^*T\mathcal{N}$: 
the fiber over each point $(x,m)\in |\Omega|\times \mathcal{N}$ is the
tangent space $T_m\mathcal{N}$. For any $I\in \I^q_2$, we choose a
smooth section $\xi_I$ of $\pi^*T\mathcal{N}$ over $|\Omega|\times \mathcal{N}$ and we consider the $\mathbb{R}[\eta^1,\cdots ,\eta^q]^0$-valued vector field
\[
\Xi:= \sum_{I\in\I^q_2} \xi_I \eta^I.
\]
Alternatively $\Xi$ can be seen as a smooth family $\left(\Xi_x\right)_{x\in
|\Omega|}$ of smooth tangent vector fields  on $\mathcal{N}$ with
coefficients in $\mathbb{R}[\eta^1,\cdots ,\eta^q]^0$. So each $\Xi_x$ defines
a first order differential operator which acts on the algebra $\mathcal{C}^\infty(\mathcal{N})\otimes \mathbb{R}[\eta^1,\cdots ,\eta^q]^0$, i.e. the set of smooth functions on $\mathcal{N}$ with values in $\mathbb{R}[\eta^1,\cdots ,\eta^q]^0$, by the relation
\[
\forall f\in \mathcal{C}^\infty(\mathcal{N})\otimes \mathbb{R}[\eta^1,\cdots ,\eta^q]^0, \quad
\Xi_x f = \sum_{I\in\I^q_2} ((\xi_I)_x  f) \eta^I.
\]
We now define (letting $\Xi^0 = 1$)
\[
e^\Xi := \sum_{n=0}^\infty {\Xi^n\over n!} = \sum_{n=0}^{[q/2]} {\Xi^n\over
  n!},
\]
which can be considered again as a smooth family parametrized by $x\in
|\Omega|$ of differential operators of order at
most $[q/2]$ acting on $\mathcal{C}^\infty(\mathcal{N})\otimes
\mathbb{R}[\eta^1,\cdots ,\eta^q]^0$. Now we choose a smooth map $\varphi:|\Omega|
\longrightarrow \mathcal{N}$ and we consider the map
\[
\begin{array}{cccc}
1\times \varphi: & |\Omega| & \longrightarrow & |\Omega|\times \mathcal{N}\\
& x & \longmapsto & (x,\varphi(x))
\end{array}
\]
which parametrizes the graph of $\varphi$. Lastly we construct the following linear operator on $\mathcal{C}^\infty(\mathcal{N})\subset \mathcal{C}^\infty(\mathcal{N})\otimes \mathbb{R}[\eta^1,\cdots ,\eta^q]^0$:
\[
\mathcal{C}^\infty(\mathcal{N})\ni f\longmapsto (1\times \varphi)^*\left(
  e^\Xi f\right) \in \mathcal{C}^\infty(\Omega),
\]
defined by
\[
\forall x\in |\Omega|,\quad
(1\times \varphi)^*\left( e^\Xi f\right)(x):= \left( e^{\Xi_x}
  f\right)(\varphi(x)) = \sum_{n=0}^{[q/2]} \left({(\Xi)_x^n\over
  n!}f\right)(\varphi(x)).
\]
We will often abusively write $(1\times \varphi)^*\left( e^\Xi f\right)(x)\simeq \varphi^*\left( e^\Xi f\right)(x)$. We observe that actually, for any $x\in |\Omega|$, we only need to define $\Xi_x$ on
a neighbourhood of $\varphi(x)$ in $\mathcal{N}$, i.e. it suffices to
define the section $\Xi$ on a neighbourhood of the graph of $\varphi$
in $|\Omega|\times \mathcal{N}$ (or even on their Taylor expansion in $m$
at order $[q/2]$ around $\varphi(x)$).
\begin{theo}\label{theo1}\cite{helein}
The map $f\longmapsto (1\times \varphi)^*\left( e^\Xi f\right)$ is a morphism from $\mathcal{C}^\infty(\mathcal{N})$ to $\mathcal{C}^\infty(|\Omega|\times \mathbb{R}^{0|q})^0$, i.e. satisfies assumptions (\ref{1.somme}) and (\ref{1.produit}).

Conversely for any morphism $\Phi^*:\mathcal{C}^\infty(\mathcal{N})$ to $\mathcal{C}^\infty(|\Omega|\times \mathbb{R}^{0|q})^0$, there exists a map $\varphi:|\Omega|\longrightarrow \mathcal{N}$ and a $\mathbb{R}[\eta^1,\cdots ,\eta^q]^0$-valued vector field $\Xi$ on $\mathcal{N}$ along $\varphi$ (i.e. a section of $\pi^*(T\mathcal{N}\otimes_\mathcal{N}\mathbb{R}[\eta^1,\cdots ,\eta^q]^0)$ over a neighbourhood of the graph of $\varphi$ in $|\Omega|\times \mathcal{N}$) such that $\Phi^*:f\longmapsto (1\times \varphi)^*\left( e^\Xi f\right)$.

Furthermore one can assume that the vector fields $\xi_I$ used to build $\Xi$ commute, i.e. that $[\xi_I,\xi_J] = 0$, for all $I,J$. Moreover if we assume that the image of $\varphi$ is contained in the domain $V$ of a single coordinate chart $y:V\longrightarrow \mathbb{R}^n$, then it is possible to construct the vector fields $\xi_I$ in such a way that $\xi_I   \xi_J   y = 0$.
\end{theo}

\section{Maps with flesh between supermanifolds}
Recall that, according to the definition given in \S \ref{newdef} (following the point of view adopted in \cite{FreedDeligne} and \cite{Freed}), a map with flesh $\Phi: \mathcal{M}^{m|k}\longrightarrow \mathcal{N}^{n|\ell}$ is defined through even morphisms $\Phi^*:\mathcal{C}^\infty(\mathcal{N}^{n|\ell})
\longrightarrow \mathcal{C}^\infty(\mathcal{M}^{m|k}\times \mathbb{R}^{0|L})\simeq
\mathcal{C}^\infty(\mathcal{M}^{m|k})\otimes \mathbb{R}[\eta^1,\cdots ,\eta^L]$. 

\subsection{Examples}
The aim is here to illustrate the differences between a map with flesh and a skeletal one.
\subsubsection{Even maps $\Phi$ from $\mathbb{R}^{0|1}\times\mathbb{R}^{0|2}$ to  $\mathcal{N}^{n|0}\simeq \mathcal{N}$}\label{example4}
It corresponds to a map with flesh from $\mathbb{R}^{0|1}$ to $\mathcal{N}$ (compare with example \ref{example2}.) We look at even maps $\Phi$ from $\mathbb{R}^{0|1}\times\mathbb{R}^{0|2}$ to an ordinary manifold $\mathcal{N}$, i.e. at even morphisms $\Phi^*:\mathcal{C}^\infty(\mathcal{N})\longrightarrow 
\left(\mathbb{R}[\theta]\otimes \mathcal{C}^\infty(\mathbb{R}^{0|2})\right)^0$, where $\mathcal{C}^\infty(\mathbb{R}^{0|2}) = \mathbb{R}[\eta^1,\eta^2]$.
Setting $\Phi^*f = A(f) + \theta B(f)$, where $A(f)$ is even and $B(f)$ is odd,
we again find that (using that $A(f)$ and $A(g)$ commute with all other terms):
\begin{equation}\label{A=AA}
A(fg) = A(f)A(g)\quad \hbox{and}\quad B(fg) = A(f)B(g) + B(f) A(g).
\end{equation}
We also know that $A(f) = a(f) + \eta^1\eta^2a_{12}(f)$ and $B(f) = \eta^1b_1(f) + \eta^2b_2(f)$.
The first relation in (\ref{A=AA}) gives thus
\[
a(fg) = a(f)a(g)\quad \hbox{and}\quad a_{12}(fg) = a(f)a_{12}(g) +
a_{12}(f)a(g), 
\]
and hence there exists $m\in \mathcal{N}$ and $\zeta\in T_m\mathcal{N}$ such that
$a(f) = f(m)$ and $a_{12}(f) = df_m(\zeta)$. The second relation gives
\[
\eta^1\left[a(f)b_1(g) + b_1(f)a(g)\right] + \eta^2\left[a(f)b_2(g) +
  b_2(f)a(g)\right] 
= \eta^1b_1(fg) + \eta^2b_2(fg).
\]
Hence $\exists \xi_1,\xi_2\in T_m\mathcal{N}$ such that $b_1(f) = df_m(\xi_1)$ and $b_2(f) = df_m(\xi_2)$. So
\[
\begin{array}{ccl}
\Phi^*f & = & \left( f(m) + \eta^1\eta^2df_m(\zeta)\right) + \theta
\left( \eta^1 df_m(\xi_1) + \eta^2 df_m(\xi_2) \right) \\
& = & f(m) + \eta^1\eta^2df_m(\zeta) + \theta df_m\left(\eta^1\xi_1 +
  \eta^2\xi_2\right)\\
& = & \left( \left(1+\eta^1\eta^2\zeta\right)   \left(1+\theta(\eta^1\xi_1 + \eta^2\xi_2) \right)   f \right)(m),
\end{array}
\]
where  $1$ stands for the identity operator acting on $\mathcal{C}^\infty(\mathcal{N})$ and we note, e.g., $\zeta  f:= \hbox{Lie}_\zeta f$. If we introduce the $\mathbb{R}^{0|2}$ valued vector fields $\Xi_\emptyset:= \eta^1\eta^2\zeta$ and $\Xi_1:= \eta^1\xi_1 + \eta^2\xi_2$, we can write the preceding relation as:
\begin{equation}\label{Phi*f0}
\Phi^*f = \left( \left(1+\Xi_\emptyset\right)   \left(1+\theta \Xi_1\right)   f \right)(m).
\end{equation}
We also write (\ref{Phi*f0}) symbolically\footnote{These notations are actually inspired by a system of notations proposed by M. Kawski and H. Sussmann \cite{kawskisussmann}.} as:
\begin{equation}\label{Phi*f}
\Phi = m\ \left(1+\Xi_\emptyset\right)  \left(1+\theta \Xi_1\right).
\end{equation}
A variant consists in adding a time variable $t\in \mathbb{R}$, i.e. to substitute $\mathbb{R}^{1|1}$ to $\mathbb{R}^{0|1}$. Then we would obtain by a similar reasoning that a morphism $\Phi^*$ from $\mathcal{C}^\infty(\mathcal{N})$ to 
$\left(\mathcal{C}^\infty(\mathbb{R})[\theta,\eta^1,\eta^2]\right)^0$ has the form
\begin{equation}\label{Phi*f}
\Phi^*f = \varphi^*\left( \left(1+\Xi_\emptyset\right)   \left(1+\theta \Xi_1\right)   f \right),
\end{equation}
where $\varphi\in \mathcal{C}^\infty(\mathbb{R},\mathcal{N})$ and $\Xi_\emptyset$ and $\Xi_1$ have the same meaning as before. We write this symbolically as
\begin{equation}\label{symbolic}
\Phi = \varphi\ \left(1+\Xi_\emptyset\right)  \left(1+\theta \Xi_1\right)
= \left(\varphi + \varphi \ \Xi_\emptyset\right) + \theta \left(\varphi\ \Xi_1\right).
\end{equation}

\subsubsection{Even maps $\Phi$ from $\mathbb{R}^{0|2}\times\mathbb{R}^{0|2}$ to  $\mathcal{N}^{n|0}\simeq \mathcal{N}$}
It corresponds to a map with flesh from $\mathbb{R}^{0|2}$ to $\mathcal{N}$ (compare with \ref{example3}.) We look at even morphisms
$\Phi^*:\mathcal{C}^\infty(\mathcal{N})\longrightarrow
\left(\mathbb{R}[\theta^1,\theta^2]\otimes \mathbb{R}[\eta^1,\eta^2]\right)^0$. One finds that $\exists m\in \mathcal{N}$, $\exists \zeta,\chi,\xi_{\alpha\beta},\pi\in
T_m\mathcal{N}$ (for $1\leq \alpha,\beta\leq 2$) such that
\begin{equation}\label{Phi*f1}
\begin{array}{ccl}
\Phi^*f & = & \left( f(m) + \eta^1\eta^2df_m(\zeta)\right) +
\theta^1df_m\left( \eta^1\xi_{11} + \eta^2\xi_{12}\right) 
+ \theta^2df_m\left( \eta^1\xi_{21} + \eta^2\xi_{22}\right) \\
 & & +\ \theta^1\theta^2\left( df_m(\chi) + \eta^1\eta^2(Pf(m) +
   df_m(\pi))\right). 
\end{array}
\end{equation}
Here $P$ is the second order differential operator
$P := \zeta   \chi - {\xi_{11}}  {\xi_{22}} +
{\xi_{12}}  {\xi_{21}}$, where we have used arbitrary extensions of the vectors $\zeta,\chi,\xi_{\alpha\beta},\pi\in T_m\mathcal{N}$ to vectors fields defined on a neighbourhood of $m$ (for example we could choose the extensions in such a way that $[\zeta,\chi] = [\xi_{11},\xi_{22}] = [\xi_{12},\xi_{21}] = 0$). Clearly the operator $P$ depends on the way we have extended the vectors fields. However two different choices of extensions lead to two operators $P$ and $P'$ such that $Pf(m)- P'f(m) = df_m(V)$, for some $V\in T_m\mathcal{M}$. Hence changing the extensions is just equivalent to change $\pi\in T_m\mathcal{N}$. The right hand side of (\ref{Phi*f1}) can factorized as
\begin{equation}\label{Phi*f2}
\Phi^*f = \left( \left( 1 +\Xi_\emptyset \right) \left( 1 + \theta^1\Xi_1\right) 
\left( 1 + \theta^1\Xi_2\right) 
\left( 1 + \theta^1\theta^2\Xi_{12}\right)   f\right)(m),
\end{equation}
where $\Xi_\emptyset:= \eta^1\eta^2\zeta$, $\Xi_1:= \eta^1\xi_{11} + \eta^2\xi_{12}$, $\Xi_2:= \eta^1\xi_{11} + \eta^2\xi_{12}$ and $\Xi_{12}:= \chi+ \eta^1\eta^2\pi$. We can summarize this result by the symbolic relation:
\begin{equation}\label{heur}
\Phi =  m\ \left( 1 +\Xi_\emptyset \right)\left( 1 + \theta^1\Xi_1\right)
\left( 1 + \theta^1\Xi_2\right)
\left( 1 + \theta^1\theta^2\Xi_{12}\right),
\end{equation}

\subsection{The general description of maps with flesh}

\subsubsection{Even maps $\Phi$ from $\Omega\times \mathbb{R}^{0|L}$ to an ordinary manifold $\mathcal{N}$}\label{definitifMap}
Here $\Omega$ is an open subset of $\mathbb{R}^{m|k}$. This case is a straightforward application of Theorem \ref{theo1}. Indeed we have $\Omega \simeq |\Omega|\times \mathbb{R}^{0|k}$, where $|\Omega|$ is an open subset of $\mathbb{R}^n$ and thus $\Omega\times \mathbb{R}^{0|L}\simeq |\Omega|\times \mathbb{R}^{0|q}$, with $q:= k+L$. Hence we just need to label by $(\theta^1,\cdots ,\theta^k,\eta^1,\cdots , \eta^L)$ the odd coordinates on $\mathbb{R}^{0|k+L}$ and to apply Theorem \ref{theo1}. This requires adapted notations: we note $\A^k(0) = \{\emptyset \}$ and for any $j\in \N^*$,
$\A^k(j):= \{(a_1,\cdots a_j)\in [\![1,k]\!]^j| a_1<\cdots <a_j\}$. We
denote by $A=(a_1,\cdots a_{j})$ an element of 
$\A^k(j)$ and write $\theta^A:= \theta^{a_1}\cdots
\theta^{a_j}$. We let $\A^k:= \cup_{j=0}^k\A^k(j)$, $\A^k_0:=
\cup_{j=0}^{[k/2]}\A^k(2j)$, $\A^k_1:= \cup_{j=0}^{[(k-1)/2]}\A^k(2j+1)$,
$\A^k_2:= \cup_{j=1}^{[k/2]}\A^k(2j)$ and $\A^k_+:= \A^k_1\cup \A^k_2$. Lastly we set $\AI:=
\{AI|A\in \A^k,I\in \I^L\}$ and, defining the length of $AI$ to be the
some of the lengthes of $A$ and $I$, we define similarly $\AI(j)$, $\AI_0$,
$\AI_1$ and $\AI_2$. Hence any (even) function $f\in
\mathcal{C}^\infty(\Omega\times \mathbb{R}^{0|L})$ (where $\Omega$ is an open subset of $\mathbb{R}^{p|k}$) can be decomposed as $f = \sum_{AI\in \AI_0}
\theta^A\eta^If_{AI}$,
where $f_{AI}\in \mathcal{C}^\infty(|\Omega|)$, $\forall AI\in
\AI_0$.

Then Theorem \ref{theo1} implies that for any morphism
$\Phi^*$ from $\mathcal{C}^\infty(\mathcal{N})$ to $\mathcal{C}^\infty(\Omega\times \mathbb{R}^{0|L})^0$, there exists a smooth map
$\varphi\in \mathcal{C}^\infty(|\Omega|,\mathcal{N})$ and  a smooth family
$\left(\xi_{AI}\right)_{AI\in \AI_2}$ of sections of $\pi^*T\mathcal{N}$
defined on a neighbourhood of the graph of $\varphi$ in
$|\Omega|\times \mathcal{N}$ such that if $\Xi:= \sum_{AI\in
  \AI_2}\xi_{AI}\theta^A\eta^I$ then $\forall f\in \mathcal{C}^\infty(\mathcal{N})$, $\phi^*f = (1\times f)^*\left(e^\Xi
  f\right)$. We decompose $\Xi$ as
\[
\Xi = \sum_{A\in \A^k}\theta^A\Xi_A = \Xi_\emptyset + \sum_{a\in
  \A^k(1)} \theta^a\Xi_a + \sum_{(a_1,a_2)\in \A^k(2)}
\theta^{a_1} \theta^{a_2}\Xi_{a_1a_2} + \cdots ,
\]
where $\forall A\in \A^k_1$, $\Xi_A = \sum_{I\in
\I^L_1}\xi_{AI}\eta^I$ and $\forall A\in \A^k_0$, $\Xi_A =\sum_{I\in
\I^L_2}\xi_{AI}\eta^I$. In particular $\Xi_\emptyset = \sum_{I\in
\I^L_2}\xi_{\emptyset I}\eta^I$ and we see that $\Xi_A$ is odd if $A$ is odd and is even if $A$ is even. Recall that the vector
fields $\left(\xi_{AI}\right)_{AI\in \AI_2}$ can be chosen in such a way that they commute pairwise. If so the relations $[\xi_{AI},\xi_{A'I'}] = 0$ imply that the vector fields $\Xi_A$ {\em supercommute} pairwise, i.e.
\[
\forall A\in \A^k(j),\forall A'\in \A^k(j'),\quad
\Xi_A\Xi_{A'} - (-1)^{jj'} \Xi_{A'}\Xi_A = 0.
\]
This is equivalent to the
fact that $\forall A,A'\in \A^k$,  $[\theta^A\Xi_A,
\theta^{A'}\Xi_{A'}] = 0$. This last commutation relation implies that
\[
e^\Xi = e^{\sum_{A\in \A^k}\theta^A\Xi_A} = e^{\Xi_\emptyset}
\prod_{A\in \A^k_+} e^{\theta^A\Xi_A}.
\]
Hence
\begin{equation}\label{finalenxi}
\forall f\in\mathcal{C}^\infty(\mathcal{N}),\quad
\phi^*f = (1\times \varphi)^*\left(e^{\Xi_\emptyset}
\prod_{A\in \A^k_+} e^{\theta^A\Xi_A}f\right).
\end{equation}
We thus obtain a generalization of (\ref{Phi*f}) and (\ref{Phi*f2}). We could alternatively write (\ref{finalenxi}) by the symbolic relation
\begin{equation}\label{finalsymbol}
\Phi = \varphi\ e^{\Xi_\emptyset} \prod_{A\in \A^k_+} e^{\theta^A\Xi_A},
\end{equation}
which generalizes (\ref{symbolic}) and (\ref{heur}).

\subsubsection{Even maps from $\mathcal{M}^{m|k}\times \mathbb{R}^{0|L}$ to a supermanifold $\mathcal{N}^{n|\ell}$}\label{3.3.3}
First assume that $\mathcal{M}^{m|k} = \Omega$ is an open subset of $\mathbb{R}^{m|k}$. We also suppose that $\Omega$ is sufficiently small, so that its image by $\Phi$ is contained in some open subset $U$ of $\mathcal{N}^{n|\ell}$ on which there exists a local chart $Y: U\longrightarrow \mathbb{R}^{n|\ell}$. By this we assume that $Y^*: \mathcal{C}^\infty(Y(U)) \longrightarrow \mathcal{C}^\infty(U)$ is an isomorphism of super algebras. Hence for any function $f\in \mathcal{C}^\infty(U)$, there exists an unique function $F \in \mathcal{C}^\infty(Y(U))$ such that $f = Y^*F$. Denote by $(z,\zeta) = (z^1,\cdots ,z^n,\zeta^1,\cdots ,\zeta^\ell)$ the canonical coordinates on $\mathbb{R}^{n|\ell}\supset Y(U)$. Using the decomposition $F = \sum_{J\in \I^\ell}F_J\zeta^J$ and the fact that $Y$ is a morphism we deduce that
\[
f = Y^*F = \sum_{J\in \I^\ell}f_J\psi^J = \sum_{J\in \I^\ell} f_J\psi^{j_1}\cdots \psi^{j_k},
\]
where $f_J:= Y^*F_J$, $\psi^j:= Y^*\zeta^j$ and we recall that $\I^\ell$ is the set of multi-indices $J=(j_1,\cdots,j_r)\in [\![1,\ell]\!]^r$ such that $j_1<\cdots <j_r$, for $0\leq r\leq \ell$. Observe that $f_J$ is a smooth real valued function defined on $|U|$ of $\mathbb{R}^n$. Then by the morphism property,
\[
\Phi^*f = \sum_{J\in \I^\ell}(\Phi^*f_J)
(\Phi^*\psi^J) = \sum_{J\in \I^\ell}(\Phi^*f_J)(\Phi^*\psi^{j_1})\cdots (\Phi^*\psi^{j_r}).
\]
Hence we conclude that the morphism $\Phi^*$ is completely determined as soon as we know all $\Phi^*\psi^j$'s, for $j\in [\![1,m]\!]$, and all pull-back images $\Phi^*f_J$, for $J\in \I^\ell$. However each $\chi^j:= \Phi^*\psi^j$ is an odd function on $\Omega$ which can be choosed arbitrarily and, since each $f_J$ is a function on $|U|\subset |\mathcal{N}|$, its pull-back by $\Phi$ is again described by Theorem \ref{theo1}.

The case of an even map $\Phi$ from $\mathcal{M}^{m|k}\times \mathbb{R}^{0|L}$ to $\mathcal{N}^{n|\ell}$, where $\mathcal{M}^{m|k}$ is a supermanifold, can be treated similarly: by taking the restriction of $\Phi$ to an open subset of $\mathcal{M}^{m|k}$ times $\mathbb{R}^{0|L}$ and using a local chart on it, we are left to the previous situation.

\subsection{Application: how to read $\Phi = x+\theta\psi$ ?}\label{thetruth}
We have claimed in \S \ref{example4} that, when writing $\Phi = x+\theta\psi$ for a map from $\mathbb{R}^{1|1}\times\mathbb{R}^{0|2}$ to an ordinary manifold $\mathcal{N}$, one should read `$\varphi+\varphi\ \eta^1\eta^2\zeta$' instead of `$x$' and `$\varphi\ (\eta^1\xi_1 + \eta^2\xi_2)$' instead of `$\psi$'. By the results expounded in the previous section the generalization of this to a map $\Phi$ from $\mathbb{R}^{1|1}\times\mathbb{R}^{0|L}$ to  $\mathcal{N}$ (where $L$ is arbitrary) is (using $(\theta)^2=0$)
\[
\Phi = \varphi\ e^{\Xi_\emptyset}e^{\theta\Xi_1} = \varphi\ e^{\Xi_\emptyset} \left( 1 + \theta\Xi_1\right)
=  \varphi\ e^{\Xi_\emptyset} + \theta \left(\varphi\ e^{\Xi_\emptyset}\Xi_1\right),
\]
so that one should interpret that `$x$' $= \varphi\ e^{\Xi_\emptyset}$ and `$\psi$' $= \varphi\ e^{\Xi_\emptyset}\Xi_1$.

We can interpret similarly the decomposition of a supersymmetric $\sigma$-model
\begin{equation}\label{sigmamodel1}
\Phi = \phi + \theta^1\psi_1 + \theta^2\psi_2 +  \theta^1\theta^2 F
\end{equation}
from $\mathbb{R}^{m|2}$ to an ordinary manifold $\mathcal{N}$. Understanding this field as an even map $\Phi$ from $\mathbb{R}^{m|2}\times \mathbb{R}^{0|L}$ to $\mathcal{N}$, we can write
by Theorem \ref{theo1}
\begin{equation}\label{howto}
\Phi = \varphi\ e^{\Xi_\emptyset} \left( 1 + \theta^1\Xi_1\right)\left( 1 + \theta^2\Xi_2\right)\left( 1 + \theta^1\theta^2\Xi_{12}\right),
\end{equation}
meaning that $\forall f\in\mathcal{C}^\infty(\mathcal{N})$,
\begin{equation}\label{Phi*y}
\Phi^*f = \varphi^*\left(e^{\Xi_\emptyset}  \left( 1 + \theta^1\Xi_1\right)  \left( 1 + \theta^2\Xi_2\right)  \left( 1 + \theta^1\theta^2\Xi_{12}\right)  f\right).
\end{equation}
We can now understand the meaning of the usual expression (\ref{sigmamodel1}) as follows. Somehow in this relation one implicitely assumes that $\varphi$ takes values in an open set $V$ of $\mathcal{N}$ on which a coordinate chart $y = (y^1,\cdots,y^n):V\longrightarrow \mathbb{R}^n$ is defined. Then the right hand side of (\ref{sigmamodel1}) is $y^*\Phi$. So let us use (\ref{Phi*y}) with $f = y^i$ ($1\leq i\leq n$) and exploit the fact that, thanks to Theorem \ref{theo1}, we can chose all vector fields $\xi_I$ such that $\xi_I  \xi_J  y = 0$. Then, setting $\varphi^i:= y^i\circ \varphi$, $\Xi_1^i:= \Xi_1 \varphi^i$, $\Xi_2^i:= \Xi_2 \varphi^i$, $\Xi_{12}^i:= \Xi_{12} \varphi^i$ and $\phi^i:= \varphi^i + \Xi_\emptyset  \varphi^i$, (\ref{Phi*y}) implies
\begin{equation}\label{sigmamodel2}
\begin{array}{ccl}
\Phi^i \simeq \Phi^*y^i & = & \left(\varphi^i + \Xi_\emptyset  \varphi^i\right) + \theta^1\left(\Xi_1  \varphi^i\right) + \theta^2\left(\Xi_2  \varphi^i\right) + \theta^1\theta^2\left(\Xi_{12}  \varphi^i\right)
\\
& = & \phi^i + \theta^1\Xi_1^i  + \theta^2\Xi_2^i  + \theta^1\theta^2\Xi_{12}^i .
\end{array}
\end{equation}
A comparison with (\ref{sigmamodel1}) gives us: $\phi^i = \varphi^i + \Xi_\emptyset  \varphi^i$, $\psi_1^i = \Xi_1^i$, $\psi_2^i = \Xi_2^i$ and $F^i = \Xi_{12}^i$. One the other hand, for an arbitrary \emph{non linear} function $f\in\mathcal{C}^\infty(\mathcal{N})$, (\ref{Phi*y}) developps as
\[
f\circ \Phi\simeq \Phi^*f = \varphi^*\left(e^{\Xi_\emptyset}  \left(f + \theta^1\Xi_1  f + \theta^2\Xi_2  f + \theta^1\theta^2(\Xi_{12} - \Xi_1  \Xi_2)  f\right)\right).
\]
We hence deduce the relation in local coordinates
\[
\begin{array}{cl}
f(\phi + \theta^1\psi_1 + \theta^2\psi_2 +  \theta^1\theta^2 F) & \displaystyle = f(\phi) + \theta^1\sum_{i=1}^n{\partial f\over \partial y^i}(\phi)\psi_1^i + \theta^2\sum_{i=1}^n{\partial f\over \partial y^i}(\phi)\psi_2^i \\
& \displaystyle +\ \theta^1\theta^2 \left( \sum_{i=1}^n{\partial f\over \partial y^i}(\phi)F^i - \sum_{i,j=1}^n{\partial^2 f\over \partial y^i \partial y^j}(\phi)\psi_1^i\psi_2^j\right).
\end{array}
\]

\subsection{Vector fields on a supermanifold with flesh}\label{derivatives}
Consider the superspace $\mathbb{R}^{m|k}\times \mathbb{R}^{0|L}$ and denote by $x^1,\cdots,x^m,\theta^1,\cdots ,\theta^k$ the coordinates on $\mathbb{R}^{m|k}$ and by $\eta^1,\cdots ,\eta^L$ the coordinates on $\mathbb{R}^{0|L}$. For $1\leq i \leq m$, the action of ${\partial \over \partial x^i}$ on $\mathcal{C}^\infty(\mathbb{R}^m)$ can be extended in a straightforward way to $\mathcal{C}^\infty(\mathbb{R}^{m|k}\times \mathbb{R}^{0|L})$: for any function $f\in \mathcal{C}^\infty(\mathbb{R}^{m|k}\times \mathbb{R}^{0|L})$, decompose $f = \sum_{AI\in \AI} f_{AI}\theta^A\eta^I$ (see \S \ref{definitifMap}), where, $\forall A\in \A^k$, $\forall I\in \I^L$, $f_{AI}\in \mathcal{C}^\infty(\mathbb{R}^m)$ and just set
\begin{equation}\label{Xextended}
{\partial f\over \partial x^i}:= \sum_{AI\in \AI}{\partial f_{AI}\over \partial x^i}\theta^A\eta^I.
\end{equation}
We define the odd derivative ${\partial \over \partial \theta^a}$ as the operator acting on $\mathcal{C}^\infty(\mathbb{R}^{m|k}\times \mathbb{R}^{0|L})$ by the rule
\[
{\partial f\over \partial \theta^j} = \sum_{AI\in \AI} (-1)^{\hbox{gr}(f_{AI})}f_{AI}{\partial \theta^A\over \partial \theta^a}\eta^I,
\]
where, if $A$ does not contain $a$, ${\partial \theta^A\over \partial \theta^a} = 0$ and, if $A$ contains $a$, say $A= (a_1,\cdots, a_{\gamma-1},a,a_{\gamma+1}\cdots ,a_k)$, ${\partial \theta^A\over \partial \theta^a} = (-1)^{\gamma-1} \theta^1\cdots \theta^{a_{\gamma-1}}\theta^{a_{\gamma+1}}\cdots \theta^{a_k}$. More generally, if $\Omega$ is an open subset of $\mathbb{R}^{m|k}$, we consider differential operators $\mathcal{D}:\mathcal{C}^\infty(\Omega\times \mathbb{R}^{0|L}) \longrightarrow \mathcal{C}^\infty(\Omega\times \mathbb{R}^{0|L})$ of the form
\[
\mathcal{D} = \sum_{i=1}^ma_i{\partial \over \partial x^i} + \sum_{a=1}^kb_a{\partial \over \partial \theta^a},
\]
where $a_i$ and $b_a$ belongs to $\mathcal{C}^\infty(\Omega\times \mathbb{R}^{0|L})$. In the following we assume that the $a_i$'s are even and the $b_a$'s are odd so that $\mathcal{D}$ is a derivation, i.e. it satisfies the Leibniz rule $\mathcal{D}(fg) = (\mathcal{D}f)g + f(\mathcal{D}g)$, $\forall f,g\in \mathcal{C}^\infty(\mathbb{R}^{m|k}\times \mathbb{R}^{0|L})$. For example on $\mathbb{R}^{1|1}\times \mathbb{R}^{0|L}$ we consider in Section \ref{particlebis} the vector field ${\partial \over \partial t}$ and, letting $\eta\in \mathcal{C}^\infty(\mathbb{R}^{0|L})^1$ ($\eta$ is odd), the (even) vector fields $\eta D = \eta{\partial \over \partial \theta} - \eta\theta{\partial \over \partial t}$ and $\eta\tau_Q = \eta{\partial \over \partial \theta} + \eta\theta{\partial \over \partial t}$.

For the following discussion it is useful to present an alternative description of the set of \emph{even} maps $\mathcal{C}^\infty(\Omega\times \mathbb{R}^{0|L})^0$, where $\Omega\subset \mathbb{R}^{m|k}$. Let $q:= k+L$ and replace $(\theta^1,\cdots,\theta^k,\eta^1,\cdots,\eta^L)$ by $(\eta^1,\cdots,\eta^q)$ to simplify the notations. Let $\Lambda^{2*}_+\simeq \mathbb{R}^{2^{q-1}-1}$ be the vector subspace of even elements of positive degree of the exterior algebra $\Lambda^*\mathbb{R}^q$, denote by $\mathfrak{s} = \left(\mathfrak{s}^I\right)_{I\in \I^q_2}$ the canonical coordinates on $\Lambda^{2*}_+$ and define the `canonical' $\mathbb{R}[\eta^1,\cdots,\eta^q]$-valued vector field $\vartheta$ on $\Lambda^{2*}_+$ by $\vartheta:= \sum_{I\in \I^q_2}\eta^I{\partial \over \partial \mathfrak{s}^I}$. Then, as shown in \cite{helein}, for any $f\in \mathcal{C}^\infty(\Omega\times \mathbb{R}^{0|L})^0$, there exists a map $\mathfrak{f}\in \mathcal{C}^\infty(|\Omega|\times \Lambda^{2*}_+)$ such that $\forall x\in |\Omega|$, $f(x,\eta) = \left(e^\vartheta \mathfrak{f}\right)(x,0)$,
or, by using the canonical embedding $\iota:|\Omega|\longrightarrow |\Omega|\times \Lambda^{2*}_+$, $x\longmapsto (x,0)$,
\[
f= \iota^*\left(e^\vartheta \mathfrak{f}\right).
\]
Note that $\mathfrak{f}$ is not unique: two maps  $\mathfrak{f}$ and $\mathfrak{f}'$ produces the same function $f$ if and only if their difference $\mathfrak{f}-\mathfrak{f}'$ lies in the ideal $\mathcal{I}^q(|\Omega|)$ spanned over $\mathcal{C}^\infty(|\Omega|)$ by the polynomials $\mathfrak{s}^{I_1}\cdots \mathfrak{s}^{I_j} - \epsilon_I^{I_1\cdots I_j}\mathfrak{s}^I$, for $I_1,\cdots ,I_j\in \I^q_2$, where $\epsilon_I^{I_1\cdots I_j} = 0$ if there exists some index $i\in [\![1,q]\!]$ which appears at least two times in the list $(I_1,\cdots ,I_j)$ and $\epsilon_I^{I_1\cdots I_j} = \pm 1$ in the other cases (then $\epsilon_I^{I_1\cdots I_j}$ is the signature of the permutation $(I_1,\cdots ,I_j)\longmapsto I$), see \cite{helein}. One hence deduces the algebra isomorphism $\mathcal{C}^\infty(\Omega\times \mathbb{R}^{0|L})^0 \simeq \mathcal{C}^\infty(|\Omega|\times \Lambda^{2*}_+) / \mathcal{I}^q(|\Omega|)$.

Thanks to the isomorphism $f \longmapsto \mathfrak{f}$ mod $\mathcal{I}^q(|\Omega|)$, we can represent each vector field $X$ acting on $\mathcal{C}^\infty(\Omega\times \mathbb{R}^{0|L})^0$ by a (non unique) vector field $\mathfrak{X}$ on $|\Omega|\times \Lambda^{2*}_+$, such that
\begin{equation}\label{fgothic}
\forall f\in \mathcal{C}^\infty(\Omega\times \mathbb{R}^{0|L})^0, \quad
Xf = \iota^*\left(e^\vartheta \mathfrak{X}\mathfrak{f}\right).
\end{equation}
Below is a list of examples.
\begin{enumerate}

\item Any vector field $X = \sum_{i=1}^mX_i(x){\partial \over \partial x^i}$ tangent to $|\Omega|$ has a straightforward extension to $|\Omega|\times \mathbb{R}^{0|q}$, induced by (\ref{Xextended}), that we still denote by $X$. Similarly $X$ can be extended to $|\Omega|\times \Lambda^{2*}_+$ into an unique vector field, also denoted by $X$, such that $X\mathfrak{s}^I = 0$ and $[X,{\partial \over \partial \mathfrak{s}^I}] = 0$. Then
$\forall f\in \mathcal{C}^\infty(\Omega\times \mathbb{R}^{0|L})^0$,  $Xf = \iota^*\left(e^\vartheta X\mathfrak{f}\right)$, i.e. we can choose $\mathfrak{X} = X$.

\item Let $X$ be a vector field as in the previous example and consider $Y = \eta^1\eta^2X$. Then
\[
\forall f\in \mathcal{C}^\infty(|\Omega|\times \mathbb{R}^{0|q})^0,\quad  \eta^1\eta^2Xf = \iota^*\left(e^\vartheta \mathfrak{s}^{12}X\mathfrak{f}\right),
\]
i.e. we can choose $\mathfrak{Y} = \mathfrak{s}^{12}X$. This is a consequence of the following. Let $\I^{{/}\!\!\!1{/}\!\!\!2q}_2$ be the subset of multi-indices in $\I^q_2$, where the values $1$ and $2$ are forbidden for the indices. For $I\in \I^{{/}\!\!\!1{/}\!\!\!2q}_2$, let $12I\in\I^q_2$ be the multi-index obtained by concatening $1,2$ and $I$. Then $\mathfrak{s}^{12}\mathfrak{s}^I = \mathfrak{s}^{12I}$ mod $\mathcal{I}^q(|\Omega|)$.

\item For $Z = \eta^1{\partial \over \partial \eta^2}$, we have, setting $\sum_* \mathfrak{s}^{1*}{\partial \over \partial \mathfrak{s}^{2*}}:= \sum_{I\in \I^{{/}\!\!\!1{/}\!\!\!2q}_2}\mathfrak{s}^{1I}{\partial \over \partial \mathfrak{s}^{2I}}$,
\[
\forall f\in \mathcal{C}^\infty(|\Omega|\times \mathbb{R}^{0|q})^0,\quad  \eta^1{\partial f\over \partial \eta^2} = \iota^*\left(e^\vartheta \sum_* \mathfrak{s}^{1*}{\partial \mathfrak{f}\over \partial \mathfrak{s}^{2*}}\right),
\]
i.e. we can choose $\mathfrak{Z} = \sum_* \mathfrak{s}^{1*}{\partial \over \partial \mathfrak{s}^{2*}}$.
\end{enumerate}

We can now make sense of odd derivatives of even maps into an ordinary manifold $\mathcal{N}$ in the sense of \S \ref{definitifMap} as follows. We first present a consequence of Theorem \ref{theo1} derived in \cite{helein}. Recall that, by Theorem \ref{theo1}, any morphism $\Phi^*:\mathcal{C}^\infty(\mathcal{N}) \longrightarrow \mathcal{C}^\infty(|\Omega|\times \mathbb{R}^{0|q})^0$ has the form $\Phi^*f = (1\times \varphi)^*\left(e^\Xi f\right)$, where, in the decomposition $\Xi = \sum_{I\in \I^q_2}\xi_I\eta^I$, we can assume that all vector fields $\xi_I$'s \emph{commute}. Hence we can integrate simultaneously these vector fields and we obtain a smooth (ordinary) map $\mathfrak{F}: |\Omega|\times \Lambda^{2*}_+ \longrightarrow \mathcal{N}$ such that:
\begin{equation}\label{Fgothic}
\mathfrak{F}(x,0) = \varphi(x)\quad \hbox{and}\quad
{\partial \mathfrak{F}\over \partial \mathfrak{s}^I} (x,\mathfrak{s}) = \xi_I\left(\mathfrak{F}(x,\mathfrak{s}) \right).
\end{equation}
Conversely one can recover $\Phi^*$ (and hence $\Phi$) from $\mathfrak{F}$ through the formula
\begin{equation}\label{recoverPhi}
\forall f\in \mathcal{C}^\infty(\mathcal{N}),\quad
\Phi^*f = \iota^*\left(e^\vartheta (f\circ \mathfrak{F})\right).
\end{equation}
Now if $X$ is a tangent vector on $|\Omega|\times \mathbb{R}^{0|q}$, we would like to define $X\Phi = \Phi_*X$ as a section of $\Phi^*T\mathcal{N}$ or alternatively (in order to avoid the definition of the fiber bundle $\Phi^*T\mathcal{N}$ over $|\Omega|\times \mathbb{R}^{0|q}$) through the map $(\Phi,X\Phi)$ from $|\Omega|\times \mathbb{R}^{0|q}$ to $T\mathcal{N}$. We do it indirectly by defining $\alpha_\Phi(X\Phi)$, for any section $\alpha$ of $T^*\mathcal{N}$ on $\mathcal{N}$. We first consider the case where $\alpha$ is exact, i.e. $\alpha = df$, for some $f\in \mathcal{C}^\infty(\mathcal{N})$. We then have, by using (\ref{recoverPhi}) and (\ref{fgothic})
\[
df_\Phi(X\Phi) = X(f\circ \Phi) = X(\Phi^*f) = X \iota^*\left(e^\vartheta (f\circ \mathfrak{F})\right) = \iota^*\left(e^\vartheta \mathfrak{X}(f\circ \mathfrak{F})\right) = 
\iota^*\left(e^\vartheta df_\mathfrak{F}(\mathfrak{X}\mathfrak{F})\right).
\]
(Note that $e^\vartheta$ and $\mathfrak{X}$ commute). We deduce from this identity a similar formula for a 1-form $gdf$, where $f,g\in \mathcal{C}^\infty(\mathcal{N})$ by  writing
\[
(g\circ \Phi) (df_\Phi(X\Phi))= \left(\iota^*\left(e^\vartheta (g\circ \mathfrak{F})\right)\right) \left( \iota^*\left(e^\vartheta df_\mathfrak{F}(\mathfrak{X}\mathfrak{F})\right) \right) = 
\iota^*\left(e^\vartheta (g\circ \mathfrak{F}) df_\mathfrak{F}(\mathfrak{X}\mathfrak{F})\right).
\]
And since any 1-form on $T\mathcal{N}$ can be written as a finite sum of forms of the type $gdf$, we obtain:
\[
\alpha_\Phi(X\Phi) = 
\iota^*\left(e^\vartheta \alpha_\mathfrak{F}(\mathfrak{X}\mathfrak{F})\right).
\]
Hence $X\Phi$ can be represented by the section $\mathfrak{X}\mathfrak{F}$ of $\mathfrak{F}^*T\mathcal{N}$ or, equivalentely $(\Phi, X \Phi)$ is a map from $\Omega$ to $T\mathcal{N}$ which can be represented by the map $(\mathfrak{F},\mathfrak{X} \mathfrak{F})$ from $|\Omega|\times \Lambda^{2*}_+$ to $T\mathcal{N}$. 

\section{An alternative theory}\label{dewittrogers}

\subsection{Another consequence of the morphism property}\label{Panother}
In this paragraph we still use the same point of view as before but, as a motivation for the following, we expound another consequence of the morphism properties (\ref{1.somme}) and (\ref{1.produit}). Assume that $\mathcal{M}^{m|k}$ is a (skeletal, for simplicity) supermanifold of dimension $m|k$ and let $U\subset \mathcal{M}^{m|k}$ an open subset and $X: U\longrightarrow \Omega$ be a chart, where $\Omega$ is an open subset of $\mathbb{R}^{m|k}$. Let $(z,\zeta) = (z^1,\cdots ,z^m,\zeta^1,\cdots ,\zeta^k)$ be the canonical coordinates on $\mathbb{R}^{m|k}\supset \Omega$. As in \S \ref{3.3.3}, for any function $f\in \mathcal{C}^\infty(U)$ there exists an unique function $F\in \mathcal{C}^\infty(\Omega)$, such that $f = X^*F$ and, if we decompose $F = \sum_{I\in \I^k}F_I\zeta^I$, for some functions $F_I\in \mathcal{C}^\infty(|\Omega|)$, and if we set $\theta^i:= X^*\zeta^i$, we obtain $f =  X^*F = \sum_{I\in \I^k}(X^*F_I)\theta^I$. Moreover $X^*F_I$ can be estimated by the following: since $F_I\in \mathcal{C}^\infty(|\Omega|)$, we can deduce from a Taylor expansion (with integral rest) that, for any $z,z_0\in |\Omega|$,
\[
F_I(z) = \sum_{|r|\leq k}{\partial^r F_I\over (\partial z)^r}(z_0){(z-z_0)^r\over r!} + \sum_{|r| = k+1}G_{I,r}(z)(z-z_0)^r,
\]
for some smooth functions $G_{I,r}$. Using (\ref{1.somme}) and (\ref{1.produit}) and setting $x^i:= X^*z^i$ and  $x_0^i:= X^*z_0^i$, this implies
\[
X^*F_I = \sum_{|r|\leq k}{\partial^r F_I\over (\partial z)^r}(z_0){(x-x_0)^r\over r!} + \sum_{|r| = k+1}G_{I,r}(y)(x-x_0)^r.
\]
Now in the case where $z-z_0$ (i.e. $x-x_0$) is nilpotent, the last term in the right hand side vanishes and, setting $f_I:= X^*F_I$ and $X_*{\partial \over \partial x^i} = {\partial \over \partial z^i}$, we are left with
\begin{equation}\label{preparation}
f_I(x)  = \sum_{|r|\leq k}{\partial^r f_I\over (\partial x)^r}(x_0){(x-x_0)^r\over r!}\quad \hbox{if }x-x_0\hbox{ is nilpotent}.
\end{equation}

\subsection{The theory of DeWitt and Rogers}
Another theory of supermanifolds was developped by B. DeWitt and A. Rogers \cite{Rogers}. The model space here is
\begin{equation}\label{defidewitt}
\mathbb{R}^{m|k}_{L'}:= \left(\Lambda_{L'}^0\right)^m\times \left(\Lambda_{L'}^1\right)^k,
\end{equation}
where $\Lambda_{L'}$ is a superalgebra isomorphic to $\mathcal{C}^\infty(\mathbb{R}^{0|L'})\simeq \Lambda^*\mathbb{R}^{L'}$ and $L'\in \mathbb{N}$. (We may choose $L'=\infty$, but in the following we assume that $L'$ is finite for simplicity.) We define the \emph{body} map $\beta: \Lambda_{L'}\longrightarrow \mathbb{R}$ which, to each $z\in \Lambda_{L'}$, associates the only real number $\beta(z)$ such that $z-\beta(z)$ is nilpotent (we also call $\sigma(z):= z - \beta(z)$ the \emph{soul} map). We extend $\beta$ to a map $\beta: \mathbb{R}^{m|k}_{L'} \longrightarrow \mathbb{R}^m$ which, to each $(z^1,\cdots ,z^m,\zeta^1,\cdots ,\zeta^k)\in \mathbb{R}^{m|k}_{L'}$, associates $(\beta(z^1),\cdots ,\beta(z^m), \beta(\zeta^1),\cdots ,\beta(\zeta^k))\simeq (\beta(z^1),\cdots ,\beta(z^m))$. Of course, as a set, the right hand side in the definition (\ref{defidewitt}) looks too large. Its too large dimension is compensated by
\begin{itemize}
\item the choice of a (non Hausdorff) topology on it: open subsets of $\mathbb{R}^{m|k}_{L'}$ are of the form $\left(|\Omega|\times \left(\Lambda_{L'}^{*0}\right)^m\right)\times \left(\Lambda_{L'}^1\right)^k$, where $|\Omega|$ is an open subset of $\mathbb{R}^m$ and $\Lambda_{L'}^{*0}$ is the subset of nilpotent elements of $\Lambda_{L'}^0$. Alternatively open subsets of $\mathbb{R}^{m|k}_{L'}$ are inverse images of open subsets of $\mathbb{R}^m$ by $\beta$. 
\item a choice of restrictions on the class of functions on $\mathbb{R}^{m|k}_{L'}$ we are working with.
\end{itemize}
We are going to present classes of functions on $\mathbb{R}^{m|k}_{L'}$ which depend only on what is happening in a formal neighbourhood of $\mathbb{R}^m$ inside $\mathbb{R}^{m|k}_{L'}$. These classes of functions will be built starting from the coordinate functions
\[
\begin{array}{cccc}
x^i: & \mathbb{R}^{m|k}_{L'} & \longrightarrow & \Lambda_{L'}\\
& (z^1,\cdots ,z^m,\zeta^1,\cdots ,\zeta^k) & \longmapsto & z^i
\end{array}
\]
and
\[
\begin{array}{cccc}
\theta^i: & \mathbb{R}^{m|k}_{L'} & \longrightarrow & \Lambda_{L'}\\
& (z^1,\cdots ,z^m,\zeta^1,\cdots ,\zeta^k) & \longmapsto & \zeta^i.
\end{array}
\]
We observe that we should at least suppose that $L'\geq k$ in order to ensure that $\theta^1\cdots \theta^k\neq 0$.\\

\noindent
\textbf{The class $H^\infty$} --- For any open subset $\Omega\subset \mathbb{R}^{m|k}_{L'}$, a function $F: \Omega\longrightarrow \Lambda_{L'}$ belongs to $H^\infty(\Omega)$ if
\begin{equation}\label{HinfiniF}
F = \sum_{I\in \I^k}\theta^I\widehat{f}_I(x),
\end{equation}
where $\theta^I := \theta^{i_1}\cdots \theta^{i_j}$, for $I = (i_1,\cdots ,i_j)$ and, for each $I\in \I^k$, there exists a (\emph{real} valued !) function $f_I:\beta(\Omega)\longrightarrow \mathbb{R}$ such that, for any $z\in \Omega$,
\begin{equation}\label{Hinfinif}
\widehat{f}_I(z):= \sum_{r=(r_1,\cdots ,r_m)} {\partial^rf_I(\beta(z))\over (\partial \beta(z))^r}{(z-\beta(z))^r\over r!}.
\end{equation}
Note that the right hand side is a finite sum, since the components of $z-\beta(z)$ are nilpotent. One can motivate conditions (\ref{HinfiniF}) and (\ref{Hinfinif}) by Relation (\ref{preparation}) obtained in \S \ref{Panother}.

However the theory of $H^\infty$-functions is limited and its use for physical models leads to the same type of inconsistencies as the theory of skeletal algebro-geometric supermanifold. \\

\noindent
\textbf{The class $G^\infty$} --- Functions $f\in G^\infty(\Omega)$ can be defined in exactly the same way as functions in $H^\infty(\Omega)$ with the exception that, in (\ref{Hinfinif}), the function $f_I$ takes values in $\Lambda_{L'}$ instead of $\mathbb{R}$. Hence the set $G^\infty(\Omega)$ is much larger than $H^\infty(\Omega)$ and allows to perform many computations on physical models that are forbidden within $H^\infty(\Omega)$.\\

\noindent
\textbf{The class $GH^\infty$} --- Since however $G^\infty(\Omega)$ may be too big, one can use as a variant the set $GH^\infty(\Omega)$ of maps where one assumes that the functions $f_J$ in (\ref{Hinfinif}) takes values in $\Lambda_L$, for $L<L'$: it can be shown that, for many purposes is is enough to choose the integers $L$ and $L'$ such that $2k< L'\leq 2L$ (see \cite{Rogers07}). The resulting theory is similar to the algebro-geometric theory of supermanifolds with flesh.\\

\noindent
\textbf{Comparisons} --- 
Each of these notions of superfunctions leads to a particular different notion of supermanifolds, by gluing open subsets of $\mathbb{R}^{m|k}_{L'}$ though $H^\infty$-, $G^\infty$- or $GH^\infty$-diffeomorphisms, see \cite{Rogers07}. Again most of the `physical' applications requires the use of $G^\infty$- or $GH^\infty$-supermanifolds. One can compare the various theories as follows:
\begin{enumerate}
\item any $H^\infty$-supermanifold is a $G^\infty$- or a $GH^\infty$-supermanifold, a consequence of the trivial facts that any  $H^\infty$-function is a $GH^\infty$-function, and any  $GH^\infty$-function is a $G^\infty$-function;
\item conversely any $G^\infty$-supermanifold can be endowed with a (non unique) $H^\infty$-superstructure, a consequence of 3. and 4. below;
\item any $G^\infty$-supermanifold has a split structure (i.e. is of the form $S(|\mathcal{M}|,E)$, where $E$ is a vector bundle over $|\mathcal{M}|$) \cite{batchelor79};
\item any split manifold $S(|\mathcal{M}|,E)$ has a $H^\infty$-structure;
\item any $H^\infty$-supermanifold is a (skeletal) algebro-geometric supermanifold \cite{Rogers07};
\item conversely any (skeletal) algebro-geometric supermanifold is a $H^\infty$-supermanifold \cite{batchelor80}.
\end{enumerate}
See \cite{Rogers07} (and \cite{bahraini} about 5 and 6) for a complete overview.

\section{The superparticle: second version}\label{particlebis}
We now present the superparticle model using the superfield $\Phi = x+\theta\psi$, a map from $\mathbb{R}^{1|1}$ to an ordinary Riemannian manifold $\mathcal{N}$. We need for that purpose to know how to integrate a function a supermanifold.

\subsection{The Berezin integral}\label{integration}
The theory for integrating a function on a supermanifold is due to Berezin. In the following we just explain how to define such an integral on an open subset of $\mathbb{R}^{m|k}$. We do not expound the complete theory, which rests on a delicate change of variable formula. We refer to \cite{Manin,FreedDeligne,Rogers07} for a complete exposition.

Let $\Omega$ be an open subset of $\mathbb{R}^{m|k}$ and $f\in \mathcal{C}^\infty(\overline{\Omega})$ be a smooth function on the closure of $\Omega$. Let us write $f = \sum_{I\in \I^k} \theta^If_I$, where $f_I\in \mathcal{C}^\infty(|\overline\Omega|)\otimes \mathcal{C}^\infty(\mathbb{R}^{0|L})$, $\forall I\in \I^k$, for some $L\in \mathbb{N}$. Then we define the Berezin integral of $f$ over $\Omega$ by
\[
\int_\Omega  dx^1\cdots dx^md\theta^1\cdots d\theta^kf := \int_{|\Omega|} dx^1\cdots dx^mf_{1\cdots k}.
\] 
So we have selected the highest degree coefficient $f_{1\cdots k}$ in the decomposition of $f$ and we just integrate it over $|\Omega|$. Note that this integral takes values in $\mathcal{C}^\infty(\mathbb{R}^{0|L})$. It also enjoys the following property
\begin{equation}\label{integraltranslation}
\int_\Omega  dx^1\cdots dx^md\theta^1\cdots d\theta^k f(x,\eta+\zeta) = \int_\Omega dx^1\cdots dx^md\theta^1\cdots d\theta^k f(x,\eta),
\end{equation}
which reflects the invariance by translation in the odd variables of the Berezin integral. Indeed (\ref{integraltranslation}) just follows by expanding $(\eta^1+\zeta^1)\cdots (\eta^k+\zeta^k)$ and using the definition of the Berezin integral. The `infinitesimal' version of (\ref{integraltranslation}) is:
\begin{equation}\label{intinfitranslation}
\forall j\in [\![1,k]\!],\quad
\int_\Omega  dx^1\cdots dx^md\theta^1\cdots d\theta^k {\partial f\over \partial \theta^j} = 0.
\end{equation}
This can be shown by observing that the action of ${\partial \over \partial \theta^j}$ decreases by one the degree of all monomials $\theta^I$. For instance ${\partial \over \partial \theta^j}(\theta^1\cdots \theta^k) = (-1)^{j-1}\theta^1\cdots \theta^{j-1}\theta^{j+1}\cdots \theta^k$, so that $\int_\Omega dx^1\cdots dx^md\theta^1\cdots d\theta^k {\partial \over \partial \theta^j}(\theta^1\cdots \theta^k) = 0$.

\subsection{The supertime $\mathbb{R}^{1|1}$}\label{supertime}

The simplest non trivial example of a superspace is the `supertime' $\mathbb{R}^{1|1}$. We denote by $t$ (the time) the even coordinate and by $\theta$ the odd coordinate on it. We will consider three vector fields on $\mathbb{R}^{1|1}$: the time derivative ${\partial \over \partial t}$ acting on even functions on $\mathbb{R}^{1|1}\times \mathbb{R}^{0|L}$ which is the extension of the usual operator as described in \S \ref{derivatives} and:
\[
D:= {\partial \over \partial \theta} - \theta{\partial \over \partial t}
\quad \hbox{and}\quad
\tau_Q:= {\partial \over \partial \theta} + \theta{\partial \over \partial t}.
\]
Clearly $D$ and $\tau_Q$ are odd operators, i.e. they change the parity of the maps on which they act. So in particular, if $\Phi:\mathbb{R}^{1|1}\times \mathbb{R}^{0|L}\longrightarrow \mathcal{N}$ is an even map, $D\Phi$ and $\tau_Q\Phi$ are not even. There are then two ways to deal with these derived maps: either we introduce the supermanifold $\Pi T\mathcal{N}$ and we can make sense of these derivatives directly. Or we allow only to consider $\eta D\Phi$ and $\eta\tau_Q\Phi$, where $\eta$ is odd, and we are then in the setting expounded in \S \ref{derivatives} (i.e. we implicitely assume that we are working with maps on $\mathbb{R}^{1|1}\times \mathbb{R}^{0|L}$). Note that $D$ (resp. $\tau_Q$) can be considered as a left (resp. right) invariant odd vector field on $\mathbb{R}^{1|1}$ (we shall see why in Section \ref{superpoincare}).

Operators ${\partial \over \partial t}$, $D$ and $\tau_Q$ satisfy the following relations:
\[
D^2 = -{\partial \over \partial t},\quad \tau_Q^2 = {\partial \over
  \partial t},\quad D\tau_Q + \tau_QD = 0, 
\]
\[
D{\partial \over \partial t} - {\partial \over \partial t}D =
\tau_Q{\partial \over \partial t} - {\partial \over \partial t}\tau_Q
= 0. 
\]
For example
\[
D^2(f(t)+\theta g(t)) = D d\left(g(t) - \theta\dot{f}(t)\right) = - \dot{f}(t) - \theta\dot{g}(t).
\]
These relations can be summarized together by introducing the supercommutator
$[\cdot,\cdot]$ of two homogeneous operators $A$ and $B$: if $A$ has parity
$\hbox{gr}(A)$ ($=0$ if $A$ is even, $=1$ if $A$ is odd) and $B$ has parity $\hbox{gr}(B)$, then
\[
[A,B]:= A\circ B - (-1)^{\hbox{gr}(A)\hbox{gr}(B)}B\circ A.
\]
We thus have
\[
[D,D] = -2{\partial \over \partial t},\quad \left[\tau_Q,\tau_Q\right]
= 2{\partial \over \partial t}, \quad 
\left[D,\tau_Q\right] = \left[D,{\partial \over \partial t}\right] =
\left[\tau_Q,{\partial \over \partial t}\right] = 0. 
\]
We see that the space $\mathfrak{m}^{1|1}$ `spanned' by ${\partial \over \partial t}$, $D$ and $\tau_Q$ is a Lie super algebra, the super Lie algebra of supertranslations of $\mathbb{R}^{1|1}$. Note that, as pointed out in \S \ref{otheralgebras}, Lie super brackets can be replaced by ordinary Lie brackets, which are easier to handle with, by tensoring $\mathfrak{m}^{1|1}$ with $\mathcal{C}^\infty(\mathbb{R}^{0|L}) = \mathbb{R}[\eta^1\cdots \eta^L]$. One then manipulates only linear combinations of ${\partial \over \partial t}$, $D$ and $\tau_Q$ with coefficients in $\mathbb{R}[\eta^1\cdots \eta^L]$ which are even. For instance
$\eta^1D$ and $\eta^2\tau_Q$ are even and their \emph{commutator} is:
\[
\left[\eta^1D,\eta^2\tau_Q\right] = \left(\eta^1D\right) \left(\eta^2\tau_Q\right) - \left(\eta^2\tau_Q\right) \left(\eta^1D\right) = 
- \eta^1\eta^2\left(D \tau_Q + \tau_Q D\right) =
-2\eta^1\eta^2\left[D, \tau_Q \right], 
\]
which is equal to $0$ because the \emph{anticommutator} $\left[D,\tau_Q\right]$ vanishes.

Note also that if we consider $\mathfrak{m}^{1|1}\otimes \mathbb{R}[\theta]$, the space spanned by ${\partial \over \partial t}$, $D$ and $\tau_Q$ over $\mathbb{R}[\theta]$, then these three generators are not linearly independant, because of the relation $\tau_Q-D = 2\theta {\partial \over \partial t}$.

\subsection{The superfield formulation of the superparticle model}
Now we come back to the superparticle: instead of working on fields $(x,\psi)$ we consider even maps $\Phi = x + \theta\psi$ from $\mathbb{R}^{1|1}\times \mathbb{R}^{0|L}$ to $\mathcal{N}$, with the meaning discussed in \S \ref{thetruth} (i.e. $\Phi = x(1+\theta \psi)$). Recall that the vector field $Q = -\eta\tau_Q$ was defined by
\[
Q(x,\psi) = (-\eta \psi, \eta \dot{x}).
\]
It is straightforward to check that this coincides with the action of $-\eta\tau_Q$ on $\Phi$ since: 
\[
\eta\tau_Q(\Phi) =  \eta \left({\partial \over \partial \theta} + \theta{\partial \over \partial t}\right) (x + \theta\psi) = \eta \left( \psi + \theta{\partial x\over \partial t}\right) = \eta\psi - \theta \left(\eta{\partial x\over \partial t}\right). 
\]
Furthermore the action $\mathcal{L}[x,\psi]$ can be written in terms of
the field $\Phi$ 
as follows. Consider the following action, given by a Berezin integral (see \S \ref{integration}):
\[
\mathcal{L}[\Phi]:= -{1\over 2}\int\int_{\mathbb{R}^{1|1}} dt\,d\theta\left\langle D\Phi, {\partial \Phi\over \partial t}\right\rangle .
\]
Through the substitution $\Phi = x+\theta\psi$ one finds
\begin{equation}\label{DPhidPhi}
-{1\over 2}\left\langle D\Phi,{\partial \Phi\over
 \partial t}\right\rangle = -{1\over 2}\left\langle \psi -
 \theta\dot{x},\dot{x}+\theta\nabla_{\dot{x}}\psi\right\rangle 
 = -{1\over 2}\left\langle \psi,\dot{x}\right\rangle
 + {\theta\over 2} 
 \left(  |\dot{x}|^2 + \left\langle \psi,
 \nabla_{\dot{x}}\psi\right\rangle \right). 
\end{equation}
Thus we recover
\[
\mathcal{L}[\Phi] = {1\over 2}\int_{\mathbb{R}}dt \left(  |\dot{x}|^2 + \left\langle \psi, \nabla_{\dot{x}}\psi\right\rangle \right) 
= \mathcal{L}[x,\psi],
\]
The superfield formulation allows an alternative proof of the invariance of $\mathcal{L}[\Phi]$ under the action of $Q$. For that purpose we need to compute $\mathcal{L}[\Phi - \eta\tau_Q\Phi]$. First we have (using the fact that $\eta$ anticommutes with $D$):
\[
\left\langle D(\Phi - \eta\tau_Q\Phi), {\partial \over \partial t}(\Phi - \eta\tau_Q\Phi) \right\rangle = \left\langle D\Phi, {\partial \Phi\over \partial t}\right\rangle + \eta \left\langle D^\nabla \tau_Q\Phi, {\partial \Phi\over \partial t}\right\rangle + \eta \left\langle D\Phi, {\partial^\nabla \over \partial t}\tau_Q\Phi\right\rangle,
\]
where $D^\nabla$ and ${\partial^\nabla \over \partial t}$ denote the covariant derivatives for the pull-back of the Levi-Civita connection on $\mathcal{N}$ by $\Phi$ (or by $\mathfrak{F}$, see \S \ref{derivatives}). Now we use the anticommutation relation $[D,\tau_Q] = 0$, the commutation relation $\left[{\partial \over \partial t},\tau_Q\right] = 0$ and the fact that $\nabla$ is torsion free to get:\\

$\displaystyle \left\langle D(\Phi - \eta\tau_Q\Phi), {\partial \over \partial t}(\Phi - \eta\tau_Q\Phi) \right\rangle $
\[
\begin{array}{ccl}
 & = & \displaystyle  \left\langle D\Phi, {\partial \Phi\over \partial t}\right\rangle + \eta \left\langle -\tau_Q^\nabla D \Phi, {\partial \Phi\over \partial t}\right\rangle + \eta \left\langle D\Phi, \tau_Q^\nabla{\partial \over \partial t}\Phi\right\rangle\\
& = & \displaystyle  \left\langle D\Phi, {\partial \Phi\over \partial t}\right\rangle - \eta\tau_Q \left\langle D\Phi, {\partial \Phi\over \partial t}\right\rangle .
\end{array}
\]
We hence deduce an expression for $\mathcal{L}[\Phi - \eta\tau_Q\Phi]$. By a comparison with the relation $\mathcal{L}[\Phi - \eta\tau_Q\Phi] = \mathcal{L}[\Phi] -\eta\tau_Q\iN \delta \mathcal{L}[\Phi]$, we deduce
\[
\eta\tau_Q\iN \delta \mathcal{L}[\Phi] = -{1\over 2}\int\int_{\mathbb{R}^{1|1}} dt\,d\theta\,\eta\tau_Q\left\langle D\Phi,
{\partial \Phi\over \partial t}\right\rangle .
\]
At this point we can conclude that $-\eta\tau_Q$ induces a symmetry since $\eta\tau_Q\iN \delta \mathcal{L}[\Phi]$ reduces to a boundary integral and because of (\ref{intinfitranslation}). However it is interesting to compute explicitely this term in view of obtaining the Noether charge of this symmetry. We observe that if we note $\left\langle D\Phi,{\partial \Phi\over \partial t}\right\rangle = f+\theta g$, then using the fact that  $f$ is odd (see (\ref{DPhidPhi})),
\[
\int\int_{\mathbb{R}^{1|1}} dt\,d\theta \,\eta\tau_Q(f + \theta g) = \int\int_{\mathbb{R}^{1|1}} dt\,d\theta \,\left(\eta g -\theta\eta{\partial f\over \partial t} \right) = -\int_\mathbb{R} dt\,\eta{\partial f\over \partial t}.
\]
Hence, in view of (\ref{DPhidPhi}),
\begin{equation}\label{avantNoether}
\eta\tau_Q\iN \delta \mathcal{L}[\Phi] = {1\over 2}\eta\int_\mathbb{R} dt {\partial \over \partial t}\left\langle \psi,\dot{x}\right\rangle ,
\end{equation}
and we obtain the same result as in Section \ref{toy}. We now recover Noether's theorem. It relies on redoing the preceding calculation but $\eta\tau_Q$ is now replaced by a smooth compactly supported `modulation' of it, i.e. we compute $\chi\eta\tau_Q\iN\delta\mathcal{L}[\Phi]$, where $\chi\in \mathcal{C}^\infty_c(\mathbb{R})$. It gives us:
\[
\mathcal{L}[\Phi -\chi\eta\tau_Q\Phi] = \mathcal{L}[\Phi] - \delta^{(0)}_{\chi\eta\tau_Q}[\Phi] - \delta^{(1)}_{\chi\eta\tau_Q}[\Phi],
\]
where we have splitted $\chi\eta\tau_Q\iN\delta\mathcal{L}[\Phi]$ into the sum of $\delta^{(0)}_{\chi\eta\tau_Q}[\Phi]$, which is a linear expression in $\chi$, and of $\delta^{(1)}_{\chi\eta\tau_Q}[\Phi]$, which is a linear expression in ${d\chi\over dt}$. However $\delta^{(0)}_{\chi\eta\tau_Q}[\Phi]$ is simply obtained by inserting $\chi$ in the right hand side of (\ref{avantNoether}), i.e.
\[
\delta^{(0)}_{\chi\eta\tau_Q}[\Phi] = {1\over 2}\eta\int_\mathbb{R} dt \chi{\partial \over \partial t}\left\langle \psi,\dot{x}\right\rangle.
\]
On the other hand
\[
\delta^{(1)}_{\chi\eta\tau_Q}[\Phi] = {1\over 2}\int_\mathbb{R} dt\,d\theta \left[D\chi \left\langle \eta\tau_Q\Phi,{\partial \Phi\over \partial t}\right\rangle + {d\chi\over dt} \left\langle D\Phi,\eta\tau_Q\Phi\right\rangle \right].
\]
We compute then $D\chi \left\langle \eta\tau_Q\Phi,{\partial \Phi\over \partial t}\right\rangle = - \theta\eta {d\chi\over dt} \langle \psi,\dot{x}\rangle$ and $\left\langle D\Phi,\eta\tau_Q\Phi\right\rangle = - 2\theta\eta\langle \psi,\dot{x}\rangle$, so that
\[
\delta^{(1)}_{\chi\eta\tau_Q}[\Phi] = {3\over 2} \int_\mathbb{R}dt\,\eta {d\chi\over dt} \langle \psi,\dot{x}\rangle.
\]
By integrating by part the last expression we then conclude that $\chi\eta\tau_Q\iN\delta\mathcal{L}[\Phi] = -\eta\int_\mathbb{R} dt \chi{\partial \over \partial t}\left\langle \psi,\dot{x}\right\rangle$, as in Section \ref{toy}.

\section{Superfields on $\mathbb{R}^{3|2}$}
\subsection{The super Minkowski space $M^{(1,2)|2}\simeq \mathbb{R}^{3|2}$}\label{super3+2}
Another example of superspace is $\mathbb{R}^{3|2}$, a supersymmetric extension of the Minkowski space of dimension $1+2$. We denote the even coordinates by $t,x$ and $y$ and the odd coordinates by $\theta^1$ and $\theta^2$. Beside the translation vector fields $\partial/\partial t$, $\partial/\partial x$ and $\partial/\partial y$, we define the (right invariant) supertranslation generators
\begin{equation}\label{R32tau}
\tau_1 :=  {\partial \over
\partial \theta^1} + \theta^1 \left({\partial \over \partial t}+
{\partial \over \partial x}\right) + \theta^2{\partial \over
\partial y},
\quad
\tau_2:=  {\partial \over
\partial \theta^2} + \theta^1{\partial \over \partial y} +
\theta^2\left({\partial \over \partial t} - {\partial \over
\partial x}\right),
\end{equation}
which satisfy the supercommutation relations
\begin{equation}\label{R32taucom}
[\tau_1,\tau_1]  = 2\left({\partial \over \partial t}
+ {\partial \over \partial x}\right),
\quad
[\tau_1,\tau_2] = 2{\partial \over \partial y},
\quad
[\tau_2,\tau_2] = 2\left({\partial \over \partial t} -
{\partial \over \partial x}\right).
\end{equation}
It may be convenient to embedd the Minkowski space $\mathbb{R}^3$ in the space $M(2,\mathbb{R})$ of $2\times 2$ real matrices with the coordinates $v^{11}, v^{12}, v^{21}$ and $v^{22}$ though the map
\[
\begin{array}{ccc}
\mathbb{R}^3 & \longrightarrow & M(2,\mathbb{R})\\
(t,x,y) &\longmapsto & v = \left(\begin{array}{cc}v^{11} & v^{12}\\
v^{21} & v^{22}\end{array}\right) =  \left(\begin{array}{cc}(t+x)/2 & y/2\\
y/2 & (t-x)/2\end{array}\right),
\end{array}
\]
i.e. to view $\mathbb{R}^3$ as the hyperplane $v^{12}-v^{21} = 0$ or the space $\mathfrak{h}_2(\mathbb{R})$ of symmetric $2\times 2$ real matrices. We then recover the Minkowski norm by the relation $t^2-x^2-y^2 = 4\hbox{ det }v$. Functions on $\mathbb{R}^3$ can then be identified with functions on $M(2,\mathbb{R})$ annihilated by ${\partial \over \partial v^{12}} - {\partial \over \partial v^{21}}$ and we have 
\[
{\partial \over \partial t} = {1\over 2}\left( {\partial \over \partial v^{11}} + {\partial \over \partial v^{22}}\right),\quad
{\partial \over \partial x} = {1\over 2}\left( {\partial \over \partial v^{11}} - {\partial \over \partial v^{22}}\right),\quad
{\partial \over \partial y} = {1\over 2}\left( {\partial \over \partial v^{12}} + {\partial \over \partial v^{21}}\right).
\]
Then, setting $\partial_a:= {\partial \over \partial \theta^a}$ and $\partial_{ab}:= {\partial \over \partial v^{ab}}$, (\ref{R32tau}) can be written
\[
\tau_a := \partial_a + \theta^b\partial_{ab},
\quad
\hbox{for }a,b = 1,2.
\]
Similarly (\ref{R32taucom}) becomes
\[
[\tau_a,\tau_b] = 2 \partial_{ab}.
\]
We define also the (left invariant) odd vector fields
\begin{equation}\label{R32D}
D_a := \partial_a - \theta^b\partial_{ab},
\quad
\hbox{for }a,b = 1,2.
\end{equation}
which satisfy the supercommutation relations $[D_a,\tau_b]= 0$ and 
\begin{equation}\label{R32Dcom}
[D_a,D_b]= - 2 \partial_{ab}.
\end{equation}

\subsection{An example of supersymmetric model}
As an elementary example of a supersymmetric model we consider `scalar' fields $\Phi$ on $\mathbb{R}^{3|2}$ which are critical points of the functional
\[
\mathcal{A}[\Phi]:= \int_{\mathbb{R}^{3|2}}d^3xd^2\theta 
\left({1\over 4}\epsilon^{ab}D_a\Phi D_b\Phi + \Phi^*h\right), 
\] 
where $d^3x:= dtdxdy$, $d^2\theta:= d\theta^1d\theta^2$ and  $h\in \mathcal{C}^\infty(\mathbb{R},\mathbb{R})$. We note $\Phi = \phi + \theta^\alpha \psi_\alpha + {1\over 2}\epsilon_{\alpha\beta}\theta^\alpha \theta^\beta F = \phi +
\theta^1 \psi_1 + \theta^2 \psi_2 + \theta^1\theta^2F$. Then, by setting
\[
({\cal D}\psi)_1 = \partial_{12} \psi_1 -
\partial_{11} \psi_2 \quad \hbox{and} \quad
({\cal D}\psi)_2 = \partial_{22} \psi_1 -
\partial_{12} \psi_2,
\]
we have
\[
D_1\Phi = \psi_1 - \theta^1\partial_{11} \phi +
\theta^2\left(F - \partial_{21} \phi\right) +
\theta^1\theta^2({\cal D}\psi)_1,
\]
\[
D_2\Phi = \psi_2 - \theta^1\left(F + \partial_{12} \phi\right) - \theta^2 \partial_{22} \phi +
\theta^1\theta^2({\cal D}\psi)_2.
\]
We hence find that, denoting $\psi{\cal D}\psi := \epsilon^{ab}\psi_a({\cal D}\psi)_b = \psi_1({\cal D}\psi)_2 - \psi_2({\cal D}\psi)_1$,
\[
\begin{array}{ccl}
\displaystyle \int d^2\theta \left({1\over
4}\epsilon^{ab}D_a\Phi \cdot D_b\Phi\right) & = & \displaystyle
{1\over 2}{\partial \phi\over \partial y^{11}} {\partial \phi\over
\partial y^{22}} - {1\over 2} \left({\partial \phi\over \partial
y^{12}}\right)2
+ {1\over 2}\psi{\cal D}\psi + {1\over 2}F^2\\
& = & \displaystyle {1\over 2}\left({\partial \phi\over \partial
t}\right)^2 - {1\over 2}\left({\partial \phi\over \partial
x}\right)^2 - {1\over 2}\left({\partial \phi\over \partial
y}\right)^2 + {1\over 2}\psi{\cal D}\psi + {1\over 2}F^2.
\end{array}
\]
The computation of $\Phi^*h$ can be done as in \S \ref{thetruth}: it is better to understand the decomposition of $\Phi$ as $\Phi = \phi (1+ \theta^1 \psi_1)(1 + \theta^2 \psi_2)(1 + \theta^1\theta^2F)$, where $\psi_1$, $\psi_2$ and $F$ are derivation operators. This gives us
\[
\Phi^*h = h\circ \phi + (h'\circ \phi)\left( \theta^1\psi_1 +
\theta^2\psi_2 + \theta^1\theta^2F\right) - (h''\circ \phi)
\theta^1\theta^2\psi_1 \psi_2,
\]
so that
\[
\int d\theta^1d\theta^2 \Phi^*h = (h'\circ \phi)F - (h''\circ
\phi)\psi_1 \psi_2.
\]
Hence we conclude that
\[
{\cal A}[\Phi] = \int_{\mathbb{R}^3}d^3x\left[ {1\over 2}\left({\partial \phi\over \partial t}\right)^2 - {1\over 2}|\nabla
\phi|^2 + {1\over 2}\psi{\cal D}\psi - h''(\phi)\psi_1
\psi_2 + {1\over 2}F^2+ h'(\phi)F\right],
\]
where $|\nabla \phi|^2:= 
\left({\partial \phi\over \partial x}\right)^2 + \left({\partial
\phi\over \partial y}\right)^2$. Another way to write this action is
\[
\mathcal{A}[\Phi] = {1\over 2}\int_{\mathbb{R}^3}d^3x\left[
\left({\partial \phi\over \partial t}\right)^2 -|\nabla \phi|^2 + \psi{\cal D}\psi - 2h''(
\phi)\psi_1 \psi_2 - (h'(\phi))^2\right] + \mathcal{Q}[\Phi],
\]
where $\mathcal{Q}[\Phi]:= {1\over 2}\int_{\mathbb{R}^3}d^3x (F+h'(
\phi))^2$. This makes clear how to eliminate the \emph{auxiliary field} $F$. The Euler--Lagrange system of equations is 
\begin{equation}\label{eulerM32}
\left\{
\begin{array}{ccl}
\displaystyle \square \phi + h''(\phi)h'(\phi)
+ h'''(\phi)\psi_1 \psi_2 & = & 0\\

\displaystyle {\cal D}\psi -(h''\circ \phi)\psi & = & 0\\

F+ h'(\phi) & = & 0,
\end{array}\right.
\end{equation}
where $\square := {\partial ^2\over (\partial t)^2} - {\partial
^2\over (\partial x)^2}-{\partial ^2\over (\partial y)^2}$. We see that the equation on $F$ is not coupled with the other equations ($F$ has no influence on $\phi$ et $\psi$).

\subsection{Bogomolny'i--Prasad--Sommerfeld solutions}\label{Bogomolnyi}
It is interesting to look at fields $\Phi= \phi + \theta^\alpha \psi_\alpha + {1\over 2}\epsilon_{\alpha\beta}\theta^\alpha \theta^\beta F$ which satisfies the condition
\begin{equation}\label{BPS}
\left\{ \begin{array}{ccc}
(\cos \alpha \tau_1 + \sin \alpha \tau _2)\cdot \Phi & = & 0\\
\psi & = & 0\\
F+h'(\phi) & = & 0,
\end{array} \right.
\end{equation}
where $\alpha \in \mathbb{R}$.

\begin{prop} Any super field $\Phi$ which satisfies (\ref{BPS}) is a solution
of the Euler--Lagrange system (\ref{eulerM32}).
\end{prop}
\emph{Proof} --- The first condition in (\ref{BPS}) gives us (by taking into account the fact that $\psi = 0$)
\[
\begin{array}{ccc}
\displaystyle 
\theta^1\left[\cos \alpha
\left({\partial \phi\over \partial t} + {\partial \phi\over
\partial x}\right) + \sin \alpha \left({\partial \phi\over \partial y}
- F\right)\right] & & \\
\displaystyle + \theta^2\left[\sin \alpha \left({\partial
\phi\over \partial t} - {\partial \phi\over \partial x}\right) +
\cos \alpha \left({\partial \phi\over \partial y} + F\right)\right]
& = & 0.
\end{array}
\]
So that $\Phi$ satisfies:
\[
\left\{ \begin{array}{ccc} \displaystyle \cos \alpha {\partial
\phi\over \partial t} + \cos \alpha {\partial \phi\over \partial
x} + \sin \alpha
{\partial \phi\over \partial y} & = &\sin \alpha F\\
\displaystyle \sin \alpha {\partial \phi\over \partial t} - \sin
\alpha {\partial \phi\over \partial x} + \cos \alpha {\partial
\phi\over \partial y} & = & -\cos \alpha F.
\end{array}\right.
\]
Denote $X:= \cos 2\alpha {\partial \over \partial x} + \sin
2\alpha {\partial \over \partial y}$ and $Y:= -\sin 2\alpha
{\partial \over \partial x} + \cos 2\alpha {\partial \over
\partial y}$. By multiplying the first equation by $\cos \alpha$ and the second one by $\sin \alpha$, we obtain
\begin{equation}\label{firstequation}
{\partial \phi\over \partial t} + X \phi = 0.
\end{equation}
By multiplying the first equation by $-\sin
\alpha$ and the second one by $\cos \alpha$, we obtain  $Y
\phi = -F$ which, taking into account the relation $F+ h'(\phi) = 0$, gives us
\begin{equation}\label{secondequation}
Y \phi = h'(\phi).
\end{equation}
Then on the one hand equation (\ref{firstequation}) implies $\left( {\partial ^2\over (\partial t)^2} -X^2\right) \phi =
\left( {\partial \over \partial t} -X\right) \left( {\partial
\over \partial t} +X\right) \phi = 0$, or equivalentely
\begin{equation}\label{firstequation1}
{\partial ^2\phi\over (\partial t)^2} -\cos^22\alpha {\partial
^2\phi\over (\partial x)^2} -2\cos 2\alpha\sin 2\alpha{\partial
^2\phi\over \partial x\partial y} -\sin^22\alpha {\partial
^2\phi\over (\partial y)^2} = 0.
\end{equation}
On the other hand equation (\ref{secondequation}) gives $
Y^2 \phi = Y (h'\circ \phi) = (h''\circ \phi)Y \phi =
(h''\circ \phi)(h'\circ \phi)$, or equivalentely
\begin{equation}\label{secondequation1}
\sin^22\alpha {\partial ^2\phi\over (\partial x)2} -2\cos
2\alpha\sin 2\alpha{\partial ^2\phi\over \partial x\partial y}
+\cos^22\alpha {\partial ^2\phi\over (\partial y)2} = h''(\phi)h'(\phi).
\end{equation}
By substracting (\ref{firstequation1}) and (\ref{secondequation1}) we find that
\[
\square \phi + h''(\phi)h'(\phi) = 0.
\]
Particularly interesting cases are: (a) if $\alpha = \pi/4$, then we find  ${\partial \phi\over \partial t} + {\partial \phi\over \partial y} = 0$ and
${\partial \phi\over \partial x} = - h'(\phi)$; (b) if $\alpha = -\pi/4$, then we find ${\partial \phi\over \partial t} - {\partial \phi\over \partial y} = 0$ and
${\partial \phi\over \partial x} =  h'(\phi)$. If we further impose the condition ${\partial \phi\over \partial t} ={\partial \phi\over \partial y} = 0$ we then find the so-called \emph{Bogomolny'i} solutions: they are stationary solutions which minimizes the energy
\[
E[\phi]:= {1\over 2}\int_{\mathbb{R}}dx\left[\left({\partial \phi\over \partial t}\right)^2 + \left({\partial \phi\over \partial x}\right)^2 + (h'\circ \phi)^2\right]
\]
among stationary fields with the same asymptotic data at infinity, a consequence of the identity
\[
E[\phi] = {1\over 2}\int_{\mathbb{R}}dx\left[\left({\partial \phi\over \partial t}\right)^2 + \left({\partial \phi\over \partial x}\mp h'(\phi)\right)^2 \pm 2{\partial \over \partial x}(h\circ \phi)\right].
\]
Moreover the energy is equal to the `topological charge' $\pm [ h(\phi(\infty)) - h(\phi(-\infty))]$. The interest of these solutions is also that, by imposing a invariance condition on $\Phi$, we necessarily obtain a dynamical solution. However the condition of being invariant by a supersymmetry translation is stable by quantization, i.e. it makes sense to speak of quantum states which are invariant by such a supertranslation. Such states are called \emph{BPS states}, after Bogomolny'i, Prasad and Sommerfeld.

\section{Other super Minkowski spaces}\label{superpoincare}
We present here the super Minkowski spaces (with minimal supersymmetry) and the associated super Poincar{\'e} groups for the dimensions $1+2$, $1+3$, $1+5$ and $1+9$: they admit particularly beautiful representations since each of these dimensions is associated with respectively: $\mathbb{R}$, $\mathbb{C}$, the quaternions $\mathbb{H}$ and the octonions (or Cayley numbers) $\mathbb{O}$. They are the only normed division algebras and have increasing dimensions and complexity: $\mathbb{R}$ and $\mathbb{C}$ are commutative fields, $\mathbb{H}$ is a non commutative field and $\mathbb{O}$ is not associative nor commutative. A good presentation of octonions can be found in \cite{Baez} (see also \S\ref{reduction}). In the following we set $\mathbb{K} = \mathbb{R}$, $\mathbb{C}$, $\mathbb{H}$ or $\mathbb{O}$ and $k:= \hbox{dim }\mathbb{K}$.

\subsection{The Lorentz group in dimensions $3$, $4$, $6$ and $10$}
We first observe that the Minkowski space $\mathbb{R}^{1,1+k}$ can be identified with $\mathbb{R}^{1,1}\times \mathbb{K}$ which, itself, can be identified with the space $\mathfrak{h}_2(\mathbb{K})$ of Hermitian $2\times 2$ matrices over $\mathbb{K}$ by
\begin{equation}\label{defimu}
\begin{array}{cccc}
h: & \mathbb{R}^{1,1}\times \mathbb{K} & \longrightarrow & \mathfrak{h}_2(\mathbb{K})\\
& (t,x,z) & \longmapsto & {1\over 2}\left(\begin{array}{cc}
t+x & z \\ \overline{z} & t-x
\end{array}\right).
\end{array}
\end{equation}
The Minkowski scalar product on $\mathbb{R}^{1+1}\times \mathbb{K}$ has a simple expression by using $h$, since
\[
t^2 - x^2 - |z|^2 = 4 \hbox{ det }h(t,x,z).
\]
This leads to the identification $\hbox{Spin}(1,1+k) = SL(2,\mathbb{K})$. For $k=1$ or $2$ (i.e. respectively $\mathbb{K} = \mathbb{R}$ or $\mathbb{K} = \mathbb{C}$), this is easy to see. Indeed, if $\mathbb{K}=\mathbb{R}$ or $\mathbb{C}$, it is straightforward to check that, for any $g\in SL(2,\mathbb{K})$,
\begin{equation}\label{RgK}
\begin{array}{cccc}
R_g^{\mathfrak{h}_2(\mathbb{K})}: & \mathfrak{h}_2(\mathbb{K}) & \longrightarrow & \mathfrak{h}_2(\mathbb{K})\\
& m & \longmapsto & gmg^\dagger,
\end{array}
\end{equation}
where $g\dagger := \overline{g}^t$,
makes sense and is a linear transformation which preserves the Minkowski norm $\hbox{ det }m$. Moreover the map $g \longmapsto R_g^{\mathfrak{h}_2(\mathbb{K})}$ is a 2:1 group morphism which is the Spin covering of $SO_0(1,1+k)$. The advantage of this construction is that it comes naturally with the spinor representation: the natural action of $SL(2,\mathbb{K})$ on $\mathbb{K}^2$.

The cases where $\mathbb{K} = \mathbb{H}$ or $\mathbb{O}$ are similar but require some care, since there are no direct definitions of $SL(2,\mathbb{H})$ and $SL(2,\mathbb{O})$. Let us consider first the situation at the level of Lie algebras. We denote by $\mathfrak{tf}(2,\mathbb{K})$ the vector space of trace free $2\times 2$ matrices over $\mathbb{K}$. We naively expect that $\mathfrak{sl}(2,\mathbb{K})$, the Lie algebra of $SL(2,\mathbb{K})$, coincides with $\mathfrak{tf}(2,\mathbb{K})$ and it is so for $\mathbb{K} = \mathbb{R}$ or $\mathbb{C}$ but not for $\mathbb{K} = \mathbb{H}$ and $\mathbb{O}$. Moreover\footnote{The difference between the dimensions is $3 = \hbox{dim}\mathfrak{so}(3) = \hbox{dim}\mathfrak{so}(\hbox{Im }\mathbb{H})$, see below.} $\hbox{dim}_\mathbb{R}\mathfrak{tf}(2,\mathbb{H}) = 12 < 15 = \hbox{dim Spin}(1,5) = \hbox{dim}\mathfrak{so}(1,5)$ and, similarly\footnote{The difference between the dimensions is $21 = \hbox{dim}\mathfrak{so}(7) = \hbox{dim}\mathfrak{so}(\hbox{Im }\mathbb{O})$, see below.} $\hbox{dim}_\mathbb{R}\mathfrak{tf}(2,\mathbb{O}) = 24 < 45 = \hbox{dim Spin}(1,9) = \hbox{dim}\mathfrak{so}(1,9)$. Actually, for $\mathbb{K} = \mathbb{H}$ or $\mathbb{O}$ we need to define $\mathfrak{sl}(2,\mathbb{K})$, the Lie algebra of $SL(2,\mathbb{K})$, as the Lie algebra of endomorphisms over $\mathbb{R}$ of $\mathbb{K}^2$ \emph{spanned} by the action of $\mathfrak{tf}(2,\mathbb{K})$ or, equivalentely, to define $\mathfrak{sl}(2,\mathbb{K})$ as the Lie algebra of real endomorphisms of $\mathfrak{h}_2(\mathbb{K})$ spanned by linear maps
\[
\begin{array}{cccc}
\rho_\sigma^{\mathfrak{h}_2(\mathbb{K})}: & \mathfrak{h}_2(\mathbb{K}) & \longrightarrow & \mathfrak{h}_2(\mathbb{K})\\
& m & \longmapsto & {1\over 2}\left(\sigma m + m \overline{\sigma}^t\right),
\end{array}
\]
for all values of $\sigma \in \mathfrak{tf}(2,\mathbb{K})$. (Note that $\sigma \longmapsto \rho_\sigma^{\mathfrak{h}_2(\mathbb{K})}$ is (${1\over 2}$ times) the linearization of $g\longmapsto R_g^{\mathfrak{h}_2(\mathbb{K})}$ at $g=1$.) So let $\rho_{\mathfrak{tf}(2,\mathbb{K})}^{\mathfrak{h}_2(\mathbb{K})}:= \{\rho_\sigma^{\mathfrak{h}_2(\mathbb{K})}|\ \sigma\in \mathfrak{tf}(2,\mathbb{K})\} \subset \hbox{End}_\mathbb{R}(\mathfrak{h}_2(\mathbb{K}))$ and let us define $\mathfrak{sl}(2,\mathbb{K})$ to be \emph{the Lie subalgebra of $\left( \hbox{End}_\mathbb{R}(\mathfrak{h}_2(\mathbb{K})), [\cdot,\cdot]\right)$ spanned by $\rho_{\mathfrak{tf}(2,\mathbb{K})}^{\mathfrak{h}_2(\mathbb{K})}$}. Then we claim that $\mathfrak{sl}(2,\mathbb{K})$ coincides with $\mathfrak{so}(1,1+k)$. This can be checked as follows. Let $(u_2,\cdots ,u_k)$ be an orthonormal basis of $\hbox{Im }\mathbb{K}$. Then a basis of the Minkowski space $\mathfrak{h}_2(\mathbb{K})$ is $(e_{-1},\cdots,e_{k})$, where
\[
e_{-1}:= \left(\begin{array}{cc}1 & 0 \\ 0 & 1 \end{array}\right), \quad
e_0:= \left(\begin{array}{cc}1 & 0 \\ 0 & -1 \end{array}\right), \quad
e_1:= \left(\begin{array}{cc}0 & 1 \\ 1 &  0\end{array}\right), \quad
e_j:= \left(\begin{array}{cc}0 & u_j \\ -u_j & 0 \end{array}\right)
\]
and where $2\leq j\leq k$. Note that (in conflict with the usual convention) $e_{-1}$ is time-like, whereas the other vectors are space-like. Then, for $0\leq j\leq k$, we define the infinitesimal boost $B_j$ to be the endomorphism of $\mathfrak{h}_2(\mathbb{K})$ such that $(e_{-1},e_j) \longmapsto (e_j,e_{-1})$ (all the other vectors are maps to 0) and, for $0\leq i< j\leq k$, the infinitesimal rotation $A_{ij}$ to be defined by $(e_i,e_j) \longmapsto (e_j,-e_i)$ (idem). Recall that $\left((B_j)_{0\leq j\leq k},(A_{ij})_{0\leq i<j\leq k}\right)$ forms a basis of $\mathfrak{so}(1,1+k)$. We now claim that $\rho_{\mathfrak{tf}(2,\mathbb{K})}^{\mathfrak{h}_2(\mathbb{K})}$ is the vector space spanned by \emph{all elements of this basis, excepted the infinitesimal rotations\footnote{In fact these linear maps keep fixed the imaginary part of $z$ in (\ref{defimu}).} $A_{ij}$ such that $2\leq i< j\leq k$}. Indeed the following table gives, for $C \in \left( (B_j)_{0\leq j\leq k},(A_{0j})_{1\leq j\leq k}, (A_{1j})_{2\leq j\leq k}\right)$, the matrix $\sigma\in  \mathfrak{tf}(2,\mathbb{K})$ such that $\rho_\sigma^{\mathfrak{h}_2(\mathbb{K})} = C$.
\[
\begin{array}{|c|c|c|c|}
\hline
C & j = 0 & j = 1 & 2 \leq j \\
\hline
B_j & \begin{array}{cc}1 & 0 \\ 0 & -1\end{array} &
 \begin{array}{cc}0 & 1 \\ 1 & 0\end{array} &
 \begin{array}{cc}0 & u_j \\ -u_j & 0\end{array} \\
\hline
A_{0j} & & \begin{array}{cc}0 & -1 \\ 1 & 0\end{array} &
 \begin{array}{cc}0 & -u_j \\ -u_j & 0\end{array} \\
\hline
A_{1j} & & & \begin{array}{cc}u_j & 0 \\ 0 & -u_j\end{array}\\
\hline
\end{array}
\]
However the missing rotations can be obtained as commutators of elements of $\rho_{\mathfrak{tf}(2,\mathbb{K})}^{\mathfrak{h}_2(\mathbb{K})}$, thanks to the relation $A_{ij} = - [B_i,B_j]$. Hence we deduce that $\mathfrak{sl}(2,\mathbb{K}) = \mathfrak{so}(1,1+k)$.

The (real) Lie groups $SL(2,\mathbb{H})$ and $SL(2,\mathbb{O})$, their respective spinor representations on $\mathbb{H}^2$ and $\mathbb{O}^2$ and their respective vector representations on $\mathfrak{h}_2(\mathbb{H})$ and $\mathfrak{h}_2(\mathbb{O})$ can then be defined by exponentiating these Lie algebras (see \cite{tachibana,schray,Baez,Deligne}). In the following we use the notation (\ref{RgK}) to represent the action of $SL(2,\mathbb{K})$ on $\mathfrak{h}_2(\mathbb{K})$ (even in the cases where $gmg^\dagger$ does make sense as a matrix product).

\subsection{The super translation Lie algebras and Lie groups}
We will now see that the spinor and the vector representations of $\mathfrak{so}(1,1+k)$ can be glued together to obtain the Lie super algebra $\mathfrak{m}^{(1,1+k)|2k}\simeq \mathfrak{h}_2(\mathbb{K})\times \Pi \mathbb{K}^2$ ($\Pi$ is the parity inversion functor) of super translations of the super Minkowski space $M^{(1,1+k)|2k}$. First we construct $\Pi \mathbb{K}^2$ as follows: consider the Clifford algebra $C(\mathbb{L}) = \{a+b\epsilon|\ a,b,\in \mathbb{R}\}$ defined in \S \ref{otheralgebras}, where $\epsilon$ is an odd variable such that $\epsilon^2 = -1$, and let $\Pi \mathbb{K}^2:= \mathbb{K}^2\otimes_\mathbb{R} C(\mathbb{L})^1$ (recall that $C(\mathbb{L})^1$ is the subspace of odd elements of $C(\mathbb{L})$). Second we embedd $\mathfrak{h}_2(\mathbb{K})\times \Pi \mathbb{K}^2$ as a subspace of $M(5,\mathbb{K})\otimes_{\mathbb{R}} C(\mathbb{L})$, where $M(5,\mathbb{K})$ is the space of $5\times 5$ matrices over $\mathbb{K}$, by the mapping
\[
\left({1\over 2}
\left(\begin{array}{cc}t+x & z \\ \overline{z} & t-x \end{array}\right),\epsilon\lambda_1,\epsilon\lambda_2\right) \longmapsto 
\left(\begin{array}{ccccc}
0 & 0 & \epsilon\lambda_1 & {1\over 2}(t+x) & {1\over 2}z\\
0 & 0 & \epsilon\lambda_2 & {1\over 2}\overline{z} & {1\over 2}(t-x)\\
0 & 0 & 0 & \epsilon\overline{\lambda_1} & \epsilon\overline{\lambda_2} \\
0 & 0 & 0 & 0 & 0\\
0 & 0 & 0 & 0 & 0
\end{array}\right).
\]
It is convenient to define, for $\lambda_1,\lambda_2\in \mathbb{K}$, the \emph{odd} matrices
\begin{equation}\label{Q1Q2}
Q_1^{\lambda_1}:= \epsilon\left(\begin{array}{ccccc}
0 & 0 & \lambda_1 & 0 & 0\\
0 & 0 & 0 & 0 & 0\\
0 & 0 & 0 & \overline{\lambda_1} & 0 \\
0 & 0 & 0 & 0 & 0\\
0 & 0 & 0 & 0 & 0
\end{array}\right), \quad
Q_2^{\lambda_2}:= \epsilon\left(\begin{array}{ccccc}
0 & 0 & 0 & 0 & 0\\
0 & 0 & \lambda_2 & 0 & 0\\
0 & 0 & 0 & 0 & \overline{\lambda_2}\\
0 & 0 & 0 & 0 & 0\\
0 & 0 & 0 & 0 & 0
\end{array}\right).
\end{equation}
Moreover we introduce the notation
\[
\left.\begin{array}{|cc}v^{11} & v^{12}\\ v^{21} & v^{22}\\
\hline \end{array}\right):= 
\left(\begin{array}{ccccc}
0 & 0 & 0 & v^{11} & v^{12}\\
0 & 0 & 0 & v^{21} & v^{22}\\
0 & 0 & 0 & 0 & 0 \\
0 & 0 & 0 & 0 & 0\\
0 & 0 & 0 & 0 & 0
\end{array}\right)
\]
and define
\[
X_{11}:= \left.\begin{array}{|cc}1 & 0\\ 0 & 0\\
\hline \end{array}\right),\ 
X_{12}:= \left.\begin{array}{|cc}0 & 1\\ 0 & 0\\
\hline \end{array}\right),\ 
X_{21}:= \left.\begin{array}{|cc}0 & 0\\ 1 & 0\\
\hline \end{array}\right),\ 
X_{22}:= \left.\begin{array}{|cc}0 & 0\\ 0 & 1\\
\hline \end{array}\right).
\]
We then have the following anti-commutation relations: $\forall \lambda,\mu\in \mathbb{K}$, $\forall a,b, = 1,2$,
\begin{equation}\label{qq}
[Q_a^\lambda,Q_b^{\mu}] := Q_a^\lambda Q_b^{\mu} + Q_b^{\mu} Q_a^\lambda = - \lambda\overline{\mu}\, X_{ab} - \mu\overline{\lambda}\,X_{ba}.
\end{equation}
In this computation we have assumed that $\lambda$ and $\mu$ are even variables (in particular they commute with $\epsilon$, although $\lambda$ and $\mu$ may not commute together if $\mathbb{K}$ is not commutative). Alternatively if, for any $\zeta\in \mathbb{K}$, we write
\[
\Re_{(ab)}(\zeta):= (\hbox{Re}\zeta) (X_{ab}+ X_{ba})/2 = (\zeta + \overline{\zeta})(X_{ab}+ X_{ba})/4
\]
and
\[
\Im_{[ab]}(\zeta):= (\hbox{Im}\zeta)(X_{ab}- X_{ba})/2 = (\zeta - \overline{\zeta})(X_{ab}- X_{ba})/4
\]
(note that $\Re_{(ab)}(\zeta) = \Re_{(ba)}(\zeta)$ and $\Im_{[ab]}(\zeta) + \Im_{[ba]}(\zeta) = 0$), we have
\begin{equation}\label{qqbis}
[Q_a^\lambda,Q_b^{\mu}] = -2 \left(\Re_{(ab)}(\lambda\overline{\mu}) +\Im_{[ab]}(\lambda\overline{\mu})\right).
\end{equation}
Now let $\left(u_\alpha\right)_{1\leq \alpha\leq k}$ be an orthonormal basis of $\mathbb{K}$ over $\mathbb{R}$ such that $u_1 = 1$ (hence $u_2,\cdots ,u_k$ are imaginary) and denote:
\begin{itemize}
\item $\Re_{(ab)}:= \Re_{(ab)}(1)$, for $(a,b) = (1,1)$, $(1,2) $ or $(2,2)$ ;
\item $\Im_\alpha:= \Im_{[12]}(u_\alpha) = {1\over 2}\left.\begin{array}{|cc}0 & u_\alpha\\ -u_\alpha & 0\\
\hline \end{array}\right)$ for $\alpha = 2,\cdots ,k$;
\item $Q^\alpha_a:= Q^{u_\alpha}_a$;
\item $\Gamma^{[\alpha\beta]\gamma}$ such that $(u_\alpha\overline{u_\beta} - u_\beta\overline{u_\alpha})/2 = \Gamma^{[\alpha\beta]\gamma}u_\gamma$ and $\epsilon_{11} = \epsilon_{22} = 0$, $\epsilon_{12} = - \epsilon_{21} = 1$.
\end{itemize}
Then (\ref{qqbis}) can also be written as
\begin{equation}\label{qqter}
[Q_a^\alpha,Q_b^{\beta}] = -2 \left(\delta^{\alpha\beta}\Re_{(ab)} + \Gamma^{[\alpha\beta]\gamma}\epsilon_{ab}\Im_\gamma\right).
\end{equation}
Note that all matrices $\Re_{(ab)}$ and $\Im_\alpha$ commute together and with the $Q_a^\alpha$'s. All that can be summarized by saying that the vector space $\mathfrak{m}^{(1,1+k)|2k}$ spanned by $\Re_{(ab)}$, $\Im_\alpha$ and $Q_a^\alpha$ is a super Lie algebra, the super Lie algebra of super translations on the super Minkowski space of dimension $(1,1+k)|2k$. The even part has the basis $\left(\Re_{(11)},\Re_{(12)},\Re_{(22)},\Im_2,\cdots,\Im_k\right)$ and coincides with the trivial Lie algebra of translations. A basis of the odd part is $\left(Q_1^\alpha,Q_2^\alpha\right)_{\alpha=1,\cdots,k}$.

The super Minkowski space itself can be identified with the super Lie group $M^{(1,1+k)|2k}$ obtained by exponentiating $\mathfrak{m}^{(1,1+k)|2k}$. This requires some extra flesh on $\mathfrak{m}^{(1,1+k)|2k}$: we introduce other odd variables $\eta^1,\cdots,\eta^q$ (some of these variables may be labelled $\theta^a_\alpha$ later on) such that $\Lambda_q:= \mathbb{R}[\eta^1,\cdots,\eta^q]$ is a free super commutative super algebra (i.e. in particular $\eta^i\eta^j + \eta^j\eta^i = 0$), which all anticommute with $\epsilon$. For that purpose, it suffices to construct $\mathbb{R}[\epsilon, \eta^1,\cdots,\eta^q] \simeq \mathbb{R}[\epsilon]\otimes \Lambda_q$ as the Clifford algebra $C(\mathbb{L}\oplus E^q)$ over the vector space $\mathbb{L}\oplus E^q$ with basis $(\epsilon_0,\epsilon_1\cdots ,\epsilon_q)$ and the bilinear map $B$ such that $B(\epsilon_0,\epsilon_0) = 1$ and $B(\epsilon_a,\epsilon_b) = 0$ if $(a,b)\neq (0,0)$. Then $\left(\mathfrak{m}^{(1,1+k)|2k}\otimes \Lambda_q\right)^0$ is a Lie algebra and can be exponentiated as follows. Consider any $v^{(ab)}\Re_{(ab)} + v^\alpha \Im_\alpha + \theta_\alpha^aQ^\alpha_a\in \left(\mathfrak{m}^{(1,1+k)|2k}\otimes \Lambda_q\right)^0$, where $v^{(ab)}, v^\alpha\in \Lambda_q^0$ and $\theta_\alpha^a\in \Lambda_q^1$. In the following we use repeatedly that, as matrices, $X_{ab}X_{cd} = X_{ab}Q^\gamma_c = Q^\gamma_cX_{ab} = Q^\alpha_aQ^\beta_bQ^\gamma_c = 0$. This implies
\[
e^{v^{(ab)}\Re_{(ab)} + v^\alpha \Im_\alpha  + \theta_\alpha^aQ^\alpha_a} =
1 + v^{(ab)}\Re_{(ab)} + v^\alpha \Im_\alpha + \theta_\alpha^aQ^\alpha_a 
+ {1\over 2}\left(\theta_\alpha^aQ^\alpha_a\right) \left(\theta_\beta^bQ^\beta_b\right).
\]
or, if we set for shortness $V:= v^{(ab)}\Re_{(ab)} + v^\alpha \Im_\alpha$ and $\Theta:= \theta_\alpha^aQ^\alpha_a$,
\[
e^{V + \Theta} = 1 + V + \Theta  + {1\over 2}\Theta^2.
\]
We can thus evaluate the product of two elements of $M^{(1,1+k)|2k}$ in the coordinates given by the exponential chart.
Let $W + \Psi:= w^{(ab)}\Re_{(ab)} + w^\alpha \Im_\alpha  + \psi_\alpha^aQ^\alpha_a$ be another element of $\left(\mathfrak{m}^{(1,1+k)|2k}\otimes \Lambda_q\right)^0$, then
\[
\begin{array}{ccl}
\displaystyle e^{V + \Theta} e^{W + \Psi} & = & \displaystyle \left(1 + V + \Theta + {1\over 2}\Theta^2\right) \left(1 + W + \Psi + {1\over 2}\Psi^2\right)\\
& = & \displaystyle  1 + (V+W) + (\Theta+\Psi) + {1\over 2}\Theta^2 + \Theta\Psi + {1\over 2}\Psi^2.
\end{array}
\]
However the quadratic terms can be written
\[
{1\over 2}\left(\Theta^2 + 2\Theta\Psi + \Psi^2\right) = {1\over 2}(\Theta+\Psi)^2 + {1\over 2}\left[ \Theta,\Psi\right].
\]
We hence deduce
\begin{equation}\label{expexp}
\hbox{exp}\left(V + \Theta\right)\hbox{exp}\left(W + \Psi\right) =
\hbox{exp}\left((V+W) + (\Theta+\Psi) + {1\over 2}\left[ \Theta,\Psi\right]\right),
\end{equation}
where, by using (\ref{qqter}),
\[
\left[ \Theta,\Psi\right] = \left[ \theta_\alpha^a Q^\alpha_a,\psi_\beta^b Q^\beta_b\right] = -\theta_\alpha^a\psi_\beta^b\left[ Q^\alpha_a,Q^\beta_b\right] = 2\theta_\alpha^a\psi_\beta^b\left(\delta^{\alpha\beta}\Re_{(ab)} + \Gamma^{[\alpha\beta]\gamma}\epsilon_{ab}\Im_\gamma\right).
\]
A particular case of (\ref{expexp}) is for $W=0$: we can then write
\[
e^{V + \Theta} e^{\Psi} = e^{V + \Theta + D_{\Psi}(V + \Theta)},
\]
where
\[
D_{\Psi}(V + \Theta) = \Psi + {1\over 2}\left[ \Theta,\Psi\right] =  \psi_\beta^b\left(Q^\beta_b - \theta_\alpha^a
\left(\delta^{\alpha\beta}\Re_{(ab)} + \Gamma^{[\alpha\beta]\gamma}\epsilon_{ab}\Im_\gamma\right)\right).
\]
We remark that
\[
D_{\Psi}(V + \Theta) = \psi_\beta^b\left( {\partial \over \partial \theta_\beta^b} - \theta_\alpha^a
\left(\delta^{\alpha\beta}{\partial \over \partial v^{(ab)}} + \Gamma^{[\alpha\beta]\gamma}\epsilon_{ab}{\partial \over \partial v^\gamma}\right)\right)(V + \Theta),
\]
so that $D_{\Psi}$ can be identified with a differential operator. Letting
\[
D^\alpha_a:= {\partial \over \partial \theta_\alpha^a} -  \theta_\beta^b\left(\delta^{\alpha\beta}{\partial \over \partial v^{(ab)}} + \Gamma^{[\alpha\beta]\gamma}\epsilon_{ab}{\partial \over \partial v^\gamma}\right),
\]
we have $D_{\Psi} = \psi_\beta^bD^\beta_b$. We can also introduce the notations: $\partial^{[\alpha\beta]} = - \partial^{[\beta\alpha]} := \sum_{\gamma=2}^k\Gamma^{[\alpha\beta]\gamma}{\partial \over \partial v^\gamma}$ for $1\leq \alpha,\beta\leq k$, $\partial_{(ab)} = \partial_{(ba)}:= {\partial \over \partial v^{(ab)}}$ for $1\leq a,b\leq 2$ and $\partial_a^\alpha:= {\partial \over \partial \theta^a_\alpha}$ for $1\leq a\leq 2$ and $1\leq \alpha\leq k$ and set
\[
D^\alpha_a = \partial^\alpha_a - \theta^b_\beta(\delta^{\alpha\beta} \partial_{(ab)} + \epsilon_{ab}\partial^{[\alpha\beta]}),
\]
or $D^\alpha_1 = \partial^\alpha_1 - (\theta^1_\alpha\partial_{(11)} + \theta^2_\alpha\partial_{(12)} + \theta^2_\beta\partial^{[\alpha\beta]})$ and $D^\alpha_2 = \partial^\alpha_2 - (\theta^1_\alpha\partial_{(12)} + \theta^2_\alpha\partial_{(22)} - \theta^1_\beta\partial^{[\alpha\beta]})$.

Similarly, for $\Phi:= \phi_\beta^bQ^\beta_b$, we have $e^{\Phi}e^{V + \Theta} = e^{V + \Theta + \tau_{\Phi}(V + \Theta)}$, where $\tau_\Phi(V + \Theta) = \Phi + {1\over 2}[\Phi,\Theta]$. Thus we can write $\tau_\Phi = \phi_\alpha^a \tau^\alpha_a$, where
\[
\tau^\alpha_a = \partial^\alpha_a + \theta^b_\beta(\delta^{\alpha\beta} \partial_{(ab)} + \epsilon_{ab}\partial^{[\alpha\beta]}).
\]
Operators $D^\alpha_a$ and $\tau^\alpha_a$ are respectively the left invariant and the right invariant vector fields on $M^{(1,1+k)|2k}$. By writing $\left(e^\Phi e^{V+\Theta}\right)e^\Psi = e^\Phi \left(e^{V+\Theta}e^\Psi\right)$, we deduce that $[\tau^\alpha_a,D^\beta_b] = 0$, for all $a,b = 1,2$ and $\alpha,\beta = 1,\cdots ,k$. We moreover have
\[
\left[\tau^\alpha_a,\tau^\beta_b\right] = 2\left(
 \delta^{\alpha\beta} \partial_{(ab)} + \epsilon_{ab}\partial^{[\alpha\beta]}\right)
\]
and
\[
\left[D^\alpha_a,D^\beta_b\right] = -2\left(
 \delta^{\alpha\beta} \partial_{(ab)} + \epsilon_{ab}\partial^{[\alpha\beta]}\right).
\]
\begin{rema}
Let us set $Q^{(\lambda_1,\lambda_2)}:= Q_1^{\lambda_1} + Q_2^{\lambda_2}$.
Then Relation (\ref{qq}) is equivalent to the fact that, for any $(\lambda_1,\lambda_2)\in \mathbb{K}^2$,
\[
\left[Q^{(\lambda_1,\lambda_2)},Q^{(\lambda_1,\lambda_2)}\right] = -2X^{(\lambda_1,\lambda_2)},
\quad \hbox{where }X^{(\lambda_1,\lambda_2)}:= 
\left.\begin{array}{|cc}|\lambda_1|^2 & \lambda_1\overline{\lambda_2}\\
\lambda_2\overline{\lambda_1} & |\lambda_2|^2\\
\hline \end{array}\right).
\]
We observe that $X^{(\lambda_1,\lambda_2)}$ is a \textbf{null vector} (a vector in the light cone $\mathcal{C}:= \{m\in \mathfrak{h}_2(\mathbb{K})|\ \hbox{\emph{det} }m = 0\}$) with a \textbf{nonnegative time coordinate} $t = |\lambda_1|^2 + |\lambda_1|^2 \geq 0$. These two properties are important since they are the reason for the fact that supersymmetric Yang--Mills theories exists only in dimensions $(1,1+k)$ (see e.g. \cite{Freed}, \S5). Moreover this construction provides us with an orientation of the time, an important property of supersymmetric theories. Also the set $\mathcal{C}/\mathbb{R}$ of real lines contained in $\mathcal{C}$, with its canonical conformal structure, can be identified with the sphere $S^k$ (the heavenly sphere at infinity) or, equivalentely with the projective line $\mathbb{K}P$. Indeed the map $\varphi: \mathcal{C}\setminus \{0\}\longrightarrow S^k\subset \mathbb{R}\times \mathbb{K}$ defined by $(t,x,z)\longmapsto (x/t,z/t)$ induces a conformal diffeomorphism between $\mathcal{C}/\mathbb{R}$ and $S^k$. Note also that the image of $X^{(\lambda_1,\lambda_2)}$ by $\varphi$ is $\left({|\lambda_1|^2 - |\lambda_1|^2\over |\lambda_1|^2 + |\lambda_1|^2}, {\lambda_1\overline{\lambda_2}\over |\lambda_1|^2 + |\lambda_1|^2}\right)$, i.e.  is nothing but the image by the Hopf fibration of $(\lambda_1,\lambda_2)$.

These observations are related to the \textbf{R-symmetries}: these are automorphisms of the super Lie algebra of supertranslations which leaves all space-time translations $\mathfrak{h}_2(\mathbb{K})$ invariant (i.e. they act only on $\Pi\mathbb{K}^2$). It is clear from the previous discussion that these symmetries can be identified with the automorphisms of $\mathbb{K}^2$ which maps each fiber of the Hopf fibration onto itself. Hence
\begin{itemize}
\item if $\mathbb{K} = \mathbb{R}$, the R-symmetry group is $\{\pm1\}$ ($(\lambda_1,\lambda_2) \longmapsto \pm (\lambda_1,\lambda_2)$);
\item if $\mathbb{K} = \mathbb{C}$, the R-symmetry group is $U(1)$ ($(\lambda_1,\lambda_2) \longmapsto (e^{i\theta}\lambda_1,e^{i\theta}\lambda_2)$, for $\theta\in \mathbb{R}$);
\item if $\mathbb{K} = \mathbb{H}$, the R-symmetry group is $\hbox{\emph{Spin} }3 \simeq SU(2)$ ($(\lambda_1,\lambda_2) \longmapsto (\lambda_1\alpha,\lambda_2\alpha)$, for $\alpha\in \mathbb{H}$ such that $|u| = 1$);
\item if $\mathbb{K} = \mathbb{O}$, the R-symmetry group is $\{\pm1\}$ ($(\lambda_1,\lambda_2) \longmapsto \pm (\lambda_1,\lambda_2)$)\footnote{Indeed assume that there exists some real endomorphism $(\varphi_1,\varphi_2):\mathbb{O}^2\longrightarrow \mathbb{O}^2$ such that $\forall \lambda_1,\lambda_2\in \mathbb{O}$, (i) $|\varphi_1(\lambda_1,\lambda_2)|^2 = |\lambda_1|^2$, (ii)  $\varphi_1(\lambda_1,\lambda_2)\overline{\varphi_2(\lambda_1,\lambda_2)} = \lambda_1\overline{\lambda_2}$ and (iii) $|\varphi_2(\lambda_1,\lambda_2)|^2 = |\lambda_2|^2$. Then: 
(a) we remark that (i) implies in particular that $\varphi_1(0,\lambda_2) = 0$ and, since $\varphi_1$ is linear, this forces that $\varphi_1(\lambda_1,\lambda_2) = \varphi_1(\lambda_1)$; (b) similarly (iii) implies that $\varphi_2(\lambda_1,\lambda_2) = \varphi_2(\lambda_2)$; (c) furthermore (i) and (iii) implies that $\varphi_1$ and $\varphi_2$ are real isometries of $\mathbb{O}$; (d) 
by testing (ii) with $\lambda_2 = 1$, we deduce that we must have $\varphi_1(\lambda) = \lambda \alpha_1$, where $\alpha = \varphi_2(1)$; (e) similarly by testing (ii) with $\lambda_1 = 1$ we deduce that $\varphi_2(\lambda) = \lambda \alpha_2$, where $\alpha = \varphi_1(1)$; (f) by testing (ii) with $\lambda_1 = \lambda_2 = 1$ we obtain that $\alpha_1 = \alpha_2 = \alpha$; (g) we can set w.l.g. $\alpha = \cos \theta + u_2\sin\theta$ and choose $\lambda_1 = u_3$ and $\lambda_2 = u_5$, where $(u_2,u_3,u_5)$ forms an orthonormal family of imaginary octonions such that $u_2u_3\perp u_5$, then one checks that the relation $(u_3\alpha)\overline{(u_5\alpha)} = u_3\overline{u_5}$ is possible only if $\alpha = \pm 1$.}.
\end{itemize}

\end{rema}

\subsection{The super Poincar{\'e} group}
Lastly we can extend the super Lie group $M^{(1,1+k)|2k}$ (respectively its super Lie algebra $\mathfrak{m}^{(1,1+k)|2k}$) by gluing it with the spin cover of the Lorentz group $\hbox{Spin}(1,1+k)\simeq SL(2,\mathbb{K})$ (respectively with $\mathfrak{so}(1,1+k)\simeq \mathfrak{sl}(2,\mathbb{K})$). We obtain the super Poincar{\'e} group $P(1,1+k|2k)$. A matricial representation of an element $g\in P(1,1+k|2k)$ is
\[
g = \left(\begin{array}{ccc}
\begin{array}{c} S \\ \\ \begin{array}{cc} 0 & 0\end{array}\end{array}
& \begin{array}{c}\theta^1\epsilon \\ \theta^2\epsilon \\ 1\end{array} 
& \begin{array}{cc}(t+x)/2 & z/2\\
 \overline{z}/2 & (t-x)/2\\
 \overline{\theta}^1\epsilon & \overline{\theta}^2\epsilon\end{array} \\
\begin{array}{cc}0&0 \\ 0&0\end{array}  & \begin{array}{c} 0 \\ 0\end{array} & (S^\dagger)^{-1}
\end{array}\right),
\]
where $S\in SL(2,\mathbb{K})\otimes \Lambda_q^0$ and $S^\dagger = \overline{S}^t$, $\theta^1,\theta^2\in \mathbb{K}\otimes \Lambda_q^1$, $t,x\in \Lambda_q^0$ and $z\in \mathbb{K}\otimes \Lambda_q^0$. Note that the conjugate $\overline{\theta}^a$ of $\theta^a$ involves the real structure of $\mathbb{K}$, i.e., if $\theta^a = u\otimes \eta$, for $u\in \mathbb{K}$ and $\eta\in \Lambda_q^1$, then $\overline{\theta}^a = \overline{u}\otimes \eta$. In particular we recover the action of an element $S\in SL(2,\mathbb{K})$ on $M^{(1,1+k)|2k}$ by $e^m \longmapsto ge^mg^{-1}$, where
\[
g = \left(\begin{array}{ccc}S & 0 & 0 \\ 0 & 1 & 0 \\ 0 & 0 &  (S^\dagger)^{-1}
\end{array}\right) \quad \hbox{and} \quad 
m = \left(\begin{array}{ccc}
\begin{array}{c} 0 \\ \\ \begin{array}{cc} 0 & 0\end{array}\end{array}
& \begin{array}{c}\theta^1\epsilon \\ \theta^2\epsilon \\ 0\end{array} 
& \begin{array}{cc}(t+x)/2 & z/2\\
 \overline{z}/2 & (t-x)/2\\
 \overline{\theta}^1\epsilon & \overline{\theta}^2\epsilon\end{array} \\
\begin{array}{cc}0&0 \\ 0&0\end{array}  & \begin{array}{c} 0 \\ 0\end{array} & 0
\end{array}\right).
\]

\subsection{Specializations}
\subsubsection{The super Minkowski space $M^{(1,2)|2}\simeq \mathbb{R}^{3|2}$}
It corresponds to the simplest case, when $\mathbb{K} = \mathbb{R}$ (see \S \ref{super3+2}).
 
\subsubsection{The super Minkowski space $M^{(1,3)|4}\simeq \mathbb{R}^{4|4}$}\label{M(1,3)4}
It corresponds to the case $k=2$, i.e. $\mathbb{K} = \mathbb{C}$. We then obtain $M^{(1,3)|4}$, the simplest non trivial supersymmetric extension of the standard Minkowsi space $M^{1,3}$. In order to connect our presentation of the super Lie algebra $\mathfrak{m}^{(1,3)|4}$ with the standard one, it is useful to embedd it in its complexification $\mathbb{C}_{\sqrt{-1}}\otimes_\mathbb{R} \mathfrak{m}^{(1,3)|4}$. Here $\mathbb{C}_{\sqrt{-1}}$ denotes a copy of $\mathbb{C}$ where the square root of $-1$ is denoted by $\sqrt{-1}$ instead of $u_2 = i$\footnote{One may set $\sqrt{-1} = i = u_2$ without trouble here, however the next cases, where $\mathbb{K} = \mathbb{H}$ or $\mathbb{O}$ require more care.}. For any $A = B + \sqrt{-1}C\in \mathbb{C}_{\sqrt{-1}}\otimes_\mathbb{R} \mathfrak{m}^{(1,3)|4}$, where $B,C\in \mathfrak{m}^{(1,3)|4}$, we note $\overline{A}^{\sqrt{-1}}:= B - \sqrt{-1}C$.  Then we set, for $a=1,2$,
\[
Q_a:= \left(Q^1_a - \sqrt{-1}Q^i_a\right)/\sqrt{2}
\quad  \hbox{and}\quad
\overline{Q}_{\dot{a}}:= \left(Q^1_a + \sqrt{-1}Q^i_a\right)/\sqrt{2} = \overline{Q_a}^{\sqrt{-1}}.
\]
We then have the relations
\[
[Q_a,Q_b] = \left[\overline{Q}_{\dot{a}},\overline{Q}_{\dot{b}}\right] = 0
\quad  \hbox{and}\quad
\left[Q_a,\overline{Q}_{\dot{b}}\right] = -2 X_{a\dot{b}},
\]
where\footnote{Again we may set $\sqrt{-1} = i$: we would then have $X_{a\dot{b}} = X_{ab}$.} $X_{a\dot{b}} =  {X_{ab}+X_{ba}\over 2} - \sqrt{-1}{iX_{ab}-iX_{ba}\over 2}$. We denote by $x^{a\dot{b}}$, $\theta^a$ and  $\overline{\theta}^{\dot{a}}$ (for $a=1,2$) the coordinates on $\mathfrak{m}^{(1,3)|4}$ in the basis $\left(X_{a\dot{b}}, Q_a,\overline{Q}_{\dot{a}}\right)$. We write $\partial_{a\dot{b}}:= \partial / \partial x^{a\dot{b}}$, $\partial_a:= \partial/\partial\theta^a$ and $\overline{\partial}_{\dot{a}}:= \partial/\partial\overline{\theta}^{\dot{a}}$ the partial derivatives with respect to these coordinates. The coordinates\footnote{This system of notations is similar to the notations introduced by R. Penrose for the twistor theory, see the text by P. Baird in this volume.} $(x^{a\dot{b}})$ are connected to the coordinates $(t,x,z^1+iz^2)\in \mathbb{R}\times \mathbb{R}\times \mathbb{C}$ by
\[
\left( \begin{array}{cc}x^{1\dot{1}} & x^{1\dot{2}} \\
x^{2\dot{1}} & x^{2\dot{2}}\end{array}\right) =
{1\over 2} \left( \begin{array}{cc}t+x & z^1+iz^2 \\
z^1-iz^2 & t-x\end{array}\right).
\]
This implies that
\[
\begin{array}{ll}
\partial_{1\dot{1}} = {\partial_t} + {\partial_x}
& \partial_{1\dot{2}} = \partial_{z^1} -i\partial_{z^2}\\
\partial_{2\dot{1}}= \partial_{z^1} +i\partial_{z^2}
& 
\partial_{2\dot{2}}= {\partial_t} - {\partial_x}.
\end{array}
\]
Lastly the left and right translations in these coordinates read respectively $D_\Psi = \psi^aD_a + \overline{\psi}^{\dot{a}}\overline{D}_{\dot{a}}$ and $\tau_\Phi  = \psi^a\tau_a + \overline{\psi}^{\dot{a}}\overline{\tau}_{\dot{a}}$ with:
\[
D_a = \partial_a - \overline{\theta}^{\dot{b}}\partial_{a\dot{b}},
\quad
\overline{D}_{\dot{a}} = \overline{\partial}_{\dot{a}} - \theta^b\partial_{b\dot{a}},
\]
and
\[
\tau_a = \partial_a + \overline{\theta}^{\dot{b}}\partial_{a\dot{b}},
\quad
\overline{\tau}_{\dot{a}} = \overline{\partial}_{\dot{a}} + \theta^b\partial_{b\dot{a}}.
\]
Very often physicists consider \emph{chiral} fields, i.e. fields $\Phi$ satisfying the constraints $\overline{D}_{\dot{1}}\Phi = \overline{D}_{\dot{2}}\Phi = 0$.\\

\subsubsection{The super Minkowski space $M^{(1,5)|8}\simeq \mathbb{R}^{6|8}$}
It corresponds to the case $k=4$, i.e. $\mathbb{K} = \mathbb{H}$. Here there is an alternative system of coordinates based on the exceptional isomorphism $SL(4,\mathbb{C})\simeq \hbox{Spin}(6)^\mathbb{C}$. This comes from the identificaton $\Lambda^2\mathbb{C}^4\simeq \mathbb{C}^6$: the natural action of $SL(4,\mathbb{C})$ on $\mathbb{C}^4$ induces an action of $SL(4,\mathbb{C})$ on $\Lambda^2\mathbb{C}^4$ which can be identified with the action of $\hbox{Spin}(6)^\mathbb{C}$ on $\mathbb{C}^6$. Let $(e_1,e_2,e_3,e_4)$ be the canonical basis of $\mathbb{C}^4$ and $(e_a\wedge e_b)_{1\leq a<b\leq 4}$ the induced basis of $\Lambda^2\mathbb{C}^4$. We denote by $\left(y^{ab}\right)_{1\leq a<b\leq 4}$ (and set $y^{ba} = - y^{ab}$) the complex coordinates on $\Lambda^2\mathbb{C}^4$ in this basis. We note that there is a canonical symmetric bilinear form $B:\Lambda^2\mathbb{C}^4\times \Lambda^2\mathbb{C}^4 \longrightarrow \mathbb{C}$ characterized by the relation $Y\wedge Y = B(Y,Y) e_1\wedge e_2\wedge e_3\wedge e_4$, $\forall Y\in \Lambda^2\mathbb{C}^4$. This bilinear form is nondegenerate and it is easy to see that the canonical action of $SL(4,\mathbb{C})$ on $\mathbb{C}^4$ induces an action on $\Lambda^2\mathbb{C}^4$ which preserves $B$, hence the identification $SL(4,\mathbb{C})\simeq \hbox{Spin}(6)^\mathbb{C}$.

Furthermore we can embedd the Minkowski space $M^{1,5}$ as the subspace $\mathfrak{M}^{1,5}$ of $\Lambda^2\mathbb{C}^4$ defined by the reality conditions $\overline{y^{12}} = y^{34}$, $\overline{y^{14}} = y^{23}$ and $y^{13}, y^{24}\in \mathbb{R}$. Alternatively $\mathfrak{M}^{1,5}$ is the image of $M^{1,5}\simeq \mathfrak{h}_2(\mathbb{H})$ by the linear map $P:{1\over 2}\left(\begin{array}{cc}t+x & z \\ \overline{z}& t-x\end{array}\right) \longmapsto \sum_{1\leq a<b\leq 4}y^{ab}e_a\wedge e_b$, where
\[
\left( \begin{array}{cccc}
0 & y^{12} & y^{13} & y^{14}\\
y^{21}& 0 & y^{23} & y^{24}\\
y^{31}& y^{32} & 0 & y^{34}\\
y^{41}& y^{42} & y^{43} & 0
\end{array}\right) = {1\over 2}
\left( \begin{array}{cccc}
0 & z^3+iz^4 & t+x & z^1 +iz^2\\
-z^3-iz^4 & 0 & z^1-iz^2 & t-x\\
-t-x & -z^1+iz^2 & 0 & z^3-iz^4\\
-z^1-iz^2 & -t+x & -z^3-iz^4 & 0
\end{array}\right).
\]
The partial derivatives $\partial_{ab}:= \partial /\partial y^{ab}$ are then related with the partial derivatives with respect to $t,x$ and $z= z^1+iz^2+jz^3+kz^4$ by:
\[
\begin{array}{lll}
\partial_{12} = \partial_{z^3} - i\partial_{z^4};
& \partial_{13} = \partial_t + \partial_x ;
& \partial_{14} = \partial_{z^1} - i\partial_{z^2}\\
& \partial_{23} = \partial_{z^1} + i\partial_{z^2};
& \partial_{24} = \partial_t - \partial_x \\
& & \partial_{34} = \partial_{z^3} + i\partial_{z^4}.
\end{array}
\]
One also observes that the restriction of $B$ to $\mathfrak{M}^{1,5}$ has the signature $(5,1)$ and, more precisely, $4B(P(t,x,z),P(t,x,z)) = - (t^2 -x^2-|z|^2)$. Hence the subgroup of $SL(4,\mathbb{C})$ which preserves $\mathfrak{M}^{1,5}$ when acting on $\Lambda^2\mathbb{C}^4$ can be identified with $\hbox{Spin}(1,5)$.

Indeed $SL(2,\mathbb{H})$ can be identified with a real form of $SL(4,\mathbb{C})$. More precisely we can identify\footnote{Most Authors use the more pleasant identification $(u^1,u^2,u^3,u^4)\longmapsto (u^1+u^3j,u^2+u^4j)$; however this convention is not that convenient in our context.} $\mathbb{C}^4$ with $\mathbb{H}^2$ through the map $T:(u^1,u^2,u^3,u^4)\longmapsto (u^1+ju^3,u^2+ju^4)$ and define $SL(2,\mathbb{H})$ as the set of transformations $g\in SL(4,\mathbb{C})$ which maps each quaternionic line into a quaternionic line, i.e. such that, for any $(\lambda_1,\lambda_2)\in \mathbb{H}^2\setminus\{0\}$, $T \circ g\circ T^{-1}$ maps the line $\{(\lambda_1\alpha, \lambda_2\alpha)|\,\alpha\in \mathbb{H}\}$ to another line $\{(\mu_1\alpha, \mu_2\alpha)|\, \alpha\in \mathbb{H}\}$. It is easy to see that a complex plane in $\mathbb{C}^4$ which contains a vector $U = (u^1,u^2,u^3,u^4)\neq 0$ is the inverse image of a quaternionic line by $T$ if and only if it contains $\sigma(U):= T^{-1}(T(U)j) = T^{-1}((u^1+ju^3)j,(u^2+ju^4)j) = (-\overline{u^3},-\overline{u^4}, \overline{u^1}, \overline{u^2})$. So $SL(2,\mathbb{H})$ is the set of transformations $g\in SL(4,\mathbb{C})$ which maps any complex plane spanned by $(U,\sigma(U))$ to a complex plane of the same type. Lastly we remark that, for any $U\in \mathbb{C}^4\setminus\{0\}$, $U\wedge \sigma(U)$ belongs to $\mathfrak{M}^{1,5}$ and the action of $SL(2,\mathbb{H})$ on the pair $(U,\sigma(U))$ is mapped to the action of $SO(1,5)$ on $\mathfrak{M}^{1,5}$. More precisely one can compute that, if $U = (u^1,u^2,u^3,u^4)\in \mathbb{C}^4$, then $U\wedge \sigma(U) = \sum_{1\leq a<b\leq 4}y^{ab}e_a\wedge e_b$, where
\[
\begin{array}{lll}
y^{12} = u^2\overline{u^3} - u^1\overline{u^4}; & y^{13} = |u^1|^2 + |u^3|^2; & y^{14} = u^1\overline{u^2} + u^4\overline{u^3};\\
& y^{23} = u^2\overline{u^1} + u^3\overline{u^4}; & y^{24} = |u^2|^2 + |u^4|^2;\\
& & y^{34} = u^3\overline{u^2} - u^4\overline{u^1},
\end{array}
\]
and we obtain that
\[
P^{-1}(U\wedge \sigma(U)) = \left(\begin{array}{cc} y^{13} & y^{14}+y^{12}j\\
y^{23}-jy^{34} & y^{24}\end{array}\right) = \left(\begin{array}{cc} |\lambda_1|^2 & \lambda_1\overline{\lambda_2} \\ \lambda_2\overline{\lambda_1} & |\lambda_2|^2 \end{array}\right),
\]
where $\lambda_1:= u^1 + ju^3$ and $\lambda_2:= u^2 + ju^4$, i.e. $(\lambda_1,\lambda_2) = T(U)$. Hence we conclude that (recall the notation $Q^{(\lambda_1,\lambda_2)}:= Q^{\lambda_1}_1+Q^{\lambda_2}_2$):
\begin{equation}\label{SL4C=SL2H}
\forall U\in \mathbb{C}^4, \hbox{ if }(\lambda_1,\lambda_2) = T(U),\quad
P\left([Q^{(\lambda_1,\lambda_2)},Q^{(\lambda_1,\lambda_2)}]\right) = -2 U\wedge \sigma(U).
\end{equation}
In other words the diagram: 
\xymatrix{
\Pi\mathbb{C}^4 \ar@{->}[r]^{\epsilon U\longmapsto -2U\wedge \sigma(U)} \ar@{->}[d]_T & \mathfrak{M}^{1,5}\\
\Pi\mathbb{H}^2 \ar@{->}[r]_{[\cdot,\cdot]} & \mathfrak{h}_2(\mathbb{H}) \ar@{->}[u]_P}
is commutative.
Note that, by polarization, (\ref{SL4C=SL2H}) implies $P\left([Q^{(\lambda_1,\lambda_2)},Q^{(\mu_1,\mu_2)}]\right) = - U\wedge \sigma(V) - V\wedge \sigma(U)$, where $T(V) = (\mu_1,\mu_2)$.

Lastly the identification $\mathfrak{m}^{(1,5)|8}\simeq \mathfrak{M}^{1,5}\times \Pi \mathbb{C}^4$ is related to an alternative simple representation of the supercommutator on $\mathfrak{m}^{(1,5)|8}$: we again complexify $\mathfrak{m}^{(1,5)|8}$ by embedding it in $\mathbb{C}_{\sqrt{-1}}\otimes_\mathbb{R} \mathfrak{m}^{(1,5)|8}$ and set
\[
\begin{array}{cl}
q_{11}:= \left(Q^1_1 - \sqrt{-1}Q^i_1\right)/\sqrt{2}, 
& q_{12}:= -\left(Q^j_1 - \sqrt{-1}Q^k_1\right)/\sqrt{2},\\
q_{21}:= \left(Q^1_2 - \sqrt{-1}Q^i_2\right)/\sqrt{2},
& q_{22}:= -\left(Q^j_2 - \sqrt{-1}Q^k_2\right)/\sqrt{2},\\
q_{31}:= \left(Q^j_1 + \sqrt{-1}Q^k_1\right)/\sqrt{2},
& q_{32}:= \left(Q^1_1 + \sqrt{-1}Q^i_1\right)/\sqrt{2}, \\
q_{41}:= \left(Q^j_2 + \sqrt{-1}Q^k_2\right)/\sqrt{2},
& q_{42}:= \left(Q^1_2 + \sqrt{-1}Q^i_2\right)/\sqrt{2}.
\end{array}
\]
We denote by $\left(\theta^{aA}\right)_{1\leq a\leq 4;A=1,2}$ the coordinates associated with these vectors and we write $\partial_{aA}:= \partial/\partial\theta^{aA}$. Then one can deduce from (\ref{qqter}) that
\[
\left[ q_{aA},q_{bB}\right] = -2\epsilon_{AB}e_a\wedge e_b, \quad
\forall A,B = 1,2, \forall a,b = 1,2,3,4,
\]
where $\epsilon_{12} = - \epsilon_{21} = 1$ and $\epsilon_{11} = \epsilon_{22} = 0$. As a consequence
\[
D_{aA} = \partial_{aA} - \epsilon_{AB}\theta^{bB}\partial_{ab}
\quad \hbox{and} \quad
\tau_{aA} = \partial_{aA} + \epsilon_{AB}\theta^{bB}\partial_{ab}.
\]

\subsection{Dimensional reductions}\label{reduction}
We have presented only particular examples of super Minkowski spaces, however one can construct new examples from this list by dimensional reduction, i.e. by considering fields which are invariant under some space-time direction. For instance one can construct super extensions of the usual Minkowski space $M^{1,3}$ with more than $4$ odd dimensions: $M^{(1,3)|8}$ (related to $N=2$ supersymmetric theories), which can be obtained by reduction of $M^{(1,5)|8}$ and $M^{(1,3)|16}$ (related to $N=4$ supersymmetric theories), which can be obtained by reduction of $M^{(1,9)|16}$. Our first example $M^{(1,3)|4}$ corresponds to $N=1$ supersymmetric theories. Similar reductions leads to various super space-times with 2 even dimensions, which are important for the theory of superstrings (see e.g. \cite{Freed}). In the following we present alternative representations of $\mathfrak{m}^{(1,5)|8}$ and $\mathfrak{m}^{(1,9)|16}$ which shows how one can produce $N=2$ and $N=4$ supersymmetry groups over $M^{1,3}$ with central extensions.\\

\noindent
\textbf{From $\mathfrak{m}^{(1,5)|8}$ to a central extension of $\mathfrak{m}^{(1,3)|8}$}\\
We embedd $\mathfrak{m}^{(1,5)|8}$ in its complexification $\mathbb{C}_{\sqrt{-1}}\otimes_\mathbb{R} \mathfrak{m}^{(1,5)|8}$ and set, for $a = 1,2$:
\[
\begin{array}{cc}
Q_{a1}:= (Q^1_a - \sqrt{-1}Q^i_a)/\sqrt{2}; &
\overline{Q}_{\dot{a}1}:= (Q^1_a + \sqrt{-1}Q^i_a)/\sqrt{2};\\
Q_{a2}:= (Q^j_a - \sqrt{-1}Q^k_a)/\sqrt{2}; &
\overline{Q}_{\dot{a}2}:= (Q^j_a + \sqrt{-1}Q^k_a)/\sqrt{2}.
\end{array}
\]
Then one deduce from (\ref{qqbis}) that, for $a,b = 1,2$ and $A,B = 1,2$,
\[
\begin{array}{ccc}
\left[Q_{aA},\overline{Q}_{\dot{b}B}\right] & = & -2 \delta_{AB} \left(\mathcal{R}_{(ab)} - \sqrt{-1} \mathcal{I}_{[ab]}(i)\right);\\
\left[Q_{aA},Q_{bB}\right] & = & 2\epsilon_{ab}\epsilon_{AB}\left(\mathcal{I}_{[12]}(j) - \sqrt{-1} \mathcal{I}_{[12]}(k)\right),
\end{array}
\]
and $[\overline{Q}_{\dot{a}A},\overline{Q}_{\dot{b}B}] = \overline{[Q_{aA},Q_{bB}]}^{\sqrt{-1}}$ and $[\overline{Q}_{\dot{a}A},Q_{bB}] = \overline{[Q_{aA},\overline{Q}_{\dot{b}B}]}^{\sqrt{-1}}$, where $\epsilon_{11} = \epsilon_{22} = 0$ and $\epsilon_{12} = -\epsilon_{21} = 1$. However using the notation $X_{a\dot{b}}$ introduced in \S\ref{M(1,3)4} and letting
\[
Z_{AB}:= - \epsilon_{AB}\mathcal{I}_{[12]}\left(j - \sqrt{-1} k\right)
\quad \hbox{and}\quad
\overline{Z}_{AB}:= - \epsilon_{AB}\mathcal{I}_{[12]}\left(j + \sqrt{-1} k\right),
\]
we can write the anticommutation relations as
\begin{equation}\label{reduc6vers4}
\begin{array}{cc}
\left[Q_{aA},\overline{Q}_{\dot{b}B}\right] = -2 X_{a\dot{b}} \delta_{AB} & [\overline{Q}_{\dot{a}A},Q_{bB}] = -2 X_{b\dot{a}} \delta_{AB}\\
\left[Q_{aA},Q_{bB}\right] = -2\epsilon_{ab}Z_{AB} & \left[\overline{Q}_{\dot{a}A},\overline{Q}_{\dot{b}B}\right] = -2\epsilon_{\dot{a}\dot{b}}\overline{Z}_{AB}.
\end{array}
\end{equation}
Upon reduction to $M^{1,3}$ the relations on the first line are the only nontrivial anticommutation relations of $\mathfrak{m}^{(1,3)|8}$ (since then we set $[Q_{aA},Q_{bB}] = \left[\overline{Q}_{\dot{a}A},\overline{Q}_{\dot{b}B}\right] = 0$). However D. Olive and E. Witten \cite{olivewitten} discovered that, for the $N=2$ supersymmetric Yang--Mills on $M^{1,3}$, the super Lie algebra of Noether charges of the Lagrangian (which is of course symmetric under the action of $\mathfrak{m}^{(1,3)|8}$) is a central extension of $\mathfrak{m}^{(1,3)|8}$, obtained by adding two central charges which correspond the the real and imaginary parts of $Z_{AB}$. The amazing point is that these charges have a topological interpretation in way similar to the example seen in \S\ref{Bogomolnyi} and are related the possibility of having an electro-magnetic duality for supersymmetric non Abelian gauge theories conjectured in \cite{olive}.\\

\noindent
\textbf{From $\mathfrak{m}^{(1,9)|16}$ to a central extension of $\mathfrak{m}^{(1,3)|16}$}\\
A similar construction starting from $\mathfrak{m}^{(1,9)|16}$ leads to a central extension of the Lie super algebra of the $N=4$ supersymmetric extension of $\mathfrak{m}^{(1,4)|4}$. We embedd $\mathfrak{m}^{(1,9)|16}$ in  $\mathbb{C}_{\sqrt{-1}}\otimes_\mathbb{R} \mathfrak{m}^{(1,9)|16}$ and set, for $a = 1,2$:
\[
\begin{array}{cc}
Q_{a1}:= (Q^1_a - \sqrt{-1}Q^2_a)/\sqrt{2}; &
\overline{Q}_{\dot{a}1}:= (Q^1_a + \sqrt{-1}Q^2_a)/\sqrt{2};\\
Q_{a2}:= (Q^3_a - \sqrt{-1}Q^4_a)/\sqrt{2}; &
\overline{Q}_{\dot{a}2}:= (Q^3_a + \sqrt{-1}Q^4_a)/\sqrt{2}; \\
Q_{a3}:= (Q^6_a - \sqrt{-1}Q^7_a)/\sqrt{2}; &
\overline{Q}_{\dot{a}3}:= (Q^6_a + \sqrt{-1}Q^7_a)/\sqrt{2}; \\
Q_{a4}:= (Q^8_a - \sqrt{-1}Q^5_a)/\sqrt{2}; &
\overline{Q}_{\dot{a}3}:= (Q^8_a + \sqrt{-1}Q^5_a)/\sqrt{2},
\end{array}
\]
where, for $\alpha = 1,\cdots ,8$ and $a= 1,2$, $Q^ \alpha_a = Q^{u_\alpha}_a$ and we recall that $(u_1,\cdots ,u_8)$ denotes an orthonormal basis of $\mathbb{O}$ such that $u_1 =1$. It is also important to precise the multiplication rule which is adopted here (since several conventions can be found in the litterature). We recall it in the following diagram:
\begin{figure}[h]\label{tableoctonionion}
\begin{center}
\includegraphics[scale=0.5]{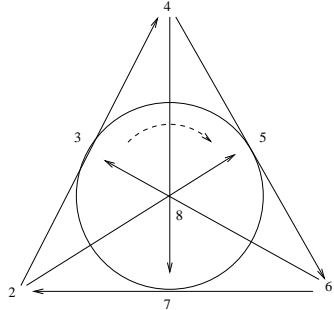}
\caption{The product of two imaginary octonions.}
\end{center}
\end{figure}
we should think the 6 oriented line segments in this diagram as being 6 oriented circles (the end point being connected to the source point by another oriented line segment), so that we actually have 7 oriented circles and seven points of intersections. Each point of intersection is labelled by $\alpha$ running from 2 to 8 and represents an imaginary element $u_\alpha$ of the basis. The product $u_\alpha u_\beta$ of two different elements $u_\alpha$ and $u_\beta$ is $\pm u_\gamma$, where $\gamma$ is the third point on the circle which contains $\alpha$ and $\beta$ and the sign $\pm$ is $+$ (resp. $-$) if the sequence $(\alpha,\beta,\gamma)$ respects (resp. does not respect) the orientation of the circle. We note that, for any $A = 1,\cdots ,4$, $Q_{aA} = (Q^\alpha_a - \sqrt{-1}Q^\beta_a)/\sqrt{2}$, where $\alpha$ and $\beta$ are chosen in such a way that $u_\alpha u_\beta = u_2$. Then we deduce from (\ref{qqbis}) that, for $a,b = 1,2$ and $A,B = 1,\cdots, 4$,
\begin{equation}\label{reduc10vers4}
\begin{array}{cc}
\left[Q_{aA},\overline{Q}_{\dot{b}B}\right] = -2 X_{a\dot{b}} \delta_{AB} & [\overline{Q}_{\dot{a}A},Q_{bB}] = -2 X_{b\dot{a}} \delta_{AB}\\
\left[Q_{aA},Q_{bB}\right] = -2\epsilon_{ab}Z_{AB} & \left[\overline{Q}_{\dot{a}A},\overline{Q}_{\dot{b}B}\right] = -2\epsilon_{\dot{a}\dot{b}}\overline{Z}_{AB},
\end{array}
\end{equation}
where the generators $Z_{AB}$ and $\overline{Z}_{AB}$ are given by 
the relations $Z_{AB} + Z_{BA} = 0$, $\overline{Z}_{AB} = \overline{Z_{AB}}^{\sqrt{-1}}$ and (setting $\mathcal{I}:= \mathcal{I}_{[12]}$ for shortness):
\[
\begin{array}{cll}
Z_{12}:= \mathcal{I}\left(-u_3 + \sqrt{-1} u_4\right) & Z_{13}:= \mathcal{I}\left(-u_6 + \sqrt{-1} u_7\right) &
Z_{14}:= \mathcal{I}\left(-u_8 + \sqrt{-1} u_5\right)\\
& Z_{23}:= \mathcal{I}\left(-u_8 - \sqrt{-1} u_5\right) &
Z_{24}:= \mathcal{I}\left(u_6 + \sqrt{-1} u_7\right)\\
& & Z_{34}:= \mathcal{I}\left(-u_3 - \sqrt{-1} u_4\right).
\end{array}
\]
We remark that, if we set $*12:= 34, *13:= 42$ and $*14:= 23$, then  $Z_{*AB} = \overline{Z}_{AB} = \overline{Z_{AB}}^{\sqrt{-1}}$. 

As shown by H. Osborn \cite{osborn} a reduction of the $N=1$ supersymmetric Yang--Mills theory on $M^{(1,9)}$ to $M^{(1,3)}$ produces an $N=4$ supersymmmetric Yang--Mills theory. The Lie algebra of Noether charges of this theory is a central extension of the pure $N=4$ supersymmetry Lie algebra and is isomorphic to $\mathfrak{m}^{(1,9)|16}$, the 6 central charges corresponding to the real and imaginary parts of the $Z_{AB}$'s. These central charges can again be interpretated as topological charges \cite{osborn}.

\section{Conclusion}
We have essentially not presented the most important thing, namely the applications of the idea of supersymmetry:
\begin{itemize}
\item
On the the physical side, its possible role in the quantum field theories for improving the renormalisation and solving the hierarchy problem (i.e. ensuring that all fundamental forces of Nature unify at very high energy, see e.g. \cite{Weinberg} III), a possible solution to the Montonen--Olive conjecture about electromagnetic duality for gauge theories by E. Witten and N. Seiberg, the theory of supergravity (see e.g. \cite{Freund} and the text by M. Eigeleh in this volume), the role of anticommuting variables in the BRST (Bechi--Rouet--Stora--Tyutin) theory of quantization of gauge theories (see e.g. \cite{Weinberg} II and \cite{KostantSternberg,Rogers07}), etc., without mentioning the theories of superstrings and branes...
\item On the mathematical side, its application by E. Witten to Morse theory (see e.g. \cite{Rogers07}), the index theorem for Dirac operators (see e.g. \cite{Witten,Alvarez,Rogers07}), the topology of 4-dimensional manifold (the Seiberg--Witten theory), the mirror symmetry for Calabi--Aubin--Yau manifolds \cite{vafa,voisin}, etc.
\end{itemize}

\end{document}